\documentclass[11pt,a4paper]{article}
\usepackage{jheppub}
\usepackage[dvipsnames]{xcolor}
\usepackage{amsfonts}
\usepackage{cancel}
\usepackage{physics}

\def\({\left(}
\def\){\right)}
\def\<{\langle}
\def\>{\rangle}
\def\cO{{\cal O}}
\def\cW{{\cal W}}
\def\cM{{\cal M}}
\newcommand{\p}{\partial}
\newcommand{\ii}{\mathrm{i}}
\def\One{1\hskip-.15cm1}
\newcommand{\nn}{\nonumber}
\usepackage{mathtools}
\usepackage[]{stackengine}
\DeclareMathOperator{\Sint}{\ensurestackMath{\stackinset{c}{0pt}{c}{0pt}{\subset}{\displaystyle\int}}}
\DeclareMathOperator{\Rint}{\ensurestackMath{\stackinset{c}{0pt}{c}{0pt}{\supset}{\displaystyle\int}}}
\newlength\correct
\renewcommand{\v}{{\rm v}}
\renewcommand{\u}{{\rm u}}

\newcommand{\beq}{\begin{equation}}
\newcommand{\eeq}{\end{equation}}
\newcommand{\beqq}{\begin{equation*}}
\newcommand{\eeqq}{\end{equation*}}
\newcommand{\la}[1]{\label{#1}}

\title{Correlators of Line Defect and Local Operator in Conformal Field Theories with a Slightly Broken Higher-Spin Symmetry}

\author[a]{Gwenaël Ferrando,}
\author[b]{Amit Sever}
\author[b]{and Elior Urisman}

\affiliation[a]{Bethe Center for Theoretical Physics, Universität Bonn, Wegelerstr. 10, D-53115, Germany}
\affiliation[b]{School of Physics and Astronomy, Tel Aviv University, Ramat Aviv 69978, Israel}

\abstract{We study three-dimensional conformal field theories with a large-$N$ limit. Leveraging the framework of slightly broken higher-spin symmetry, we bootstrap correlation functions between the single-trace, local operators and straight, conformal line defects with boundaries. These correlation functions, which depend on a single conformal cross-ratio, encapsulate all bulk-defect operator product expansion coefficients. Concentrating on the quasi-fermionic theory, we explicitly compute all correlators involving the spin-zero and spin-one conserved currents, along with an infinite family of correlators involving the higher-spin currents. Furthermore, we demonstrate that the dependence of these correlators on the defect's shape is fully determined by our bootstrap constraints.}

\begin{document}

\maketitle

\section{Introduction}

Most studies of conformal field theories (CFTs) to date have concentrated on local operators. However, CFTs also admit extended operators, such as line defects, which offer a rich structure of their own. These line defects can undergo renormalisation group (RG) flows, with their fixed points corresponding to conformal defects. Understanding the interplay between local and extended operators is therefore of significant interest.

The operator-state correspondence provides a natural bridge between these two types of operators. A compact defect can be expanded in terms of local operators, and conversely, local operators can be decomposed in terms of defect operators. The coefficients in these respective operator product expansions (OPEs) can be extracted from correlation functions involving both local and defect operators. In one OPE channel, a local operator approaches the defect, while in the other, it is taken far away from it.

In this work, we focus on a particular class of three-dimensional CFTs with a slightly broken higher-spin symmetry. These theories admit a large-$N$ limit, in which local operators naturally separate into single-trace and multi-trace classes, obeying the familiar large-$N$ factorisation. A similar structure emerges for line defects. At large $N$ they decompose into fundamental defects and those obtained via fusion of multiple fundamentals. Fundamental line defects are characterised by the factorisation of their dominant operators—at large $N$—into products of fundamental and anti-fundamental defect-changing operators, on which the fundamental defects can end. 

The prime example we have in mind is that of Chern--Simons theory, with matter in the fundamental representation. This class of CFTs is expected to be holographically dual to parity-breaking
versions of Vasiliev's higher-spin theory \cite{Vasiliev:1990en,Vasiliev:1992av,Klebanov:2002ja,Giombi:2011kc,Aharony:2011jz}. The single-trace operators in these theories are higher-spin currents, which are conserved at large $N$ and are dual to Vasiliev’s higher-spin fields in $AdS_4$. In other examples of the AdS/CFT correspondence, line operators in the fundamental representation are dual to the open-string partition function in AdS. Similarly, here, we expect the fundamental conformal lines to play a central role. In order to shed light on their relation to Vasiliev's higher-spin fields, we study their correlation functions with the local currents.

In terms of vector fields, the single-trace operators take the form $J=\bar\varphi_i\varphi_i$, with derivatives distributed among the constituent fields. The slightly broken higher-spin symmetry implies that $\p J=O(1/N)$. In \cite{Maldacena:2012sf}, this constraint was used to fix the three-point correlation functions of these operators in terms of a single parameter, $\tilde\lambda$. Similarly, defect operators factorise as $\varphi\times\bar\varphi$, again with derivatives inserted. In this case, the vector indices are contracted along the line. We refer to operators such as $\varphi$ and $\bar\varphi$ as boundary operators.

An infinite straight conformal line preserves an $SL(2,{\mathbb R})\times U(1)$ subgroup of the full conformal group. This residual symmetry is used to classify both the defect operators and the boundary operators. The slightly broken higher-spin symmetry implies the existence of so-called tilt operators on the defect, whose $SL(2,{\mathbb R})$ dimensions and transverse $U(1)$ spins remain protected at leading order in $1/N$. The spectrum of boundary operators has been bootstrapped in \cite{Gabai:2022vri,Gabai:2022mya,Gabai:2023lax} and is reviewed below. It is characterised by one continuous parameter $\Delta\in[0,1]$, and is uniquely parameterised by the transverse spin. In our convention for the sign of transverse spin, a boundary operator located at the right end point of a line along the $\hat x^3$ direction has transverse spin ${\mathfrak s}_s=s+1-\Delta$, where $s\in{1\over2}+\mathbb Z$. We denote these boundary operators as $\cO_s$, with the anti-fundamental counterparts labeled bar $\overline{\cO}_{-s}$ sharing the same spectrum.

In this paper, we combine the large-$N$ structure of single-trace operators with the properties of the conformal line defect to bootstrap their correlation functions. We focus on the quasi-fermionic theory, in which the dimension of the spin-zero current $J_0$ is 2. However, our methods are readily applicable to the quasi-bosonic case, where $\dim J_0=1$. The correlators of interest involve a conserved higher-spin current $J_{\tilde s}$ in the bulk and a straight conformal line $\cW$, stretched between fundamental and anti-fundamental boundary operators
\beq\la{corr}
\includegraphics[scale=1.5]{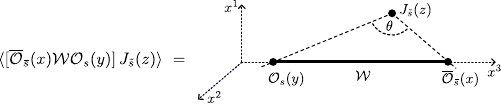}
\eeq
Here, $\tilde s=0,1,2,3,\ldots$ denotes the $SO(3)$ spin of the local bulk current. When $\tilde{s}=0$, the correlator, after factoring out a simple kinematical term accounting for conformal weights and spins, depends only on a single 
$SL(2,{\mathbb R})\times U(1)$-invariant quantity, $\cos\theta$. For $\tilde s\geqslant 1$, multiple tensor structures contribute, leading to multiple functions of the conformal invariant. We show that all such functions can be completely bootstrapped using only the large-$N$ constraints. In particular, we recover the known relation between the defect parameter $\Delta$ and the bulk parameter $\tilde\lambda$.

Concretely, we compute all correlation functions involving either $J_0$ or $J_1$ and a straight mesonic line with arbitrary boundary operator. As the spin of the bulk current increases, the correlators become increasingly intricate. When the transverse spins of the boundary operators have the same sign, we find a compact closed-form expression for the correlator. For opposite transverse spins and general bulk spin, we argue that the correlators can still be fixed in principle, though we do not find a general closed-form expression in this case.

\section{Setup}
\label{sec:setup}

Here we review the large-$N$ properties of the operators that enter the correlators we study. These are the bulk single-trace operators, the conformal lines, and the operator on which the lines end.

\subsection{The Higher-Spin Currents}

We assume that the spectrum of primary single-trace operators, $J_{\tilde s}$, consists of symmetric traceless tensors of spin $\tilde s$ for any integer spin, $\tilde s=0,1,2,3,\ldots$. The operators with $\tilde s>0$ have twist equal to one and are conserved currents, up to $O(N^{-1})$ corrections. In the quasi-bosonic theory, the twist of the scalar operator $J_0$ is also equal to one, while in the quasi-fermionic theory it is equal to two. In this paper, we have chosen to focus on the latter. In CS theory coupled to fermions, the scalar operator takes the form $J_0=\bar\psi_i\psi_i$. 

There are also theories with only even higher-spin symmetries. We expect that the correlators in these theories can be obtained from those we derive here by projecting to charge-conjugation-invariant combinations.\footnote{We thank O. Aharony for a discussion on this topic.}

\subsection{Fundamental Conformal Line Defect at Large $N$}
\label{sec:norm}

A conformal line defect along a smooth path is a line operator $\cW$ that transforms trivially under conformal transformations. Namely, 
\beq
\cW[x(\cdot)]\ \mapsto\ \cW[\tilde x(\cdot)]\,,
\eeq
where $x(\cdot)$ is some parameterisation of the path and $\tilde x(\cdot)$ is its image under the conformal transformation. In particular, a straight conformal line preserves an $SL(2,{\mathbb R})\times U(1)$ symmetry. We denote by $\Delta_\cO$ the $SL(2,{\mathbb R})$ dimension of an operator $\cO$ on the line and by $\mathfrak s_\cO$ its $U(1)$ transverse spin. The sign of the transverse spin is defined such that the spin of $x^\pm = (x^1\pm\ii x^2)/\sqrt 2$ is equal to $\pm1$ for a line that is oriented in the $\hat x^3$ direction.

A fundamental conformal line is a conformal line whose operators factorise as
\beq\label{largeN2}
\cO_{\text{line}}=\cO\times\overline\cO+O(1/N)\,,
\eeq
at large $N$. Here, $\cO$ and $\overline{\cO}$ are boundary fundamental and anti-fundamental operators. That is, these are defect-changing operators that interpolate between the oriented line defect and the trivial defect. Both of these lines are invariant under $SL(2,{\mathbb R})\times U(1)$ and therefore the boundary operators are also characterised by their $SL(2,{\mathbb R})$ dimension and their transverse spin.

As mentioned in the introduction, the higher-spin symmetry in the bulk implies the existence of tilt operators on the line. These are operators with integer $SL(2,{\mathbb R})$ dimensions and twist equal to one. The lowest in this family is the displacement operator $\mathbb D$, which exists under mild assumptions for any conformal line operator in a CFT. It has dimension two and transverse spin one. In our case it factorises into a product of fundamental and anti-fundamental boundary operators as in (\ref{largeN2}). This factorisation implies that the dimensions and transverse spins of the boundary operators are related as $\Delta_\cO+\Delta_{\bar\cO}=2$ and $|{\mathfrak s}_\cO+{\mathfrak s}_{\bar\cO}|=1$, \cite{Gabai:2022mya}. Further study of these operators in CS-matter theories has shown that for one component of ${\mathbb D}$ we have $(\Delta_\cO,{\mathfrak s}_\cO)=(\Delta,{1\over2}-\Delta)$ and $(\Delta_{\overline{\cO}},{\mathfrak s}_{\overline{\cO}})=(\Delta,\Delta-{1\over2})$ for the other. This relation can be derived algebraically by embedding the theory in a supersymmetric one such that in the large-$N$ limit the extra degrees of freedom decouple and one remains with the original theory only, \cite{Gabai:2022vri,Gabai:2022mya}. 

Based on this information, the spectrum of primary operators has been bootstrapped in \cite{Gabai:2023lax}. It is parameterised by $s\in{\mathbb Z}+{1\over2}$ and is given by
\beq\la{spectrumR}
(\Delta_s,\mathfrak{s}_s)=\left\{\begin{array}{lcl}\(\Delta-s-{1\over2},s-\Delta+1\)&\quad&s<0\\\(s-\Delta+{3\over2},s-\Delta+1\)&\quad&s>0\end{array}\right.,\qquad s\in{\mathbb Z}+{1\over2}\,.
\eeq
for either the fundamental or anti-fundamental operators. Here, the transverse spin is for an operator at the end of a line that extends from it towards the positive $\hat x^3$ direction. If the line extends from it towards the negative $\hat x^3$ direction then its sign is flipped. In what follows, we focus on a straight conformal line along the $\hat x^3$ direction, and we place the fundamental operator before the anti-fundamental one in that direction. We denote these operators by $\cO_s$ and $\overline\cO_{-s}$ respectively. In this notation, the displacement operator reads
\begin{align}\la{displace}
    \mathbb{D}_+ &= \eta_+\, \cO_{+\frac12}\times\overline \cO_{+\frac12}+{\nu_+\over N{\cal N}_1}J_{1,+}+O(1/N^2)\,,\\
    \mathbb{D}_- &= \eta_-\, \cO_{-\frac12}\times\overline \cO_{-\frac12}+{\nu_-\over N\mathcal{N}_1}J_{1,-}+O(1/N^2)\,,\nn
\end{align}
where $\eta_\pm$ and $\nu_\pm$ depend on our choice of normalisation of the operators and $\mathcal{N}_1$ is defined in \eqref{normalization Js}.\footnote{The coefficients $\sqrt{{\cal N}_1}\nu_\pm$ encode the two possible ways in which $J_1$ can appear in ${\mathbb D}^i$, $J_1^i$ and ${\epsilon^{3i}}_jJ_1^j$.} The second term of the displacement operator stands for a factorised (normal ordered) product of the line and $J_1$ at the point on the line. Even though this term is of order $1/N$, it contributes at leading order to the correlator of $\mathbb D$ with $J_1$. We will show later that $\nu_+ = -\nu_-$.

In the quasi-bosonic theory, $J_0$ has dimension one and therefore also the operator $\p_\pm J_0$ on the line has dimension two and transverse spin one. However, as opposed to $J_1$, $\p_\pm J_0$ does not transform as a primary under conformal transformations that do not preserve the line and therefore cannot appear in (\ref{displace}).

The displacement operator has a canonical normalisation such that under a small deformation, $x\to x+{\rm v}$, the line transforms as
\beq
\delta \mathcal{W} =\int \mathrm{ds} \,|\dot x(s)|\, \mathrm{v}^\mu (s) \, \mathbb{D}_\mu(x(s))\,\cW\,,
\eeq
where $x(\tau)$ is some parameterisation of the path and $\mathbb D$ is inserted at $x(\tau)$ along the line. Similarly, we can also displace the boundary operators
\begin{equation}\label{boundaryvar}
    \delta \cO = \mathrm{v}^\mu \delta_\mu \cO + \mathrm{dv}^\mu \delta_\mu^{(1)} \cO + \mathrm{ddv}^\mu \delta^{(2)}_\mu \cO + \dots\,,
\end{equation}
where $\mathrm{dv}= \mathrm{\dot{v}}/|\dot x|$ and $\mathrm{d^nv}$ are higher-order reparameterisation-invariant derivatives of $\mathrm{v}$ along the path, see \cite{Gabai:2023lax} for more details.

We denote a straight line stretched along the $\hat x^3$ axis by $\cM=\overline O\cW\cO$ or in more details, by
\begin{equation}
    \cM^{(\bar s,s)}_{\tau\sigma} = \overline\cO_{\bar s}(x_\tau)\,\cW\,\cO_s(x_\sigma)\, ,
\end{equation}
where $x_\tau=(0,0,\tau)$ and $\tau>\sigma$.

We assume that the OPE expansion of $\cM$ in terms of the local operators is dominated by the single-trace ones, i.e. that the multi-trace OPE coefficients are suppressed by additional powers of $1/N$. It follows that the correlation functions of $\cM$ and the local operators obey the same large-$N$ factorisation as that of the single-trace operators.  

Finally, note that the spectrum of boundary operators (\ref{spectrumR}) is invariant under interchanging $\Delta$ with $2-\Delta$ together with a flip of the transverse spin ${\mathfrak s}\to-{\mathfrak s}$, or equivalently a parity transformation that exchanges $x^+\leftrightarrow x^-$. The combined transformation exchanges
\beq\label{Deltasym}
    \Delta\ \rightarrow\ 2-\Delta\,,\qquad s\ \rightarrow\ -s\,.
\eeq

\subsection{Bootstrapping Strategy}

The conformal covariance of the boundary and bulk operators fixes the form of the correlators (\ref{corr}) up to a few functions of a single conformal cross-ratio. To bootstrap these functions we first impose the invariance of the correlators under translations in the directions transverse to the line, which follows from the conservation of the stress-energy tensor $J_2$. This constraint leads to a system of differential equations for the functions of the cross-ratio. They fix all correlators with a given $J_{\tilde{s}}$ up to a single coefficient $d_{\tilde{s}}$.

To relate the correlators with different $J_{\tilde s}$ we use the non-conservation of $J_3$ at order $1/N$, in the spirit of \cite{Maldacena:2012sf}. It involves double-trace operators and is of the form $\p J_3 \sim \, :\!JJ\!:/N$. We determine the explicit form of the divergence and use the large-$N$ factorisation property to derive new constraints that mix between correlators with different currents which relate all the coefficients $d_{\tilde{s}}$.

\subsection{Comments on Normalisation} \label{setup:norm}

We use the following normalisation of the currents
\begin{equation}\label{normalization Js}
    \<J_0(x) J_0(0)\> = \frac{N\mathcal{N}_0}{|x|^4}\quad \text{and}\quad \<J^{+\dots +}_{\tilde s}(x) J^{+\dots +}_{\tilde s}(0)\> = N \mathcal{N}_{\tilde s}\, \p^{2\tilde s}_{x^-}\frac1{|x|^2}\quad \text{for}\quad\tilde s\geqslant 1\,,
\end{equation}
where the normalisation constants $\mathcal{N}_{\tilde s}$ are of order $O(N^0)$. For the stress-energy tensor, $J_2$, we choose the canonical normalisation, so that the momentum generator is $P^\mu = \int \dd^2 S_\nu J_2^{\mu\nu}$. \footnote{This is the same normalisation that was used in \cite{Maldacena:2012sf}, and their parameter $\tilde{N}$ is given by $\tilde{N} = 2^{10} \pi^2 N \mathcal{N}_2$.} Similarly, the current $J_1$ can be associated with a $U(1)$ symmetry and we choose a canonical normalisation for it, such that the boundary operators have charge $\pm 1$.

For the boundary operators, we relate the normalisation of all operators with $|s|>1/2$ to those of $\overline\cO_{\pm\frac12}$ and $\cO_{\pm\frac12}$ through
\begin{equation}\label{boundary tower}
    \overline\cO_{\bar s} = \begin{cases}
\delta^{\bar s - \frac12}_+\overline\cO_{\frac12} &\quad \bar s > 0\\
\delta^{-\bar s-\frac12}_-\overline\cO_{-\frac12} & \quad \bar s< 0
\end{cases}\,,
\end{equation}
and similarly for $\cO_s$. Here, $\delta_\mu \cO$ stands for the {\it path derivative} of the boundary operator $\cO$; it is defined in \eqref{boundaryvar}. In this normalisation it follows that \cite{Gabai:2022vri}
\begin{align}\label{boundary eom}
    \delta_- \overline\cO_{\bar s +1} &= -\frac12\delta_3^2\overline\cO_{\bar s} + O(1/N)\qquad\bar s>0\,,\\
    \delta_+ \overline\cO_{\bar s -1} &= -\frac12\delta_3^2\overline\cO_{\bar s} + O(1/N)\qquad\bar s < 0\,,
\end{align}
and similarly for $\delta_{\pm}\cO_{s\mp 1}$. Here, the $1/N$ corrections are double-trace operators involving any of the bulk currents. As an example, \eqref{boundary eom} contains $:\!(\p_+^{\bar s+k} J^3_{1}) \overline{\mathcal{O}}_{-k}\!:$ for $1-\bar s<k<0$. Using these equations, one can relate the expectation values of various mesonic lines. For instance, for $\bar s >0$ we have 
\begin{equation}
    \delta_- \<\cM_{10}^{(\bar s + 1,-\bar s)}\> = 0 \qquad\Rightarrow \qquad\p^2_{x_1^3} \<\cM_{10}^{(\bar s,-\bar s)}\> = 2 \<\cM_{10}^{(\bar s+1,-\bar s-1)}\>\,.
\end{equation}
The solution of this recursion relation gives the expectation value of a straight mesonic line operator as
\begin{equation}\label{normalization M}
    \<\cM_{10}^{(\bar s,-\bar s)}\> = \frac{c_+\,2^{\frac12-\bar{s}}\,\Gamma(2\Delta_{-\bar s})}{\Gamma(2\Delta)\,  |x_1-x_0|^{2\Delta_{-\bar s}}} \,,\qquad\<\cM_{10}^{(\bar s,-\bar s)}\> = \frac{c_-\,2^{\frac12+\bar{s}}\,\Gamma(2\Delta_{-\bar s})}{\Gamma(4-2\Delta)\,  |x_1-x_0|^{2\Delta_{-\bar s}}}\,,
\end{equation}
for $\bar s>0$ and $\bar s<0$ respectively, in accordance with the symmetry (\ref{Deltasym}). Note that these expectation values are consistent with the derivatives of $\<\cM_{10}^{(1/2,-1/2)}\>$ and $\<\cM_{10}^{(-1/2,-1/2)}\>$ respectively, as if they were the derivatives of two point functions of local operators.

In this normalisation convention, the two-point function of the displacement operator, which was bootstrapped in \cite{Gabai:2022mya,Gabai:2022vri}, is given by
\begin{equation}\label{2pt displacement}
    \eta_-\eta_+\, c_+\, c_- = -\frac1{2\pi}(2\Delta - 3)(2\Delta - 2)(2\Delta - 1)\sin(2\pi\Delta)\,.
\end{equation}

\section{Correlators with $J_0$}
\label{sec:J0}

In this section, we bootstrap the correlators $\<\cM_{10}^{(\bar s,s)}\,J_0(x_2)\>$ for arbitrary boundary spins. We first use covariance under $\mathrm{SL}(2,\mathbb{R})\times \mathrm{U}(1)$ to determine each of the correlator up to a single function $f_{\bar s,s}$ of one variable. We then study the invariance of the correlator under translations in a transverse direction
\begin{equation}\label{variations J0}
    \delta_{\pm}\<\cM_{10}^{(\bar s,s)}\,J_0(x_2)\> = 0\,.
\end{equation}
We find that this constraint is sufficient to fix the functions $f_{\bar s,s}$ for every $s$ and $\bar s$.

\subsection{Constraints from Conformal Covariance}

Covariance under $\mathrm{SL}(2,\mathbb{R})\times \mathrm{U}(1)$ fixes the correlator to be of the form\footnote{Here, we have used that $x_2^\pm=x_{21}^\pm=x_{20}^\pm$. We also recall that we always assume $x_0^3 < x_1^3$.}
\begin{equation}\label{corrsbars}\<\cM_{10}^{(\bar s,s)}\,J_0(x_2)\> = \frac{\left(\frac{x_2^\pm |x_{10}|}{|x_{20}| |x_{21}|}\right)^{|\bar s + s|} }{|x_{10}|^{\Delta_{-\bar s} + \Delta_{s} - 2} |x_{20}|^{2 + \Delta_{s} - \Delta_{-\bar s}} |x_{21}|^{2 + \Delta_{-\bar s} - \Delta_{s}}}\times f_{\bar s,s}(\cos\theta)\,,
\end{equation}
where $x_{ij}=x_i-x_j$, $\pm = -\operatorname{sign}(\bar s + s)$ and $\theta$ is the $\mathrm{SL}(2,\mathbb{R})\times \mathrm{U}(1)$-invariant angle defined by
\begin{equation}\la{ccr}
    \cos\theta =u= \frac{x_{20}\cdot x_{21}}{|x_{20}||x_{21}|}\,,\qquad \sin^2\!\theta=1-u^2=\frac{2\,x_2^+x_2^- \left|x_{10}\right|^2}{\left|x_{21}\right|^2\left|x_{20}\right|^2}\,.
\end{equation}

Let us elaborate on (\ref{corrsbars}). The prefactor carries the same $SL(2,{\mathbb R})\times U(1)$ weights as those of the operators on the left-hand side. Here we have made use of the fact that the anomalous spin cancels between the left and right sides, so ${\mathfrak s}+\bar{\mathfrak s}=s+\bar s$. Therefore, $f_{\bar s,s}$ is a function of $\mathrm{SL}(2,\mathbb{R}) \times \mathrm{U}(1)$-invariant quantities. To see that there is only one independent invariant, $\cos\theta$, we may first use the $\mathrm{SL}(2,\mathbb{R})$ symmetry to fix $x_0$ and $x_1$. There remains one $\mathrm{SL}(2,\mathbb{R})$ generator that leaves these two points invariant. We then use it to set $|x_2\cdot x_{10}/x_{10}^2|=1/2$ for instance. This way, the positions of $x_0$, $x_1$ and the projection of $x_2$ to the line are all fixed. The angle of $x_2$ in the transverse plane can be fixed using the $U(1)$ symmetry, and only its radial position remains arbitrary. Another way of parameterising the conformal invariant is to take the third direction to be time-like. Then consider the one-dimensional cross-ratio built out of the points $x_0$, $x_1$ and the two intersections between the light cone of $x_2$ and the line.\footnote{Labeling these two points by $x_3$ and $x_4$, the relation between the conformal invariants is $1-u^2=(4z(1-z))^{-1}$, where $z=x_{03}\cdot x_{14}/x_{01}\cdot x_{34}$ is the one-dimensional conformal cross-ratio.} The correlator (\ref{corrsbars}) depends on our normalisation of the bulk scalar and the boundary operators. A normalisation-independent correlator can be obtained by taking ratios.

In the following subsections, we determine $f_{\bar s,s}(\cos\theta)$. We first consider the invariance of the correlator under variations \eqref{variations J0} that do not involve insertions of the displacement operator on the line. It further fixes them in terms of four constants, $f_{\pm\frac12,\pm'\frac12}(1)$. In the last subsection, we consider variations that involve the displacement operator to relate these four constants. To test our result, we perform an explicit one-loop computation of the correlators with $|s| = |\bar s| = \pm1/2$ in appendix \ref{pertapp}. We find perfect agreement with our bootstrap results. 

Note that CPT is a symmetry of the theory that maps
\beq\la{flipOOb}
\overline\cO_{\bar s}\,\cW\,\cO_s\quad\longrightarrow\quad(-1)^{s+\bar s}\ \overline\cO_{- s}\,\cW\,\cO_{-\bar s}\,.
\eeq
Because the spectrum of the fundamental and anti-fundamental operators is the same, it follows that any bootstrap solution for $\<\cM_{10}^{(\bar s,s)}\,J_0(x_2)\>$ is also a solution for $\<\cM_{01}^{(- s,-\bar s)}\,J_0(x_2)\>$. Adding to this that $\cos\theta$ is invariant under interchanging $x_0\leftrightarrow x_1$, and that the rotation (\ref{flipOOb}) interchanges the $x^+$ and the $x^-$ directions, we get 
\beq\la{sflip}
f_{\bar s, s}(u)\propto f_{- s,-\bar s}(u)\,.
\eeq
where the proportionality constant depends on our normalisation of the operators. Hence, without loss of generality, we can assume that $s+\bar s\geqslant 0$. Additionally, using the symmetry (\ref{Deltasym}) we can also assume that $\bar s>0$.

\subsection{Spins of the Same Sign}

If $s$ and $\bar s$ are both positive, then the correlators in \eqref{corrsbars} simplify to
\begin{equation}\la{equalspins}
    \<\cM_{10}^{(\bar s,s)}\,J_0(x_2)\> = \frac{(x_2^-)^{\bar s + s} |x_{10}| f_{\bar s,s}(u)}{|x_{20}|^{4-2\Delta + 2s} |x_{21}|^{2\Delta + 2\bar s}}\,.
\end{equation}
We impose that these correlators be invariant under translation in the plus direction. The linear variation of the correlator takes the form
\begin{equation}\la{plusshift}
    \delta_+\<\cM_{10}^{(\bar s,s)}\,J_0\> = \<\cM_{10}^{(\bar s + 1,s)}\,J_0\> + \<\cM_{10}^{(\bar s,s + 1)}\,J_0\> + \<\cM_{10}^{(\bar s,s)}\,\p_+J_0\> = 0\,,
\end{equation}
where we have used the fact that the displacement operator $\mathbb{D}_+$ does not contribute with this choice of boundary spins. That is, due to large-$N$ factorisation, the insertion of the displacement operator factorises as 
\begin{align}\la{displacementinsert}
\<[\overline\cO_{\bar s}\,\cW\,\mathbb{D}_+\,\cW\,\cO_s]\, J_0\>\propto\,&\<[\overline\cO_{\bar s}\,\cW\,\cO_{1\over2}]\,[\overline\cO_{1\over2}\,\cW\,\cO_s]\, J_0\>\\
=&\<\overline\cO_{\bar s}\,\cW\,\cO_{1\over2}\>\times\<[\overline\cO_{1\over2}\,\cW\,\cO_s]\, J_0\>+\<[\overline\cO_{\bar s}\,\cW\,\cO_{1\over2}]\,J_0\>\times\<\overline\cO_{1\over2}\,\cW\,\cO_s\>\nn
\end{align}
and for $s,\bar s>0$
\beq
\<\overline\cO_{\bar s}\,\cW\,\cO_{1\over2}\>=\<\overline\cO_{1\over2}\,\cW\,\cO_s\>=0\,.
\eeq
The constraint (\ref{plusshift}) translates to the following three equations:
\begin{equation}
f'_{\bar s,s}(u) = 0\,,\quad f_{\bar s +1,s} = 2(\Delta+\bar s) f_{\bar s,s}\,,\qquad f_{\bar s,s+1} = 2(2-\Delta+s) f_{\bar s,s}\,.
\end{equation}
They imply that $f_{\bar s>0,s>0}(u)=f_{\bar s,s}$ are all constants and are given by
\begin{equation}
    f_{\bar s,s}= 2^{\bar s + s - 1} \frac{\Gamma(\Delta + \bar s) \Gamma(2 -\Delta + s)}{\Gamma(\Delta + \frac12) \Gamma(\frac52 -\Delta)} f_{\frac12,\frac12}\,.
\end{equation}
The correlator \eqref{equalspins} can be written compactly as
\begin{equation}\label{ssbpositive J0}
    \<\cM_{10}^{(\bar s>0,s>0)}\,J_0(x_2)\> = f_{\frac12,\frac12} \,x_2^- |x_{10}|\, \p^{\bar s - \frac12}_{x^+_1} \p^{s - \frac12}_{x^+_0} \frac1{|x_{20}|^{5-2\Delta} |x_{21}|^{2\Delta+1}}\,.
\end{equation}

Using the symmetry \eqref{Deltasym}, the correlator with negative $s$ and $\bar s$ reads
\begin{equation}\label{ssbnegative J0}
    \<\cM_{10}^{(\bar s<0,s<0)}\,J_0(x_2)\> = f_{-\frac12,-\frac12}\,x_2^+ |x_{10}|\, \p^{-\bar s - \frac12}_{x^-_1} \p^{-s - \frac12}_{x^-_0} \frac1{|x_{20}|^{2\Delta+1} |x_{21}|^{5-2\Delta}}\,.
\end{equation}

These correlators are the same as the correlators between $J_0$ and two local primary operators of dimensions $(\Delta_{-\bar s},\Delta_s)$ and spins $(|\bar s|,|s|)$, with their projection in the third direction given by $(\bar s,s)$.\footnote{Such correlators contain several tensor structures that either vanish or are proportional to each other in the kinematical configuration we are considering.} This is because all the constraints that we have imposed to fix the correlators also apply to these three-point functions of local operations. Note that the transverse spins, $\mathfrak s$ and $\bar{\mathfrak s}$, are neither integers nor half-integers. Such spins cannot be attributed to local operators. However, their anomalous part, $\mathfrak s-s$ and $\bar{\mathfrak s}-\bar s$ have canceled in (\ref{corrsbars}). In particular, (\ref{ssbpositive J0}) and (\ref{ssbnegative J0}) do not depend on the conformal invariant $u$. One implication of this equality with the three-point function of local operators is that the bulk-line OPE is trivial.

\subsection{Spins of Opposite Signs}

Recall that without loss of generality, we can assume that $\bar s + s \geqslant 0$ and $\bar s>0$. Hence, in this case $s<0$ and the correlator (\ref{corrsbars}) reads
\begin{equation}\label{correl total spin pos}
    \<\cM_{10}^{(\bar s,s)}\,J_0(x_2)\> = \frac{(x_2^-)^{\bar s + s} f_{\bar s,s}(u)}{|x_{10}|^{2\Delta -2s - 3} |x_{20}|^2 |x_{21}|^{2 + 2\bar s + 2s}}\,.
\end{equation}
Requiring $\delta_-\<\cM_{10}^{(\bar s + 1,s)}\,J_0(x_2)\> = 0$, i.e. that the correlator be invariant under translation in the minus direction, we arrive at
\begin{equation}\la{minusvariation}
   \<\delta_- \overline{\mathcal{O}}_{\bar s + 1}(x_1) \mathcal{W} \mathcal{O}_s(x_0) \,J_0(x_2)\> + \<\cM_{10}^{(\bar s + 1,s - 1)}\,J_0(x_2)\> + \<\cM_{10}^{(\bar s + 1,s)}\,\p_-J_0(x_2)\> = 0\,,
\end{equation}
where the displacement operator contribution, analogous to (\ref{displacementinsert}), still vanishes. The variation of the left boundary operator contains two terms that contribute to this variation
\begin{equation}\label{eom}
\delta_- \overline{\mathcal{O}}_{\bar s + 1} = -\frac12\delta_3^2\overline{\mathcal{O}}_{\bar s} + \frac{\alpha_{\bar s,s}}{N} :\!(\p_+^{\bar s + s} J_0) \overline{\mathcal{O}}_{-s}\!\!: + \dots\,.
\end{equation}
The first term is the ``line equations of motion" that was bootstrapped in \cite{Gabai:2022vri}. The second term is a product of a single-trace and an anti-fundamental operator. It is analogous to a double-trace contribution and is allowed because it has the correct dimension and transverse spin. The coefficient $\alpha_{\bar s,s}$ is of order $O(N^0)$. Equation (\ref{minusvariation}) thus becomes
\begin{multline}\la{ssbarg0}
    -\frac12\p_{x_1^3}^2 \<\cM_{10}^{(\bar s,s)}\,J_0(x_2)\> + \frac{\alpha_{\bar s,s}}{N} \<\p_+^{\bar s + s} J_0(x_1)\,J_0(x_2)\> \<\cM_{10}^{(-s,s)}\> \\
    + \<\cM_{10}^{(\bar s + 1,s - 1)}\,J_0(x_2)\> + \<\cM_{10}^{(\bar s + 1,s)}\,\p_-J_0(x_2)\> = 0\,.
\end{multline}
The first term takes the form
\begin{align}
&-\frac12\p_{x_1^3}^2 \<\cM_{10}^{(\bar s,s)}\,J_0(x_2)\>\\
&= (x^-_2)^S\Bigg[\frac{(\Delta + \bar s) (1-2\Delta -2\bar s) f_{\bar s,s}}{|x_{10}|^{2\Delta - 2s - 1} |x_{20}|^2 |x_{21}|^{2 + 2S}} + \frac{2(\Delta + \bar s) [2(S+1) u f_{\bar s,s} - (1-u^2)f'_{\bar s,s}]}{|x_{10}|^{2\Delta - 2s - 1} |x_{20}| |x_{21}|^{3 + 2S}}\nn\\
&\qquad\qquad+ \frac{2(S+1)(1-2(S+2)u^2) f_{\bar s,s} + (4S+7) u(1-u^2) f'_{\bar s,s} - (1-u^2)^2 f''_{\bar s,s}}{2\,|x_{10}|^{2\Delta - 2s - 1} |x_{21}|^{4 + 2S}}\Bigg]\,,\nn
\end{align}
where we have used the shorthand notation $S = \bar s + s$, and we have omitted the argument $u$ of the functions. Similarly, we have
\begin{multline}
    \<\cM_{10}^{(\bar s + 1,s)}\,\p_-J_0(x_2)\> = (x^-_2)^S\Bigg[\frac{2(S+u^2) f_{\bar s+1,s} - u(1-u^2) f'_{\bar s+1,s}}{2\,|x_{10}|^{2\Delta - 2s - 1} |x_{20}|^2 |x_{21}|^{2 + 2S}}\\
    + \frac{(1-u^2) f'_{\bar s+1,s} - 2(S+1)u f_{\bar s+1,s}}{|x_{10}|^{2\Delta - 2s - 1} |x_{20}| |x_{21}|^{3 + 2S}}  + \frac{2((S+2)u^2-1)) f_{\bar s+1,s} - u(1-u^2) f'_{\bar s+1,s}}{2\, |x_{10}|^{2\Delta - 2s - 1} |x_{21}|^{4 + 2S}}\Bigg]\,,
\end{multline}
and
\begin{equation}
    \frac{\alpha_{\bar s,s}}{N} \<\p_+^{\bar s + s} J_0(x_1)\,J_0(x_2)\>\<\cM_{10}^{(-s,s)}\> = \frac{\tilde\alpha_{\bar s,s} (x^-_2)^S}{|x_{10}|^{2\Delta - 2s - 1} |x_{21}|^{4 + 2S}}
\end{equation}
where $\tilde\alpha_{\bar s,s}$ is a constant that can be expressed in terms of $\alpha_{\bar s,s}$ and the normalisation factors in \eqref{normalization Js} and \eqref{normalization M}. After plugging these expressions into \eqref{ssbarg0} we find three different functional structures and correspondingly three constraints. From the terms proportional to $|x_{20}|^{-1}$, we find that $g_{\bar s,s}\equiv f_{\bar s+1,s} - 2(\Delta + \bar s) f_{\bar s,s}$ satisfies
\begin{equation}
    (1-u^2)\,g_{\bar s,s}' = 2(S+1)\,u\,g_{\bar s,s}\,,
\end{equation}
whose general solution is $g_{\bar s,s}(u)\propto(1-u^2)^{-S-1}$. This function is singular as $u\to1^-$, which is the limit where $J_0$ approaches the line outside of the segment $[x_0,x_1]$. The powers of $x_2$ that appear in this OPE limit are all non-negative and therefore $g_{\bar s,s}$ must be regular there. Hence, $g_{\bar s,s}=0$ and
\begin{equation}\label{sbps}
    f_{\bar s+1,s}(u) = 2(\Delta + \bar s) f_{\bar s,s}(u)\,.
\end{equation}

Injecting this relation into the constraint that comes from the terms proportional to $|x_{20}|^0$, we obtain
\begin{multline}
    \frac12(1-u^2)^2f_{\bar s,s}'' + (\bar s + 2s + 7/2- \Delta) u(u^2-1) f_{\bar s,s}'\\
    +\left[2(S+2)(1+s-\Delta)u^2 + 2\Delta + \bar s - s - 1\right] f_{\bar s,s} = \tilde\alpha_{\bar s,s}\,.
\end{multline}
A generic solution to this second order differential equation behaves as $(1-u)^{-S-1}$ when $u\to 1^-$. Setting the coefficient of this term to zero, there remains
\begin{equation}\label{sbs}
    f_{\bar s,s}(u) = A_{\bar s,s}\, u (1-u^2)^{\Delta - s -\frac32} + f_{\bar s,s}(1)\, {}_2 F_1\!\big(2,2+2s-2\Delta;5/2+s-\Delta; (1-u)/2\big)\,,
\end{equation}
where $A_{\bar s,s}$ and $f_{\bar s,s}(1) = \tilde\alpha_{\bar s,s}/[(S+1)(2s+3-2\Delta)]$ are undetermined constants. For $\Delta < s + 3/2$ we must set $A_{\bar s,s} = 0$ so that the correlator remains finite when $u\to 1^-$. We will later show that $A_{\bar s,s} = 0$, independently of the value of $\Delta$. Note that the dependence of the solution on $\bar s$ only enters the overall normalisation dependent factor $f_{\bar s,s}(1)$. On the other hand, the solution depends on $s$ in a non-trivial way through the hypergeometric function.\footnote{This asymmetry is not in contradiction with the bootstrap symmetry that interchanges $(\bar s,s)$ with $(-s,-\bar s)$ because we have assumed that $\bar s+s>0$, so $|s|\leqslant|\bar s|$.}

Finally, the third constraint results from the terms proportional to $|x_{20}|^{-2}$ in (\ref{ssbarg0}). It reads
\begin{equation}\label{sbpsm}
    f_{\bar s + 1,s-1}(u) =(\Delta - s - 1/2 - u^2)f_{\bar s + 1,s}(u) + \frac12u(1-u^2)f'_{\bar s + 1,s}(u)\,.
\end{equation}
Since equations \eqref{sbps}, \eqref{sbs}, and \eqref{sbpsm} are valid for arbitrary $\bar s> 0$ and $s<0$ such that $\bar s + s\geqslant 0$, we can relate all the $A_{\bar s,s}$ and $f_{\bar s,s}(1)$ to their values at $(\bar s,s) = (1/2,-1/2)$. We find that 
\beq \la{relationsforAandf}
{A_{\bar s,s}\over A_{\frac12,-\frac12}} = 2^{\bar s - \frac12} \frac{\Gamma(\Delta + \bar s) \Gamma(\Delta - s)}{\Gamma\left(\Delta + \frac12\right)^2} \,,\qquad {f_{\bar s,s}(1)\over f_{\frac12,-\frac12}(1)} = 2^{\bar s - \frac12} \frac{\Gamma(\Delta + \bar s) \Gamma(\Delta - s - \frac32)}{\Gamma\left(\Delta + \frac12\right) \Gamma(\Delta  - 1)}\,.
\eeq

Using (\ref{flipOOb}), the correlator with $\bar s + s \leqslant 0$ reads
\begin{equation}
    \<\cM_{10}^{(\bar s,s)}\,J_0(x_2)\> = \frac{(x_2^+)^{-\bar s - s} f_{-s,-\bar s}(u)}{|x_{10}|^{2\Delta +2\bar s - 3} |x_{20}|^{2 - 2\bar s - 2s} |x_{21}|^2}\,,\qquad\bar s + s \leqslant 0\,,\quad\bar s>0\,.
\end{equation}

Similarly, using (\ref{Deltasym}), we find that $f_{\bar s<0,s>0}(u)$ is given by $f_{-\bar s,-s}(u)$ up to $\Delta\to2-\Delta$, $f_{\frac12,-\frac12} \to f_{-\frac12,+\frac12}$, and $A_{\frac12,-\frac12} \to A_{-\frac12,+\frac12}$.

Note that in contradistinction to \eqref{ssbpositive J0} and \eqref{ssbnegative J0}, the solution \eqref{sbs} depends on the conformal invariant $u$ and therefore is not the same as a three-point function of some local operators. The reason for this difference is that here we have used the line equation of motion \eqref{eom} that feels the line (except at the free-scalar of free-fermion points).

\subsection{Relating the Normalisations Using Variations with Displacement} \la{j0normrel}

In the previous subsections, we have been able to determine all the correlation functions involving one mesonic line and $J_0$ up to the four constants $f_{\pm\frac12,\pm'\frac12}(1)$ and the two constants $A_{\pm\frac12,\mp\frac12}$. We now relate these constants using variation equations that involve the displacement operator. 

Consider first the variation
\begin{multline}\la{minusvar}
    \delta_- \<\cM_{10}^{(\frac12,-\frac12)}\,J_0(x_2)\> = \<\delta_- \overline{\mathcal{O}}_\frac12(x_1) \mathcal{W} \mathcal{O}_{-\frac12}(x_0) \,J_0(x_2)\> + \<\cM_{10}^{(\frac12,-\frac32)}\,J_0(x_2)\>\\
    + \<\cM_{10}^{(\frac12,-\frac12)}\,\p_-J_0(x_2)\>+ \eta_- |x_{10}| \int\limits_{\tilde{\epsilon}}^{1-\tilde{\epsilon}}\!\dd s\,\<\cM_{1s}^{(\frac12,-\frac12)}\> \<\cM_{s0}^{(-\frac12,-\frac12)}\,J_0(x_2)\> = 0\,.
\end{multline}
Here, the last term represents the integration of ${\mathbb D}_-$ along the line and $\tilde\epsilon$ is a point-splitting cutoff. This integral reads
\beq\la{D-integral}
\int\limits_{\tilde{\epsilon}}^{1-\tilde{\epsilon}}\!\dd s\, \<\cM_{1s}^{(\frac12,-\frac12)}\> \<\cM_{s0}^{(-\frac12,-\frac12)}\,J_0(x_2)\>= \frac{c_+\, f_{-\frac12,-\frac12}(1)\, x_2^+}{|x_{10}|^{2\Delta - 3} |x_{20}|^{2\Delta + 1}} \!\int\limits_{\tilde{\epsilon}}^{1-\tilde{\epsilon}}\!\frac{\dd s\,s}{(1-s)^{2\Delta} |x_{2s}|^{5-2\Delta}}\,.
\eeq
For $\Delta\geqslant 1/2$ this integral contains divergences in $\tilde{\epsilon}$ that originate from the $s\to1$ region of integration. They are canceled by the first term in \eqref{minusvar}, which otherwise has no finite contribution. The finite part of the integral with $\Delta\geqslant 1/2$ can be obtained by analytic continuation from $\Delta<1/2$ and the result can be written in terms of hypergeometric functions. Instead of analysing it for general kinematics, it suffices to consider its behaviour as $2x_2^+x^-_2=r^2\rightarrow\infty$, which is
\beq
(\ref{D-integral})=\frac{c_+\, f_{-\frac12,-\frac12}(1)\, x_2^+}{2(\Delta-1)(2\Delta-1)|x_{10}|^{2\Delta - 3}}\(r^{-6}+O(r^{-8})\)\,.
\eeq
In this limit the constraint \eqref{minusvar} becomes
\begin{equation}
A_{\frac12,-\frac12}r^{2-2\Delta}\(1+O(r^{-2})\) + \left[(3-2\Delta) f_{\frac12,-\frac12}(1) - \frac{\eta_-\, c_+ f_{-\frac12,-\frac12}(1)}{(2\Delta-2)(2\Delta-1)}\right]\! \(1+O(r^{-2})\)=0\,,
\end{equation}
We conclude that
\begin{equation}\label{cf=f}
    A_{\frac12,-\frac12} = 0\,,\qquad\eta_-\, c_+\, f_{-\frac12,-\frac12}(1) = -(2\Delta - 3)(2\Delta - 2)(2\Delta - 1) f_{\frac12,-\frac12}(1)\,.
\end{equation}
This relation can be rewritten using the relation between $\Delta$ and the displacement two-point function, \eqref{2pt displacement}, as
\begin{equation}
    \sin(2\pi\Delta) f_{-\frac12,-\frac12}(1) = 2\pi\eta_+\, c_-\, f_{\frac12,-\frac12}(1)\,.
\end{equation}

The variation $\delta_+ \<\cM_{10}^{(\frac12,-\frac12)}\,J_0(x_2)\> = 0$ is treated in the same way and implies that
\begin{equation}
    \sin(2\pi\Delta) f_{\frac12,\frac12}(1) = 2\pi\eta_-\, c_-\, f_{\frac12,-\frac12}(1)\,.
\end{equation}
Similarly, the variations $\delta_\pm \<\cM_{10}^{(-\frac12,\frac12)}\,J_0(x_2)\> = 0$ imply that
\begin{equation}
    c_-\, f_{\frac12,-\frac12}(1) = - c_+\, f_{-\frac12,\frac12}(1)\,.
\end{equation}
In summary, we have
\beq
{f_{-\frac12,\frac12}(1) \over f_{\frac12,-\frac12}(1)}=-{c_-\over c_+}\,,\qquad{f_{-\frac12,-\frac12}(1)\over f_{\frac12,\frac12}(1)}={\eta_+\over\eta_-}\,,\quad\text{and}\quad{f_{\frac12,\frac12}(1)\over f_{\frac12,-\frac12}(1)}={2\pi\eta_-\, c_-\over\sin(2\pi\Delta)}\,,
\eeq
where $c_\pm$, $\eta_\pm$ and $f_{\pm\frac{1}{2},\pm\frac{1}{2}}$ are defined in (\ref{normalization M}), (\ref{spectrumR}) and (\ref{corrsbars}) respectively. These relations are in agreement with (\ref{sflip}). Additionally, we find
\begin{multline}\la{ssbnegative}
\<\cM_{10}^{(\bar s>0,s<0)}\,J_0(x_2)\> =  2^{\bar s - \frac12} \frac{\Gamma(\Delta + \bar s) \Gamma(\Delta - s - \frac32)}{\Gamma\left(\Delta + \frac12\right) \Gamma(\Delta  - 1)} \frac{(x_2^-)^{\bar s + s}\, f_{\frac12,-\frac12}(1)}{|x_{10}|^{2\Delta -2s - 3} |x_{20}|^2 |x_{21}|^{2 + 2\bar s + 2s}} \\
\times\, {}_2 F_1\!\big(2,2+2s-2\Delta;5/2+s-\Delta; (1-\cos\theta)/2\big)
\end{multline}
when $\bar s + s\geqslant 0$.

\section{Correlators With Spinning Local Operators} \label{sec:spinning}

The correlator $\<\cM_{10}^{(\bar s,s)}J_{\tilde s}\>$ has $2\tilde s+1$ independent, $\mathrm{SL}(2,\mathbb{R})\times \mathrm{U}(1)$-covariant tensor structures. In appendix \ref{app:embedding} we determine them using the embedding space formalism. Each of these structures comes multiplying a function of the conformal invariant that we wish to bootstrap.

\subsection{The Spin-one Current} \la{spin1corr}

We first bootstrap the correlator with the spin one current, $\tilde s=1$, and then proceed to the general case. The correlator with $J_1^\mu$ has three tensor structures. We choose them as follows
\begin{equation}\label{conf structures}
    Q^\mu_1=\frac{x^\mu_{21}}{|x_{21}|^2} - \frac{x^\mu_{20}}{|x_{20}|^2}\,,\quad\quad Q^\mu_2= \frac{r^\mu}{r^2} - \frac{x^\mu_{20}}{|x_{20}|^2} - \frac{x^\mu_{21}}{|x_{21}|^2}\,,\quad\quad Q^\mu_3 = \epsilon^{\mu\nu 3} \frac{x_{2,\nu}}{r^2}\,,
\end{equation}
where $r^\mu = (x^1_2,x^2_2,0)$ is the transverse position of $J_1$. In terms of these $Q$'s, the correlator takes the form
\begin{equation} \label{conformal J1}
    \<\cM_{10}^{(\bar s,s)}\,J_1^\mu(x_2)\> = \left(\frac{x_2^\pm |x_{10}|}{|x_{20}| |x_{21}|}\right)^{|\bar s + s|}  \frac{g_{\bar s,s}(u) Q^\mu_1
    + h_{\bar s,s}(u)Q^\mu_2 + k_{\bar s,s}(u)Q^\mu_3}{|x_{10}|^{\Delta_{\bar s} + \Delta_{-s} - 1} |x_{20}|^{1 + \Delta_{-s} - \Delta_{\bar s}} |x_{21}|^{1 + \Delta_{\bar s} - \Delta_{-s}}}\,.
\end{equation}

As for the scalar current, we require the invariance of this correlator under constant shifts in the transverse directions. This constraint is insufficient to fix the functions of $u$; we need to additionally impose that $J_1$ be conserved, $\p_\mu J_1^\mu=0$, and appropriate boundary conditions at 
$u\to1^-$. If, moreover, we normalise $J_1$ such that the boundary operators have charge $\pm1$ under $Q^{(1)} =  \int \dd^2 S_\mu J_1^{\mu}$, then we find that the correlator is uniquely fixed. The details of the computations are relegated to appendices \ref{app:corrwithj1} and \ref{app:can norm}. The solution for $g_{\bar s,s}$, $h_{\bar s,s}$, and $k_{\bar s,s}$ with spins of equal signs is presented in section \ref{eqlsignJ1}. They are all of the simple form $a + b/u + c\, u$. The solutions when the spins have opposite signs are nontrivial functions of $u$, as for $J_0$, for an example see \eqref{j1Ssol}.

For later use, we give here the correlator with $J_1^+$ when $\bar s,s<0$, which reads simply
\begin{equation}\label{guess j1}
    \<\cM_{10}^{(\bar s,s)}\,J^{+}_1(x_2)\> = \p^{-\frac12-\bar s}_{x^-_1} \p^{-\frac12-s}_{x^-_0}\left[d_1 \, \mathcal{B}_1 + \tilde{d}_1 \left(\p_{x^-_0} -  \p_{x^-_1}\right)\, \mathcal{B}_2\right]\,,
\end{equation}
where
\begin{equation}\label{usefulfunc}
    \mathcal{B}_1 = \p_{x_0^-}\p_{x_1^-} |x_{20}|^{1-2\Delta} |x_{21}|^{2\Delta - 3}\,,\qquad \mathcal{B}_2 = (\Delta - 1) x_2^+ |x_{20}|^{-2\Delta} |x_{21}|^{2\Delta - 4}\,.
\end{equation}
In appendix \ref{app:can norm}, we show that the coefficients $d_1$ and $\tilde d_1$ are given by
\begin{equation}\label{d1 explicit}
    \eta_-\, d_1 = \frac{\sin(2\pi\Delta)}{2\pi^2} - 2 \nu_+ \sin^2(\pi\Delta)\quad \text{and}\quad \eta_-\, \tilde{d}_1 = \frac{\sin(2\pi\Delta)}{2\pi^2} + 2 \nu_+ \cos^2(\pi\Delta)\, .
\end{equation}

We expect that the correlators with $\tilde s>1$ are also determined by our bootstrap axioms. Moreover, they follow from the correlators $\<\cM^{(\bar s,s)}\,J_0\>$, $\<\cM^{(\bar s,s)}\,J_1\>$ that we have already considered and the three-point functions $\<J_0\,J_1\,J_{\tilde s}\>$, $\<J_0\,J_0\,J_{\tilde s}\>$. To see that, consider the correlator $\<\cM^{(\bar s,s)}\,J_0(x_2)\,J_1(x_3)\>$. By taking the bulk-bulk OPE limit $J_0\times J_1\to J_{\tilde s}\to \cM^{(\bar s,s)}$ one can read all the correlators $\<J_{\tilde s}\,\cM^{(\bar s,s)}\>$. On the other hand the bulk-line OPE expansion of the same correlator 
$J_0\times\cM^{(\bar s,s)}\to\cM^{(\bar s,s')}\cM^{(s'',s)}\to\cM^{(s'',s)}J_1$ is expressed in terms of the of the bulk-line OPE coefficients that are encoded in the correlators that we have considered.

In the next subsection, we present a conjecture for all the correlators in which the boundary spins have the same sign.

\subsection{Higher-spin Currents} \label{sec:hs}

In order to encode the tensor structure of the correlators with the higher-spin currents, it is convenient to introduce a null polarisation vector $\zeta\in\mathbb{C}^3$, $\zeta\cdot\zeta = 0$, and define
\begin{equation}
    J_s(x;\zeta) = \zeta_{\mu_1}\!\cdots\zeta_{\mu_s}J_s^{\mu_1\cdots\mu_s}(x)\,.
\end{equation}
Covariance under $\mathrm{SL}(2,\mathbb{R})\times \mathrm{U}(1)$ then fixes the correlators to be of the form
\begin{multline}\label{conformal Js}
    \<\cM_{10}^{(\bar s,s)}\,J_{\tilde{s}}(x_2;\zeta)\> =  \frac{\left(\frac{x_2^\pm |x_{10}|}{|x_{20}| |x_{21}|}\right)^{|\bar s + s|}}{|x_{10}|^{\Delta_{\bar s} + \Delta_{-s} - 1} |x_{20}|^{1 + \Delta_{-s} - \Delta_{\bar s}} |x_{21}|^{1 + \Delta_{\bar s} - \Delta_{-s}}}\\
    \times \left[\sum_{i=0}^{\tilde{s}} h^{(i)}_{\bar s, s,\tilde{s}}(u) (\zeta\cdot Q_1)^i (\zeta\cdot Q_2)^{\tilde{s}-i} + \zeta\cdot Q_3 \sum_{i=0}^{\tilde{s}-1} k^{(i)}_{\bar s, s,\tilde{s}}(u) (\zeta\cdot Q_1)^i (\zeta\cdot Q_2)^{\tilde{s}-i}\right]\,,
\end{multline}
see appendix \ref{app:embedding} for details. Notice that we did not include terms with higher power of $\zeta\cdot Q_3$. That is because the identity
\begin{equation}
    (\zeta\cdot Q_3)^2 = \frac{(\zeta\cdot Q_1)^2}{u^2 - 1} - \frac{(\zeta\cdot Q_2)^2}{u^2}\,,
\end{equation}
relates them to the terms in (\ref{conformal Js}).

We expect that for a given $\tilde{s}$, the full set of functions $\{ h^{(i)}_{\bar s, s,\tilde{s}},k^{(i)}_{\bar s, s,\tilde{s}}\}_{\bar s,s,i}$ can be determined up to one overall normalisation-dependent constant, which we will denote $d_{\tilde{s}}$. These functions are expected to be relatively complicated, particularly when $\bar s$ and $s$ are of opposite sign. Here, we content ourselves with a conjecture for the case where the boundary spins are of equal sign. Without loss of generality, see \eqref{Deltasym}, we choose them to be negative, $\bar s,s<0$. We claim that the correlator of the mesonic line with $j_{\tilde{s}}\equiv J_{\tilde s}^{+\cdots +}$ for $\tilde{s}>1$ is given by\footnote{When $\nu_+ = \nu_- = 0$, this formula also applies to $j_1$. Otherwise, the coefficients in front of the two terms are different, see \eqref{guess j1} and \eqref{d1 explicit}.}
\begin{equation}\label{guess js}
    \<\cM_{10}^{(\bar s,s)}\,j_{\tilde{s}}(x_2)\> = d_{\tilde{s}}\, \p^{-\frac12-\bar s}_{x^-_1} \p^{-\frac12-s}_{x^-_0}\left[\mathcal{D}_{1,\tilde{s}}(\p_{x^-_0}, \p_{x^-_1})\, \mathcal{B}_1 + \mathcal{D}_{2,\tilde{s}}(\p_{x^-_0}, \p_{x^-_1})\, \mathcal{B}_2\right]\,,
\end{equation}
where the functions $\mathcal{B}_1$ and $\mathcal{B}_2$ have been defined in \eqref{usefulfunc}. Here, $\mathcal{D}_{i,\tilde{s}}(\p_{x^-_0}, \p_{x^-_1})$ are differential operators. They are encoded in the following generating functions
\begin{equation}\label{diff}
    e^{(u - v) t} \frac{\sin\left(2t\sqrt{u v}\right)}{2\sqrt{u v}} = \sum_{\tilde{s}=1}^{+\infty} \frac{t^{\tilde{s}}}{\tilde{s}!} \mathcal{D}_{1,\tilde{s}}(u,v)\,,\qquad e^{(u - v) t}\cos\left(2t\sqrt{u v}\right) = 1 + \sum_{\tilde{s}=1}^{+\infty} \frac{t^{\tilde{s}}}{\tilde{s}!} \mathcal{D}_{2,\tilde{s}}(u,v)\,.
\end{equation}
These generating functions are the same as those for the currents in free fermionic and bosonic CFTs in 3d \cite{Giombi:2016ejx}. The only place where the interacting nature of our theory enters is through the dimension $\Delta$ that appears in $\mathcal{B}_1$ and $\mathcal{B}_2$. In particular, for $\Delta = 1/2$ or $3/2$, $\mathcal{B}_1 = 0$ and the correlator reduces to the one in the free boson theory. Similarly, for $\Delta = 1$, $\mathcal{B}_2 = 0$ and the correlator reduces to that of the free fermion theory.

It is enough to match \eqref{guess js} with the $+\dots +$ component of \eqref{conformal Js} to fix all the functions $\{ h^{(i)}_{\bar s, s,\tilde{s}},k^{(i)}_{\bar s, s,\tilde{s}}\}_i$. Thus, we can reconstruct the full correlation function \eqref{conformal Js} from (\ref{guess js}). We expect that these functions are the only solution to the constraints\footnote{The second constraint corresponds to the conservation of $J_{\tilde{s}}$.}
\begin{equation}\la{constraints}
    \delta_-\<\cM_{10}^{(\bar s,s)}\,J_{\tilde{s}}(x_2;\zeta)\> = 0\quad\text{and}\quad \left[(2\tilde{s} - 1)\p_\zeta + \zeta\, \square_\zeta\right]\cdot\p_{x_2}\<\cM_{10}^{(\bar s,s)}\,J_{\tilde{s}}(x_2;\zeta)\> = 0
\end{equation}
that remains finite as $u\to 1^-$. For $\tilde s=2$, we have verified this expectation explicitly. For $\tilde{s} = 3$, we have checked that our conjecture is consistent with the constraints (\ref{constraints}).

As the correlator $\<\cM_{10}^{(\bar s,s)}\,J_0\>$ with boundary spins of equal signs (\ref{ssbpositive J0}), we thus expect the correlator $\<\cM_{10}^{(\bar s,s)}\,J_{\tilde s}\>$ in (\ref{guess js}) with $\tilde s>1$ to be fixed by constraints that are also satisfied by a three-point function of local primary operators. However, unlike the case of $J_0$, we cannot equate these two types of correlators. That is because the conservation constraint sets the three-point function of local operators to zero unless the operators at $x_0$ and $x_1$ have the same dimension.

Finally, we remark that equation \eqref{ssbnegative J0} can also be rewritten using the function $\mathcal{B}_1$ defined in \eqref{usefulfunc} as
\begin{equation}\label{ssbnegative J0 bis}
    \<\cM_{10}^{(\bar s<0,s<0)}\,J_0(x_2)\> = d_0 \frac{|x_{10}|}{x_2^+}\, \p^{-\bar s - \frac12}_{x^-_1} \p^{-s - \frac12}_{x^-_0} \mathcal{B}_1\, ,
\end{equation}
where we have introduced the new constant $d_0\equiv f_{-\frac12,-\frac12}/[(2\Delta - 1)(3 - 2\Delta)]$. In the next section, we use constraints from slightly broken, higher-spin symmetry to fix all the relative normalisations $d_{\tilde{s}}/d_0$. Nonetheless, let us mention that using the canonical normalisation for the stress-energy tensor fixes
\begin{equation}\label{d2 explicit}
    d_2 = -\frac{\sin(2\pi\Delta)}{8\pi^2 \eta_-}\, ,
\end{equation}
as shown in appendix \ref{app:can norm}.

\section{Constraints From Slightly Broken Higher-Spin Symmetry}\label{sec:MZ}

In this section, we impose additional constraints on the correlator $\<\cM^{(\bar s,s)}J_{\tilde s}\>$ that follow from the non-conservation of the higher-spin currents at order $1/N$. These constraints allow us to relate correlators of different spins $\tilde s$ and in this way fix the relative normalisation constants. More precisely, we will show that
\begin{equation}\label{csc0}
    16 \cos^2(\pi\Delta)\frac{d_0^2}{\mathcal{N}_0} = \frac{1}{\mathcal{N}_1}\left[d_1^2\cos^2(\pi\Delta) + \tilde{d}^2_1\sin^2(\pi\Delta)\right] = \frac{d_{\tilde{s}}^2}{\mathcal{N}_{\tilde{s}}}\,,\qquad \tilde s\geqslant 2\,.
\end{equation}

The proof relies on a bootstrap technique that was introduced in \cite{Maldacena:2012sf}. The authors of that paper considered approximately conserved currents $J_{\tilde s}$ of even spin ${\tilde s}\geqslant 2$, and used the divergence of $J_4$ to bootstrap their three-point functions at large $N$. Here, since we have currents of all spins, we have found it simpler to use the divergence of $J_3$ instead.\footnote{In theories with only even higher-spin symmetry we expect that the same result can also be obtained using $J_4$.}

The idea is to write the Ward identity that follows from the non-conservation of $J_3$ for the correlators at hand. We start from
\begin{align}\label{Wardid}
&\lim_{\epsilon\to0} \int_{\mathcal{D}_\epsilon}\!\dd^3x\, \< \mathcal{M}_{10}^{(-\frac12,-\frac12)}j_{\tilde s}\,\p^{\mu}J_{3,\mu--}(x) \>\Big|_\text{finite}\\
&\qquad\qquad\qquad\qquad\qquad\qquad=\<[Q^{(3)}_{--},\cM_{10}^{(-\frac12,-\frac12)}]\,j_{\tilde s}\>+\<\cM_{10}^{(-\frac12,-\frac12)}\,[Q^{(3)}_{--},j_{\tilde s}]\> \,,\nn 
\end{align}
where the domain of integration $\mathcal{D}_\epsilon$ is $\mathbb{R}^3$ minus a ball of radius $\epsilon$ centered on $j_{\tilde s}$ and a region of size $\epsilon$ around the segment between $x_0$ and $x_1$. Here, the {\it finite} subscript means that we only keep the finite terms in the $\epsilon\to0$ limit. The left-hand side can also be computed using the explicit form of the divergence of $J_3$, which we will determine in section \ref{sec:divJ3}. The right-hand side involves the action of the pseudo-charge $Q^{(3)}$ on the local current, which is defined as
\beq\la{localact}
    [Q^{(3)}_{--},j_{\tilde s}(0)] = \lim_{\epsilon\to0} \epsilon^2 \int_{S^2}\dd^2 n\,n^\mu J_{3,\mu--}(\epsilon n)\, j_{\tilde s}(0)\Big|_\text{finite}\,,
\eeq
where $S^2$ is the unit two sphere. Its action on the mesonic line is defined similarly by integrating $J_3$ on a small tube around the line with boundaries. Here, we divide this domain into two regions as plotted in figure \ref{tube}. The first region is a tube of size $\epsilon$ around the line that is pinched at two points close to the two boundaries. It gives the action of $Q^{(3)}$ on the line and produces the {\it tilt} operator ${\mathbb D}^{(2)}$. It takes the form
\begin{equation}\la{lineact}
    [Q^{(3)}_{--},\cW] = \lim_{\epsilon\to 0}\epsilon\!\!\int_{S^1}\dd\hat n \int\dd\tau|\dot x_\tau|\,\hat n^\mu J_{3,\mu--}(x_\tau+\epsilon\hat n)\, \cW\Big|_\text{finite}\!\!\equiv \int\!\dd\tau\,|\dot x_\tau|\, \mathbb{D}^{(2)}_{--}(x_\tau)\,\cW\,,
\end{equation}
where $\hat n$ is the unit vector parameterising an $S^1$ in the transverse space to the line. The action of $Q^{(3)}_{--}$ on the boundary operators takes the same form as in (\ref{localact}) with the line pinching the size $\epsilon$ region around the line endpoints, see figure \ref{tube}.

\begin{figure}
\centering{}\includegraphics[scale=1]{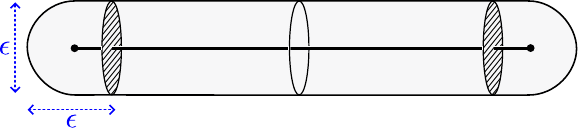}\caption{The current $J_3$ is integrated along a tube of size $\epsilon$ around the mesonic line. The tube is divided into three regions. Two regions of size $\epsilon$ around the line endpoints, and a cylinder around the line that is pinched by the line at its extremities.}
\la{tube}
\end{figure}

\subsection{Divergence of $J_3$}
\label{sec:divJ3}

Because $\Delta(J_3) = 4 + O(1/N)$, the divergence $\p_\mu J_3^{\mu\nu\rho}$ is a conformal primary at leading order in $1/N$, \cite{Giombi:2011kc,Maldacena:2012sf}. This divergence is a dimension-$5$, spin-$2$ operator. There are no such single-trace operators, but there are four double-trace ones, hence
\begin{equation}\label{divergence J3}
    \p_\mu J_3^{\mu\nu\rho} = \frac1{N}\sum_{k=1}^4 a_k \mathcal{O}^{\nu\rho}_k\,,
\end{equation}
where the coefficients $a_k$ are of order $O(N^0)$ and $\cO_i^{\mu\nu}$ are the four primaries\footnote{The operator $\mathcal{O}_1$ is traceless because $J_1$ is truly conserved, i.e. $\p_\mu J_1^\mu = 0$. We also note that the third operator could equivalently be written $\mathcal{O}^{\mu\nu}_3 = \epsilon^{\alpha\beta\mu}\left[2 J_{1,\alpha} \p^{\nu} J_{1,\beta} - J_{1,\alpha} \p_\beta J_1^{\nu}\right] + (\mu\leftrightarrow \nu) + 2\, \eta^{\mu\nu} \epsilon^{\alpha\beta\gamma} J_{1,\alpha} \p_\beta J_{1,\gamma}$.}
\begin{align}\la{Qs}
    \mathcal{O}^{\mu\nu}_1 &= J_0 \p^{\mu}\! J_1^{\nu} - \frac32 \p^{\mu}\!J_0 J_1^{\nu} + (\mu\leftrightarrow \nu) + \, \eta^{\mu\nu} \p_{\rho} J_0 J_1^{\rho}\,,\\
    \mathcal{O}^{\mu\nu}_2 &= \epsilon^{\alpha\beta\mu} J_{1,\alpha} J_{2,\beta}^{\nu}  + (\mu\leftrightarrow \nu)\,,\nn\\
    \mathcal{O}^{\mu\nu}_3 &= \epsilon^{\alpha\beta\mu}\left[J_{1,\alpha} \p^{\nu} J_{1,\beta} + J_1^{\nu} \p_{\alpha} J_{1,\beta}\right] + (\mu\leftrightarrow \nu)\,,\nn\\
    \mathcal{O}^{\mu\nu}_4 &= J_0 J_2^{\mu\nu}\,.\nn
\end{align}
Note that charge conjugation symmetry forbids the last two contributions. We will not assume it here. Still, we will see in the next subsections that $a_3 = a_4 = 0$ and that $a_2\propto (d_1 - \tilde{d}_1)$; in agreement with the findings of \cite{Giombi:2016zwa} for the matter Chern--Simons theories where $d_1 = \tilde{d}_1$.

\subsection{Action of the Pseudo-Charge}

\subsubsection{Action on the Currents}

In order to shorten the notation, we define $j_{\tilde s} = J_{\tilde s}^{+\dots+}$ with $\tilde s\ge1$. The action of the pseudo-charge on the spinning currents takes the form 
\begin{equation}\label{Q on js}
    [Q^{(3)}_{--},j_{\tilde s}(0)] = 
    \sum_{k=0}^{\min\{4,\tilde s+1\}}e_{\tilde s,\tilde s+2-k}\, \p_-^{k} j_{\tilde s+2-k}(0)\,.
\end{equation}
By acting on the two-point function of the spinning currents (\ref{normalization Js})
\begin{equation}\label{Q3jj} 
    \<[Q^{(3)}_{--},j_s(x)]\,j_{s'}(y)\>+\<j_s(x)\,[Q^{(3)}_{--},j_{s'}(y)]\> = O(1)\,,
\end{equation}
we find
\begin{equation}\label{symmetry coeffs Q}
    e_{s,s'}\, \mathcal{N}_{s'} = (-1)^{1+s'+s} e_{s',s}\, \mathcal{N}_{s}\,.
\end{equation}
In particular, note that this implies $e_{s,s} = 0$.

The cases where $(s,s')\in\{(1,1),(1,2),(2,1)\}$ need special attention. That is because the double traces $J_1J_1$ and $J_1J_2$ appear in $\cO_3$ and $\cO_2$ respectively, (\ref{divergence J3}). As a result, their contribution to (\ref{Q3jj}) is enhanced to order $N$. However, we find that these contributions sum into total derivatives and, therefore,
\begin{equation}\la{jjQ3}
    \int\dd^3z\left\langle j_1(x)  j_1(y) \mathcal{O}_3^{++}(z)\right\rangle=O(N)\,,\qquad  \int\dd^3z\left\langle j_1(x) j_2(y) \mathcal{O}_2^{++}(z)\right\rangle=O(N)\,.
\end{equation}
As a result, (\ref{Q3jj}) still holds in these cases. To derive (\ref{jjQ3}) we have used the explicit form of $\mathcal{O}_3^{++}$ and $\mathcal{O}_2^{++}$, (\ref{Qs}), as well as the two-point functions $\left\langle j_1(y)  J^3_1(z)\right\rangle$ and $\left\langle j_2(y)  J^{3+}_2(z) \right\rangle$ in (\ref{normalization Js}).

For the scalar operator $J_0$, we have
\begin{equation} \label{QonJ0}
    [Q^{(3)}_{--},J_0]=e_{0,2} \left(\p_- J_2^{3+} - \p_3 j_2\right) +e_{0,1} \p_- \left(\p_- J_1^3 - \p_3 j_1\right) + \frac{\tilde{e}_{0,1}}{N}\, j_1 j_1\,.
\end{equation}
Here, we have discarded a term proportional to $\p^2_- J_0$ because there is no double-trace term involving two $J_0$ in the divergence of $J_3$, see (\ref{divergence J3}).

\subsubsection{Action on the Mesonic Line}

The action of the pseudo-charge on the mesonic line can be divided into its action on the line and its action on the boundary operators. The action on the line in (\ref{lineact}) produces the {\it tilt} operator
\beq
    {\mathbb D}^{(2)}_{--} = \xi_1\, \cO_{-\frac32} \times \overline \cO_{-\frac12} + \xi_2\, \cO_{-\frac12} \times \overline \cO_{-\frac32} +\(\xi_3\,\p_- j_1 +\xi_4\, j_2\)/N\,,
\eeq
where the coefficients $\xi_1$ and $\xi_2$ are fixed in appendix \ref{protectedop}. Here, we will only need to consider correlation functions where the boundary spins are $s=\bar s=-1/2$, see \eqref{Wardid}. In this case, all the operators in ${\mathbb D}^{(2)}_{--}$ drop out from the correlator, so we can focus on the action of $Q^{(3)}$ on the boundary operators only. It takes the form
\beq
[Q^{(3)}_{--}, \overline \cO_{-\frac12}] = \bar q\,\overline \cO_{-\frac52}\,,\qquad [Q^{(3)}_{--},\cO_{-\frac12}]= q\,\cO_{-\frac52}\,.
\eeq

\subsection{Constraining the Correlation Functions}\la{constraining}

\subsubsection{The Correlator with $J_0$}

For ${\tilde s}=0$, the Ward identity (\ref{Wardid}) takes the form
\begin{align}\label{MZ J0}
    \< \mathcal{M}_{10}^{(-\frac12,-\frac12)} &\left[e_{0,2} \left(\p_- J_2^{3+} - \p_3 j_2\right) +e_{0,1} \p_- \left(\p_- J_1^3 - \p_3 j_1\right)\right]\> + \bar{q} \< \mathcal{M}_{10}^{(-\frac52,-\frac12)} J_0\> \\
    &+ q\langle \mathcal{M}_{10}^{(-\frac12,-\frac52)} J_0\rangle = \frac{-1}{N}\int \dd^3x\,\langle \mathcal{M}_{10}^{(-\frac12,-\frac12)} J_0 \left[a_1\mathcal{O}^{++}_1(x) + a_4\mathcal{O}^{++}_4(x) \right]\rangle \,.\nn
\end{align}
The relevant correlators and integrals are collected in appendix \ref{app:useful correlators}. Using them, we find
\begin{multline}
   \frac{-1}{N}\int\dd^3y\,\langle \mathcal{M}_{10}^{(-\frac12,-\frac12)} J_0(x_2) \mathcal{O}^{++}_1(y)\rangle\\
   = \frac{5\pi^2 \mathcal{N}_0 |x_{10}|}{2\, x_2^+ } \p_{x_2^-} \Big[\frac{4\,d_1}{\tan(\pi\Delta)} \p_{x_0^-}\p_{x_1^-} \mathcal{B}_2 + \tilde{d}_1\tan(\pi\Delta) \left(\p_{x_0^-} - \p_{x_1^-}\right) \mathcal{B}_1\Big]\,,
\end{multline}
where $d_1$, $\tilde d_1$ are defined in (\ref{guess j1}) and ${\cal B}_1$, ${\cal B}_2$ in (\ref{usefulfunc}). Similarly, 
\begin{align}\la{m12m12J2}
    \frac{1}{N}\int\dd^3y\,&\langle \mathcal{M}_{10}^{(-\frac12,-\frac12)} J_0(x_2) \mathcal{O}^{++}_4(y) \rangle = \frac{\pi^2 \mathcal{N}_0\, d_2\, |x_{10}|}{4\, x_2^+}\\
    &\times\Big[\frac{8}{\tan(\pi\Delta)} \left(\p_{x_1^-} - \p_{x_0^-}\right) \p_{x_0^-}\p_{x_1^-} \mathcal{B}_2 - \tan(\pi\Delta) \left(\p^2_{x_0^-} - 6\p_{x_0^-}\p_{x_1^-} + \p^2_{x_1^-}\right) \mathcal{B}_1\Big]\,.\nn
\end{align}
Putting everything together, we find that \eqref{MZ J0} is satisfied provided
\beq\label{constraints MZ J0}
    a_4 =e_{0,2}=0\,,\quad e_{0,1} \tilde{d}_1 = -\frac{5\pi^2\mathcal{N}_0 a_1 d_1}{\tan(\pi\Delta)} \,,\quad q = -\bar{q}\,,\quad 2qd_0=e_{0,1} d_1 - 5\pi^2 \tan(\pi\Delta) \mathcal{N}_0 a_1 \tilde{d}_1\,,
\eeq
where we assumed that $d_2\neq 0$.

\subsubsection{The Correlator with $J_1$}

For ${\tilde s}=1$, (\ref{Wardid}) takes the form
\begin{multline}\label{MZ J1}
    \< \mathcal{M}_{10}^{(-\frac12,-\frac12)} \left[ e_{1,3}\, j_3 + e_{1,2}\, \p_- j_2\right]\>  + \bar{q} \< \mathcal{M}_{10}^{(-\frac52,-\frac12)} j_1\> + q \<\mathcal{M}_{10}^{(-\frac12,-\frac52)} j_1\> \\
    =-\frac{1}{N}\int\dd^3x\< \mathcal{M}_{10}^{(-\frac12,-\frac12)} j_1 \left[a_1\mathcal{O}^{++}_1(x) + a_2\mathcal{O}^{++}_2(x) + a_3\mathcal{O}^{++}_3(x)\right]\>\,.
\end{multline}
The integrals on the right-hand side give
\begin{align}
\frac{1}{N}\int\dd^3x_3 \langle \mathcal{M}_{10}^{(-\frac12,-\frac12)} j_1(x_2) \mathcal{O}^{++}_1(x_3)\rangle &=- 5\,\mathcal{N}_1d_0\, \p^3_{x_2^-} A_1\,,\\
\frac{1}{N}\int\dd^3x_3\langle \mathcal{M}_{10}^{(-\frac12,-\frac12)} j_1(x_2) \mathcal{O}^{++}_3(x_3) \rangle &= 3\ii\,\mathcal{N}_1\, \p^2_{x_2^-} \Big[d_1 \big(\p_{x_1^-} - \p_{x_0^-}\big) A_1 + 4\, \tilde{d}_1 A_2\Big]\,,\nn\\
\frac{1}{N}\int\dd^3x_3\langle \mathcal{M}_{10}^{(-\frac12,-\frac12)} j_1(x_2) \mathcal{O}^{++}_2(x_3) \rangle
&= \ii\,\mathcal{N}_1 d_2\, \p_{x_2^-} \Big[\big(6\p_{x_0^-}\p_{x_1^-}-\p^2_{x_0^-}-\p^2_{x_1^-}\big)A_1\nn\\
&\qquad\qquad\qquad\qquad\qquad+ 8 \big(\p_{x_0^-} - \p_{x_1^-}\big) A_2\Big]\,,\nn
\end{align}
where the integrals $A_1$ and $A_2$ are defined in \eqref{star-triangle J1}. Computing them yields
\begin{equation}
    A_1 = \frac{4\pi^2}{\tan(\pi\Delta)} \mathcal{B}_2\quad \text{and}\quad A_2 = \pi^2 \tan(\pi\Delta) \mathcal{B}_1\,.
\end{equation}
Plugging these into \eqref{MZ J1}, we find that the equation is satisfied provided
\begin{align}\label{constraints MZ J1}
    &a_3 =e_{1,3} = 0\,,\quad q\, \tilde{d}_1 = -\frac{40\,\pi^2 \mathcal{N}_1 a_1 d_0}{\tan(\pi\Delta)}\,,\quad q(d_1 - \tilde{d}_1) = \frac{16\ii\pi^2\mathcal{N}_1 a_2 d_2}{\sin(2\pi\Delta)}\,,\\
    &2e_{1,2}\, d_2 = q\left[ d_1 \cos^2(\pi\Delta)+ \tilde{d}_1 \sin^2(\pi\Delta)\right]\,.\nn
\end{align}

\subsubsection{The Correlator with $J_2$}

For ${\tilde s}=2$, \eqref{Wardid} takes the form
\begin{multline}\label{MZ J2}
    \< \mathcal{M}_{10}^{(-\frac12,-\frac12)} \left[e_{2,4}\, j_4 + e_{2,3}\, \p_- j_3 + e_{2,1}\, \p^3_- j_1\right] \> +\bar{q} \< \mathcal{M}_{10}^{(-\frac52,-\frac12)} j_2\>+q\<\mathcal{M}_{10}^{(-\frac12,-\frac52)} j_2\>\\
    = -\frac{a_2}{N}\int\dd^3x_3\< \mathcal{M}_{10}^{(-\frac12,-\frac12)} j_2(x_2) \mathcal{O}^{++}_2(x_3)\>\,.
\end{multline}
The integrals on the right-hand side are of the form \eqref{star-triangle J1}. We find that the equation is satisfied provided
\begin{align}\label{constraints MZ J2}
    e_{2,4} &= 0\,,\,\qquad 2e_{2,1}\, d_1 = q\, d_2 + 8\ii\pi^2\tan(\pi\Delta) \mathcal{N}_2 a_2 \tilde{d}_1\,,\\
    2e_{2,3}\, d_3 &= q\, d_2\,,\quad 2e_{2,1}\, \tilde{d}_1 = q\, d_2 - \frac{8\ii\pi^2 \mathcal{N}_2 a_2 d_1}{\tan(\pi\Delta)}\,.\nn
\end{align}

\subsubsection{Higher Spins}

For $\tilde s\geqslant 3$ the divergence of $J_3$ in \eqref{divergence J3} does not contribute to the Ward identity \eqref{Wardid}, so it reduces to 
\beq
\sum_{k=0}^{\min(4,\tilde{s}+1)} e_{\tilde{s},\tilde{s}+2-k}\, \< \mathcal{M}^{(-\frac12,-\frac12)}\p_-^{k} j_{\tilde{s}+2-k}\> + \bar{q} \< \mathcal{M}^{(-\frac52,-\frac12)} j_{\tilde{s}}\> + q \< \mathcal{M}^{(-\frac12,-\frac52)} j_{\tilde{s}}\> = 0\,.
\eeq
Using our conjecture \eqref{guess js} and the relation $\bar q = -q$ from \eqref{constraints MZ J0}, the equation reduces to
\begin{equation}
    \sum_{k=0}^{\min(4,\tilde{s}+1)} e_{\tilde{s},\tilde{s}+2-k} \, d_{\tilde{s}+2-k} (-1)^k (u+v)^{k} \mathcal{D}_{i,\tilde{s}+2-k}(u,v) + q\, d_{\tilde{s}} (u^2-v^2) \mathcal{D}_{i,\tilde{s}}(u,v) = 0\,,
\end{equation}
for $i\in\{1,2\}$. It is equivalent to
\begin{equation}\label{qrel}
    e_{\tilde{s},\tilde{s}-2} = e_{\tilde{s},\tilde{s}} = e_{\tilde{s},\tilde{s}+2} = 0\,, \quad e_{\tilde{s},\tilde{s}+1} \,d_{\tilde{s}+1} = e_{\tilde{s},\tilde{s}-1} \,d_{\tilde{s}-1} = q\,d_{\tilde{s}}/2\,.
\end{equation}
Combining these relations with those we found previously, namely \eqref{symmetry coeffs Q}, \eqref{constraints MZ J0}, \eqref{constraints MZ J1}, and \eqref{constraints MZ J2}, we finally arrive at \eqref{csc0}.

As a consistency check, in appendix \ref{Appcsc0} we perform a perturbation theory calculation in $SU(N)_k$ Chern--Simons matter theory, where $a_2=0$ and $d_1=\tilde d_1$. We find agreement with \eqref{csc0}.

If we choose the canonical normalisation for $J_1$ and $J_2$, then $e_{2,3} = 2$, see \cite{Maldacena:2011jn,Maldacena:2012sf}, and the coefficients $d_1$, $\tilde{d}_1$, and $d_2$ can be computed, see appendix \ref{app:can norm}. Combining these results with the relations we found in this section, the divergence of $J_3$ becomes
\begin{equation}\label{div J3 final}
    \p_\mu J_3^{\mu\nu\rho} = \frac{2\eta_-d_3}{N \mathcal{N}_1}\left[\frac{\tan(\pi\Delta) + 2 \pi^2\nu_+}{5\pi^2 \eta_-d_0} \mathcal{O}_1^{\nu\rho} + \frac{16 \ii \pi^2 \nu_+}{\sin(2\pi\Delta)} \mathcal{O}_2^{\nu\rho}\right]\, .
\end{equation}

\subsubsection{Relating $\Delta$ to  $\tilde \lambda$}

In the previous subsection, we have fixed the divergence of $J_3$ in terms of line parameters. Here, we use this result when $\nu_+ = \nu_- = 0$ to relate our (line) parameter $\Delta$ to the (bulk) parameter $\tilde{\lambda}$ introduced in \cite{Maldacena:2012sf} . We find that
\beq\la{lambdatildeDelta}
\tilde{\lambda}^2 = \tan^2(\pi\Delta)\,.
\eeq
In matter Chern--Simons theories, both parameters have been computed as functions of the 't Hooft coupling $\lambda$ through resummation of Feynman diagrams \cite{Aharony:2012nh,Gur-Ari:2012lgt,Gabai:2022mya,Gabai:2022vri}, and \eqref{lambdatildeDelta} can be verified. The following derivation does not require any perturbative computation.

Using the relation between the various constants derived in the previous subsection and the fact that $e_{2,3} = 2$ when the stress-energy tensor is canonically normalised \cite{Maldacena:2011jn,Maldacena:2012sf}, one can extract the following normalisation-independent ratio
\begin{equation}\label{physical ratio}
    a^2_1\frac{\mathcal{N}_0\mathcal{N}_1}{N \mathcal{N}_3} = \frac{4\sin^2(\pi\Delta)}{25\,\pi^4 N\mathcal{N}_2}\,.
\end{equation}
This quantity is proportional to (the square) of one of the two normalised structure constants entering the 3-point function $\<J_0 J_1 J_3\>$, as can be seen by taking the divergence of $J_3$ in this correlator. It is also proportional to the first non-planar correction to the anomalous dimension of $J_3$. Although both are also computable using the bootstrap approach of \cite{Maldacena:2012sf}, only the latter is easily extracted from the literature \cite{Maldacena:2012sf,Giombi:2016zwa,Gur-Ari:2012lgt} in terms of $\tilde{\lambda}$.

Hence, let us focus on the anomalous dimension of $J_3$. The exact two-point function of this spin-3, primary operator takes the form
\begin{equation}
    \< J_3(x_1,\zeta_1) J_3(x_2,\zeta_2)\> = \frac{2^6 6!\, N\mathcal{N}_3}{|x_{12}|^{8+2\gamma_3}} \left(\frac{(\zeta_1\cdot x_{12}) (\zeta_2\cdot x_{12})}{|x_{12}|^2} - \frac{\zeta_1\cdot \zeta_2}2\right)^3\,,
\end{equation}
where $\gamma_3 = O(1/N)$, and $\zeta_1$ and $\zeta_2$ are two null polarisation vectors. On the one hand, taking the divergence of the right-hand side, we obtain, in the planar limit,
\begin{equation}
    \<\p_\mu J^{\mu++}_3(x_1)\p_\nu J^{\nu++}_3(x_2)\> \sim -2^{10}\times 63N\gamma_3\, \mathcal{N}_3 \frac{(x^+_{12})^4}{|x_{12}|^{14}}.
\end{equation}
On the other hand, using the expression \eqref{divergence J3} for the divergence of the current and the two-point functions of the spin-0 and spin-1 operators, we get
\begin{equation}
    \<\p_\mu J^{\mu++}_3(x_1)\p_\nu J^{\nu++}_3(x_2)\> =- 2^6 \times 15\, a_1^2 \mathcal{N}_0 \mathcal{N}_1 \frac{(x^+_{12})^4}{|x_{12}|^{14}}\(1+O(1/N)\)\,.
\end{equation}
Equating the previous two equations and using \eqref{physical ratio} yields
\begin{equation}
    \gamma_3 = \frac{\sin^2(\pi\Delta)}{440\,\pi^4N\mathcal{N}_2}\,.
\end{equation}
According to \cite{Giombi:2016zwa}, the same anomalous dimension is given in terms of $\tilde{N}$ and $\tilde{\lambda}$ by
\begin{equation}
    \gamma_3 = \frac{256\, \tilde\lambda ^2}{105\,\pi^2 \tilde{N}(1+\tilde\lambda ^2)}\,.
\end{equation}
In the canonical normalisation for the stress-momentum tensor, we have $\tilde{N} = 2^{10} \pi^2 N \mathcal{N}_2$, we thus arrive at \eqref{lambdatildeDelta}.

Using the divergence of $J_3$, one can also extract some structure constants that involve this operator. For example, the correlator $\<J_3 J_1 J_0\>$ must take the form \cite{Giombi:2011rz}
\begin{equation}
    \< J_3(x_1) J_1(x_2) J_0(x_3)\> = \frac{\alpha\left(2Q_1^3Q_2+4Q_1^2P_3^2\right)+\beta Q_1^2S_3}{|x_{23}|^2 |x_{31}|^2},
\end{equation}
where $\alpha$ and $\beta$ are structure constants and $Q,P$ and $S$ are the conformal structures that can be found in \cite{Giombi:2011rz} (not to be confused with our $Q$'s). Taking the divergence of $J_3$ from one side and inserting the double-trace operator \eqref{divergence J3} from the other side, we find
\begin{equation}
    a_1 = -\frac{8 \alpha }{15 N \mathcal{N}_0 \mathcal{N}_1}.
\end{equation}
Using $\eqref{physical ratio}$ we obtain
\beq
{\alpha^2\over{\cal N}_0{\cal N}_1{\cal N}_3}={3\over2{\cal N}_2}\({5 N\sin(\pi\Delta)\over8\pi^2}\)^2\,.
\eeq

Finally, note that all these relations, and (\ref{lambdatildeDelta})-(\ref{physical ratio}) in particular, are invariant under $\Delta\leftrightarrow1-\Delta$. Hence, these relations can also be taken as a bootstrap derivation that large-$N$, higher-spin theories have two different conformal line operators that are related by $\Delta\to1-\Delta$. The RG flow between these fixed points has been studied in \cite{Nagar:2024mjz}.

\section{Smooth Conformal Defects and Bulk Operator} \label{sec:smooth}

So far we have considered the correlation function $\<\cM^{(\bar s,s)}J_{\tilde s}\>$ involving a straight mesonic line. In this section, we extend our considerations to
\begin{equation}\label{triangle}
    \langle \left[\overline\cO_{\bar s}(x_\tau)\mathcal{W}[x(\cdot)]\cO_{s}(x_\sigma)\right] J_{\tilde s}(x_2)\rangle\,,
\end{equation}
where $x(\cdot)$ is an arbitrary smooth path between $x(\sigma)=x_\sigma$ and $x(\tau)=x_\tau$. 

The expectation value of such mesonic lines ($J_{\tilde s}\to\One$) has been studied in \cite{Gabai:2023lax}. In that paper, it was shown that the assumptions we have made here are sufficient to compute the expectation value of the mesonic line in a systematic expansion around the straight line. The idea is to start with a straight line and deform it smoothly as
\beq\la{deform}
x_{\text{straight}}(\sigma)\quad\rightarrow\quad x(\sigma) = x_{\text{straight}}(\sigma)+\v(\sigma)\,.
\eeq
One then writes the most general effective action on the line order by order in $\v/|x_{10}|$ and use the bootstrap assumptions to fix the coefficients. Here, we repeat their analysis in the presence of the local operator $J_{\tilde s}$. Using the same regularisation scheme, all the coefficients that have been fixed in \cite{Gabai:2023lax} are local and therefore remain the same in the presence of $J_{\tilde s}$. The reason that our task is non-trivial is that there are new terms that contribute to the expansion of the correlator (\ref{triangle}) but not to the line expectation value. Two of these have already appeared in the sections above. One was the single-trace correction to the displacement operator in (\ref{displace}), which is a parameter of the theory and is related to the nonconservation of $J_3$, (\ref{divergence J3}). The other is the single-trace correction to the boundary equation of motion (\ref{eom}) that was already fixed. Few additional new terms appear in the line effective action below.

We find that the conclusion of \cite{Gabai:2023lax} still holds. Concretely, we use our assumptions and the correlators $\<\cM^{(\bar s,s)}J_{\tilde s}\>$ that we have bootstrapped above to fix (\ref{triangle}) to second order in an expansion of $x(\cdot)$ around a straight line, (\ref{deform}). For simplicity, we choose to showcase $\bar s=-s=1/2$ and $\tilde s=0$.

\subsection{First Order}

Consider the variation of the path \eqref{deform} to first order in $\mathrm{v}$. It can be expressed as a linear combination of local operators inserted
on the line and at the two boundaries
\beq\la{firstvar}
\delta\<\overline{\mathcal{O}}\mathcal{W}\mathcal{O}J_0\>=\<\overline{\mathcal{O}}\delta\mathcal{W}\mathcal{O}J_0\>+\<\delta\overline{\mathcal{O}}\mathcal{W}\mathcal{O}J_0\>+\<\overline{\mathcal{O}}\mathcal{W}\delta\mathcal{O}J_0\>\,.
\eeq
Using the properties of the fundamental line detailed in section \ref{sec:setup} and applying the regularisation technique developed in \cite{Gabai:2023lax} we arrive at the following finite values for the terms in (\ref{firstvar})
\begin{align}\la{fo-1}
    \langle\delta\overline{\mathcal{O}}_{\frac12}(x_1)\mathcal{W}\mathcal{O}_{-\frac12}(x_0) J_0\rangle & = \mathrm{v}^+_1 \<\cM^{(\frac32,-\frac12)}_{10}J_0\>\,,\\
    \langle\overline{\mathcal{O}}_{\frac12}(x_1)\mathcal{W}\delta\mathcal{O}_{-\frac12}(x_0) J_0\rangle & = \mathrm{v}^-_0\<\cM^{(\frac12,-\frac32)}_{10}J_0\>\,,\nn\\ \langle\overline{\mathcal{O}}_{\frac12}(x_1)\delta\mathcal{W}\mathcal{O}_{-\frac12}(x_0) J_0\rangle & =\frac{\eta_+\left|x_{10}\right|}{2\ii\,\sin\left(2\pi\Delta\right)}\underset{(0,1}{\Sint}\mathrm{d}\sigma\,\v_\sigma^+\<\cM_{1\sigma}^{(\frac12,\frac12)}J_0\>\,\<\cM_{\sigma0}^{(\frac12,-\frac12)}\>\nn\\
    & + \frac{\ii\,\eta_-\left|x_{10}\right|}{2\sin\left(2\pi\Delta\right)}\underset{0,1)}{\Rint}\mathrm{d}\sigma\,\v_\sigma^-\<\cM_{1\sigma}^{(\frac12,-\frac12)}\>\,\<\cM_{\sigma0}^{(-\frac12,-\frac12)}J_0\>\nn\,,
\end{align}
where $\Sint$ and $\Rint$ stand for integrals around the cuts emerging from $\sigma = 0$ and $\sigma = 1$ respectively. For convenience, we have tuned the deformation so that it deforms the endpoints in the transverse plane only, $\v_0^3=\v_1^3=0$. Using our results for $\<\cM_{1\sigma}^{(\frac12,\frac12)}J_0\>$, $\<\cM_{\sigma0}^{(-\frac12,-\frac12)}J_0\>$ in (\ref{ssbpositive J0}), (\ref{ssbnegative J0}), the first-order variation (\ref{firstvar}) can be evaluated for a smooth deformation profile $\v$.

\subsection{Second Order}
\label{second-order}

The second order variation can be divided into line and boundary contribution as follows
\begin{equation}\label{second-order-var}
    \delta^2\! \left(\overline{\mathcal{O}} \mathcal{W} \mathcal{O}\right) = \delta^2\overline{\mathcal{O}} \mathcal{W} \mathcal{O} + \delta\overline{\mathcal{O}} \mathcal{W} \delta\mathcal{O} + \overline{\mathcal{O}} \mathcal{W} \delta^2\mathcal{O} + \delta\overline{\mathcal{O}} \delta\mathcal{W} \mathcal{O} + \overline{\mathcal{O}} \delta\mathcal{W} \delta\mathcal{O} + \overline{\mathcal{O}} \delta^2\mathcal{W} \mathcal{O}\nn\,.
\end{equation}
At the boundary of the line, the second order variation takes a similar form to the first order in \eqref{boundaryvar} and is given by 
\begin{equation}
    \delta^2\mathcal{O} = \mathrm{v}^{\mu}\mathrm{v}^{\nu} \delta_{\mu}^{(0)} \delta_{\nu}^{(0)} \mathcal{O} + \mathrm{dv}^{\mu}\mathrm{v}^{\nu} \delta_{\mu}^{(1)} \delta_{\nu}^{(0)}\mathcal{O} + \mathrm{dv}^\mu \mathrm{dv}^\nu \delta^{(1)}_\mu\delta^{(1)}_\nu\cO+...\,.
\end{equation}
In the bulk of the line, the second order line variation reads
\begin{align}\label{double-line-var}
\delta^2\mathcal{W} & =\left|x_{10}\right|^2\int\limits_0^1\mathrm{d}\sigma\,\v_\sigma^{\mu}\int\limits_0^1\mathrm{d}\tau\,\v_\tau^{\nu}\,\mathbb{D}_{\mu}\left(x_\sigma\right)\mathbb{D}_{\nu}\left(x_\tau\right)\mathcal{W}+\left|x_{10}\right|\int\limits_0^1\mathrm{d}\sigma\,\delta\mathbb{D}\left(x_\sigma\,\right)\mathcal{W}\,.
\end{align}
Here, using integration by parts, we have chosen the double integral to run over two displacement operators, with no derivatives of $\mathrm{v}$. With this choice, the operator $\delta\mathbb{D}$ takes a form similar to $\delta^2\mathcal{O}$.

We have repeated the procedure of the first order, canceling the divergences using counter terms and keeping only finite contributions. We use the notation of \cite{Gabai:2023lax} in which 
\beq\la{second-order-list}
\delta^2\! \(\overline{\mathcal{O}} \mathcal{W} \mathcal{O}\) = [\bar{B}B]+[BB]+[\bar B\bar B]+[{\rm int}]+[{\rm int}^2]\,,
\eeq
where $B$, $\bar B$, $[{\rm int}]$, $[{\rm int}^2]$ stand for left boundary, right boundary, single integral and double integral contributions, respectively. We find
\begin{align} \label{sordervarations}
    \left[\bar{B}B\right] & =\v_1^+\v_0^- \<\cM_{10}^{(\frac32,-\frac32)}J_0\>\\
    [BB] & =\Big[\gamma_0\v_0^+\v_0^-\delta_3^2+\gamma_1\left(\v_0^+\text{dv}_0^-+\v_0^-\text{dv}_0^+\right)\delta_3+\gamma_4\left(\v_0^+\text{dv}_0^--\v_0^-\text{dv}_0^+\right)\delta_3\nonumber \\
    & +\gamma_3\text{dv}_0^+\text{dv}_0^-+\gamma_2\left(\v_0^+\text{dd}\v_0^-+\v_0^-\text{ddv}_0^+\right)+\gamma_5\left(\v_0^+\text{ddv}_0^--\v_0^-\text{ddv}_0^+\right)\Big]\<\cM_{10}^{(\frac12,-\frac12)}J_0\>\nn\\
    & + \frac{\gamma_6}{N}\v_0^+\v_0^-\<\cM_{10}^{(\frac12,-\frac12)}\>\langle J_0\left(x_0\right)J_0\left(x_2\right)\rangle + \gamma_7(\v_0^-)^2 \<\cM_{10}^{(\frac12,-\frac52)}J_0\>\nn\\
    [\bar{B}\bar{B}] & =\Big[\tilde{\gamma}_0\v_1^+\v_1^-\delta_3^2+\tilde{\gamma}_1\left(\v_1^+\text{dv}_1^-+\v_1^-\text{dv}_1^+\right)\delta_3-\tilde{\gamma}_4\left(\v_1^+\text{dv}_1^--\v_1^-\text{dv}_1^+\right)\delta_3\nonumber \\
    & +\tilde{\gamma}_3\text{dv}_1^+\text{dv}_1^-+\tilde{\gamma}_2\left(\v_1^+\text{dd}\v_1^-+\v_1^-\text{ddv}_1^+\right)-\tilde{\gamma}_5\left(\v_1^+\text{ddv}_1^--\v_1^-\text{ddv}_1^+\right)\Big] \<\cM_{10}^{(\frac12,-\frac12)}J_0\>\nn\\
    & +\frac{\tilde{\gamma_6}}{N}\v_1^+\v_1^-\<\cM_{10}^{(\frac12,-\frac12)}\>\langle J_0\left(x_1\right)J_0\left(x_2\right)\rangle+\tilde{\gamma}_7(\v_1^+)^2 \<\cM_{10}^{(\frac52,-\frac12)}J_0\>\nn\\
    \left[\text{int}\right] & =\Xi_0\left|x_{10}\right|\int^1_0\!\mathrm{d}\sigma\left(\mathrm{dv}_\sigma^+\mathrm{ddv}_{s}^--\mathrm{dv}_\sigma^-\mathrm{ddv}_\sigma^+\right) \<\cM_{10}^{(\frac12,-\frac12)}J_0\>\nn\\
    & +\frac{\Xi_{\pm}}{N}\left|x_{10}\right|\<\cM^{(\frac12,-\frac12)}_{10}\>\int^1_0\! \mathrm{d}\sigma\left(\v_\sigma^+\mathrm{dv}_\sigma^-\pm\v_\sigma^-\mathrm{dv}_\sigma^+\right)\<J_0(x_\sigma) J_0(x_2)\>\nonumber \\
    &+\frac{\ii\left|x_{10}\right|}{2\sin\left(2\pi\Delta\right)}\underset{0,1)}{\Rint}\dd\sigma\,\v_\sigma^-\<\cM_{1\sigma}^{(\frac12,-\frac12)}\> \Big[\eta_-\v_0^-\<\cM_{\sigma 0}^{(-\frac12,-\frac32)}J_0\>+\Omega_-\v_\sigma^-\<\cM_{\sigma 0}^{(-\frac32,-\frac12)}J_0\>\Big]\nn\\
    & -\frac{\ii\left|x_{10}\right|}{2\sin\left(2\pi\Delta\right)}\underset{(0,1}{\Sint}\dd\sigma\,\v_\sigma^+ \Big[\eta_+\v_1^+\<\cM_{1\sigma}^{(\frac32,\frac12)}J_0\>+\Omega_+\v_\sigma^+\<\cM_{1\sigma}^{(\frac12,\frac32)}J_0\>\Big]\<\cM_{\sigma 0}^{(\frac12,-\frac12)}\>\nn \\
    \left[\text{int}^2\right] & =\frac{\ii\,\eta_-\left|x_{10}\right|}{2\sin\left(2\pi\Delta\right)}\underset{0,1)}{\Rint}\mathrm{d}\sigma\,\v_\sigma^-\Big[\< \cM_{1\sigma}^{(\frac12,-\frac12)} J_0\>\delta\<\cM_{\sigma 0}^{(-\frac12,-\frac12)}\>+\<\cM_{1\sigma}^{(\frac12,-\frac12)}\>\delta \<\cM_{\sigma 0}^{(-\frac12,-\frac12)}J_0\>\Big]\nn
\end{align}
where
\begin{align}
\delta\<\cM_{\sigma 0}^{(-\frac12,-\frac12)}\> 
&= \frac{\eta_+ |x_{10}|}{2\ii\sin(2\pi\Delta)}\underset{[0,\sigma]}{\oint}\mathrm{d}\tau\,\v_\tau^+\<\cM_{\sigma\tau}^{(-\frac12,\frac12)}\>\<\cM_{\tau0}^{(\frac12,-\frac12)}\>\,,\\
\delta \<\cM_{\sigma0}^{(-\frac12,-\frac12)} \! J_0\> &= \frac{\eta_+ |x_{10}|} {2\ii\sin(2\pi\Delta)} \! \underset{[0,\sigma]}{\oint} \!\dd\tau\,\v_\tau^+\Big[ \<\cM_{\sigma\tau}^{(-\frac12,\frac12)} \! J_0\> \<\cM_{\tau0}^{(\frac12,-\frac12)}\> \!+\! \<\cM_{\sigma\tau}^{(-\frac12,\frac12)}\> \<\cM_{\tau0}^{(\frac12,-\frac12)} \! J_0\>\Big]\,. \nn
\end{align}
Here, the $\gamma_i$'s, $\tilde \gamma_i$'s, $\Omega_i$'s and $\Xi_i$'s are coefficients that we need to fix. Some of them have already been computed in \cite{Gabai:2023lax}. Those values are local data, and therefore the appearance of the local operator cannot affect them. The new coefficients that only contribute due to the presence of the local operator are $\gamma_6$, $\gamma_7$, $\tilde \gamma_6$, $\tilde\gamma_7$, $\Omega_\pm$ and $\Xi_\pm$. To bootstrap them, we find it sufficient to only impose the symmetry of a straight line under a conformal transformation, without deforming it first (as was necessary in \cite{Gabai:2023lax}). We relegate details to appendix \ref{app:smooth} because they are very similar to \cite{Gabai:2023lax}. We find a unique solution for which
\begin{equation}\la{bootsol}
    \gamma_6 = \frac{2(1-\Delta)  f_{\frac12,-\frac12}(1)}{c_+ \mathcal{N}_0}\,,\quad  \tilde\gamma_6 =0\,,\quad  \gamma_7=\tilde\gamma_7=\frac12\,,\quad \Omega_\pm= \frac{\eta_\pm}2\,,\quad \Xi_\pm=0\,.
\end{equation}

We conclude that, at least at the second order we are working at, the correlator (\ref{triangle}) is fixed by our large-$N$ assumptions, with no free parameter.

\section{Discussion}
\label{sec:dis}

We studied a three-dimensional conformal field theory (CFT) in the large-$N$ limit, characterised by higher-spin symmetry that is broken at order $1/N$, and the presence of a fundamental conformal line defect.\footnote{Here, we have assumed higher-spin symmetry for any integer spin. We expect that by taking charge-conjugation-invariant combinations, our results also apply to theories with only even higher-spin symmetries.} Leveraging these features, we bootstrapped the correlation functions between a spin-$\tilde s$ local current and a straight conformal defect, in the presence of fundamental and anti-fundamental defect-changing operators.

The residual $SL(2,\mathbb{R})\times U(1)$ symmetry preserved by a straight line defect constrains the correlator to a sum over $2 \tilde s +1$ independent tensor structures, each multiplied by a function of a single conformal cross-ratio. When the boundary operators have transverse spins of the same sign, we found that the correlator exhibits no non-trivial dependence on this cross-ratio. Conversely, for boundary operators with opposite transverse spins, the correlator becomes a linear combination of hypergeometric functions.

Our large-$N$ bootstrap solution is characterised by two parameters. The first is the lowest boundary scaling dimension, $\Delta$. The second, denoted $a_2$, is the coefficient of $J_2J_1$ in the $1/N$ correction to the divergence of $J_3$, see (\ref{divergence J3}). Alternatively, $a_2$ can be identified as the coefficient of $J_1$ in the $1/N$ correction to the displacement operator, see (\ref{displace}) and (\ref{div J3 final}).

A key example where our assumptions apply is three-dimensional Chern–Simons theory coupled to fundamental bosonic or fermionic matter. In the planar limit, the gauge-group rank $N$ and the Chern–Simons level $k$ are both large with fixed ’t Hooft coupling $\lambda=N/k$.\footnote{In this limit, distinctions between gauge groups such as $SU(N)$ vs. $U(N)$, or the $U(1)$ level in $U(N)_{k,k'}$ become irrelevant.} These theories admit two distinct conformal line operators. In the bosonic description, they have boundary operators of dimensions $\Delta=(1\pm\lambda)/2$ respectively, \cite{Gabai:2022vri,Gabai:2022mya}. For theories with gauge groups $SU(N)$ or $U(N)$, the parameter $a_2$ vanishes, as shown by perturbative computations \cite{Giombi:2011kc}. However, consistent CFTs with non-zero $a_2$ exist. They can be realised in Chern–Simons matter theories with gauge group $SU(N)_k\times U(1)_{k'}$, where the matter is charged under the $U(1)$ factor and $k'=O(N)$ in the planar limit. Integrating out the $U(1)$ gauge field induces a non-local term of the form $\frac1{k'}(J_1\p^{-1}J_1)$. This marginal double-trace deformation does not run due to the quantisation of the $U(1)$ level and does not affect the dimension of the conserved current $J_1$. At leading order, it influences only correlators involving $J_1$, and hence contributes to $a_2$. This parameter should also emerge in a bootstrap of three-point functions involving both even and odd high-spin currents, in the spirit of \cite{Maldacena:2012sf}.

The bootstrapped correlators encode the complete set of bulk-to-line operator product expansion (OPE) coefficients. These OPE coefficients allow the line operator to be expressed as an infinite sum over local operators. In this sense, our results provide an explicit construction of two conformal line operators (which are related by $\Delta\leftrightarrow1-\Delta$) in large-$N$, higher-spin theories.

Finally, large-$N$ Chern–Simons matter theories are known to be holographically dual to Vasiliev’s higher-spin theory in $AdS_4$. Our results open the door to extending this holographic duality to include conformal line defects, offering a new window into the structure of higher-spin holography.

\section*{Acknowledgments}

We thank De-liang Zhong for collaboration at an early stage of this project. We thank O. Aharony, B. Gabai and D.-l. Zhong for useful discussions and for comments on the manuscript. AS and EU are supported by the Israel Science Foundation, grant numbers 1197/20 and 1099/24. The work of GF is funded by the Deutsche Forschungsgemeinschaft (DFG, German Research Foundation) – Projektnummer 508889767.

\begin{appendix}

\section{Perturbative Checks in the Quasi-Fermion Theory}
\label{pertapp}

In this appendix, we compare some of our bootstrap results with explicit 1-loop perturbative computations. The computations are performed in the quasi-fermionic theory, which is defined by coupling Chern--Simons theory to a fundamental fermion
\begin{equation}
    S= \frac{\ii k}{4\pi} \varepsilon^{\mu\nu\rho} \int\dd^3x \Tr\big(A_\mu\p_\nu A_\rho-\frac{2\ii}3 A_\mu A_\nu A_\rho\big) + \int\dd^3x\,  \bar\psi\gamma^\mu D_\mu \psi\,,
\end{equation}
where $D_\mu\psi= \p_\mu \psi -\ii A_\mu \psi $ and $\bar\psi = \psi^\dagger$ and the gauge group is $SU(N)$. Here, $\psi^i_a$ is a two-component spinor in the fundamental of $SU(N)$, i.e. $a\in\{1,2\}$ and $i\in\{1,\dots,N\}$. The $\gamma$ matrices in Euclidean signature are taken to be the Pauli matrices, $\gamma^\mu=\sigma^\mu$. There are no relevant or marginal operators other than the fermion mass. Hence, after tuning it to zero, the theory is conformal. We consider the CFT in the planar limit, $N\rightarrow\infty$ with $\lambda=N/k$ fixed. At large $N$, the spectrum of primary operators consists of a tower of conserved twist-1 currents and a single, twist-2 scalar operator $J_0 = \bar\psi \psi$.

The free propagators in Feynman gauge are given by
\begin{equation}
    \< \psi^i_a(x) \bar\psi^j_b(y)\>_0 = \delta^{ij}\frac{(x-y)_\mu (\gamma^\mu)_{ab}}{4\pi |x-y|^3}\,,
\end{equation}
\begin{equation}
    \< A^I_\mu(x)A^J_\nu(y)\>_0 = -\frac{\ii}{k}\delta^{IJ} \varepsilon_{\mu\nu \rho}\frac{(x-y)^\rho}{|x-y|^3}\,,
\end{equation}
where $I,J$ are adjoint, gauge-group indices. The Greek indices run from 1 to 3. Finally, we normalise the adjoint generators $T_I$ of the gauge group such that $\Tr(T_I T_J) = \delta_{IJ}/2$.

The stable conformal line operator in this theory is the standard Wilson line
\begin{equation}
    \mathcal{W}[x(\cdot)] = {\cal P}\,\exp\left[{\ii\int\dd\sigma\,\dot{x}^\mu(\sigma)} A_\mu(x(\sigma)) \right]\,.
\end{equation}
As explained in \cite{Gabai:2022vri,Gabai:2022mya}, the corresponding fundamental and anti-fundamental boundary operators for a line oriented in the $\hat x^3$ direction are
\begin{equation}\label{boundary ferm}
    \overline\cO_{\bar s} = \frac1{\sqrt{N}} \begin{cases}
D^{\bar s - \frac12}_+ \bar\psi_2 &\quad \bar s > 0\\
D^{-\bar s-\frac12}_- \bar\psi_1 & \quad \bar s< 0
\end{cases} \qquad\text{and}\qquad \cO_{s} = \frac1{\sqrt{N}} \begin{cases}
D^{s - \frac12}_+ \psi_1 &\quad s > 0\\
D^{-s-\frac12}_- \psi_2 & \quad s< 0
\end{cases}\,.
\end{equation}

\subsection{One-loop Check for $\< \mathcal{M}^{(\frac12,-\frac12)}  J_0\>$ }
\label{pert-fermion}

Given the dictionary \eqref{boundary ferm}, the mesonic line operator we consider is
\begin{equation}
    \mathcal{M}^{(\frac12,-\frac12)}_{10} = \frac1N\bar\psi_2(x_1)\mathcal{W}\, \psi_2(x_0)\,.
\end{equation}
At tree-level, the correlator is simply a product of two free propagators
\begin{equation} \label{fermiontree}
    \langle\mathcal{M}_{10}^{(\frac12,-\frac12)} J_0(x_2)\rangle_0 = \frac{\cos\theta}{(4\pi)^2 |x_{20}|^2 |x_{21}|^2}\,.
\end{equation}

At one loop, there are a few Feynman diagrams that are relevant, see figure \ref{diagrams}. In Feynman gauge, the diagram \ref{diagrams}(d) gives the anomalous spin of the boundary operators. This anomalous part is canceled between the right and left operators, so this diagram does not contribute. In the following subsections we compute the other three diagrams.

\begin{figure}
\centering{}\includegraphics[scale=1.7]{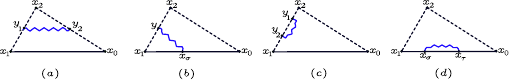}\caption{The four types of one-loop diagrams that contribute to $\<\mathcal{M}_{10}^{(\bar s,s)}J_{\tilde s}(x_2)\>$. The dashed lines are fermion propagators, the blue wiggly line is a gluon propagator, and the full line is the Wilson line. We denote these diagrams by (a) bulk-to-bulk gluon exchange, (b) bulk-to-line gluon exchange, (c) propagator corrections and (d) line-to-line gluon exchange.}
\la{diagrams}
\end{figure}

\subsubsection{Bulk-to-Bulk Gluon Exchange}

The Feynman integral, see figure \ref{diagrams}(a), is
\begin{align}
    I_1 &= \frac{\ii\lambda\, \varepsilon_{\mu\nu\rho}}{2 (4\pi)^4} (\gamma^{\eta} \gamma^{\mu} \gamma^{\delta} \gamma^{\beta} \gamma^{\nu} \gamma^{\alpha})_{22} \!\int\dd^3y_1 \dd^3y_2 \frac{(y_1-x_2)_{\beta} (x_1-y_1)_{\alpha} (x_2-y_2)_{\delta} (y_2-x_0)_{\eta} \,y_{12}^{\rho}}{|y_1-x_2|^3 |x_1-y_1|^3 |x_2-y_2|^3 |y_2-x_0|^3 |y_{12}|^3}  \nn\\
    &= \frac{-\lambda}{(4\pi)^4} \left(\gamma^{\eta} \gamma^{\rho} \gamma^{\alpha}\right)_{22} \!\int\dd^3y_1 \dd^3y_2 \frac{(y_1-x_2)\cdot(x_2-y_2) (x_1-y_1)_{\alpha} (y_2-x_0)_{\eta} \,y_{12,\rho}}{|y_1-x_2|^3 |x_1-y_1|^3 |x_2-y_2|^3 |y_2-x_0|^3 |y_{12}|^3}\,, \label{bulktobulkdiagram}
\end{align}
where we used properties of the Pauli matrices to go from the first to the second line.  We conjecture that this integral is equal to
\begin{equation}
    I_1 = \frac{\lambda\left[|x_{20}|(|x_{20}|+x_{02}^3) + |x_{21}|(|x_{21}|+x_{21}^3)\right]}{4(4\pi)^2 |x_{20}|^3 |x_{21}|^3}\,.
\end{equation}
We were unable to derive this analytically. Instead, we have checked that it agrees with numerical evaluation of the integral for a few generic values of the coordinates.

\subsubsection{Bulk-to-Line Gluon Exchange} \label{bulktolineJ0}

There are two such diagrams. One of them, see figure \ref{diagrams}(b), corresponds to
\begin{align}
    I_2 & = 
    \frac{\ii\lambda\, x_{20,\delta}}{2 (4\pi)^3 |x_{20}|^3} (\gamma^{\delta}\gamma^{\beta}\gamma^{\nu}\gamma^{\alpha})_{22} \varepsilon_{\mu\nu\rho} \int\limits_0^1\dd \sigma\,\dot{x}_\sigma^\mu \int \dd^3y \frac{(x_2-y)_{\beta}  (y-x_1)_{\alpha} (x_\sigma-y)^{\rho}}{|x_1 - y + \epsilon\hat{e}_3|^3 |y-x_2|^3 |x_\sigma-y|^3} \nn\\
    & = \frac{\ii\lambda\,|x_{10}| x_{20,\delta}}{2 (4\pi)^2 |x_{20}|^3} \varepsilon_{3\nu\rho} (\gamma^{\delta}\gamma^{\beta}\gamma^{\nu}\gamma^{\alpha})_{22} \int\limits_0^1\dd\sigma\, \p_{x_1^\alpha} \p_{x_2^\beta} \frac{\frac{(x_{1\sigma}+\epsilon\hat{e}_3)^{\rho}}{|x_{1\sigma}+\epsilon\hat{e}_3|} + \frac{x_{2\sigma}^\rho}{|x_{2\sigma}|}}{|x_{1\sigma}| + |x_{2\sigma}| + |x_{12}|}\nn\\
    &= \frac{\ii\lambda\,|x_{10}| x_{20,\delta}}{2 (4\pi)^2 |x_{20}|^3} \varepsilon_{3\nu\rho} (\gamma^{\delta}\gamma^{\beta}\gamma^{\nu}\gamma^{\alpha})_{22} \\
    &\qquad\quad \times\p_{x_2^\beta} \int\limits_0^1\frac{\dd\sigma}{|x_{1\sigma}| + |x_{2\sigma}| + |x_{12}|}\left[\frac{\delta_\alpha^\rho}{|x_{1\sigma}+\epsilon\hat{e}_3|} - \frac{x_{2\sigma}^\rho \left(\delta^3_\alpha +\frac{x_{12,\alpha}}{|x_{12}|}\right)}{|x_{2\sigma}|(|x_{1\sigma}| + |x_{2s}| + |x_{12}|)} \right]\nn \,,
\end{align}
where $x_{\sigma}=x_0+\sigma x_{10}$, and we only keep the regulator $\epsilon$ in the terms that diverge. The divergent part of the integral comes only from the region of integration near $\sigma=1$ from the first term inside the bracket; it is thus the same as
\begin{equation}
    \frac{\lambda\,|x_{10}| x_{20,\delta}}{(4\pi)^2 |x_{20}|^3} (\gamma^{\delta}\gamma^{\beta})_{22}\p_{x_2^\beta} \frac1{2|x_{12}|} \int\limits_0^1\frac{\dd\sigma}{|x_{1\sigma}+\epsilon\hat{e}_3|} = \frac{\lambda \cos\theta\,\ln \frac{\epsilon}{|x_{10}|}}{2 (4\pi)^2 |x_{20}|^2 |x_{21}|^2}+ O(\epsilon)\,.
\end{equation}
For the finite part, we only need the following two simpler integrals, which we computed using Mathematica
\begin{equation}
    V_1 = \int\limits_0^1 \frac{|x_{10}|\dd\sigma}{|x_{2\sigma}| \left(|x_{1\sigma}| + |x_{2\sigma}| + |x_{12}|\right)^2} = \frac{|x_{20}| (\cos\theta-1)}{2 (|x_{12}|+x^3_{12})^2 |x_{10}|} - \frac{\ln\!\left(\frac{|x_{20}| + |x_{21}|- |x_{10}|}{2 |x_{12}|}\right)}{(|x_{12}|+x^3_{12})^2 }
\end{equation}
and
\beq
    V_2= \int\limits_0^1\frac{\dd\sigma}{1-\sigma} \left[ \frac1{|x_{1\sigma}|+|x_{2\sigma}|+|x_{12}|} - \frac1{2|x_{21}|} \right]= -(|x_{12}|+x^3_{12}) V_1 + \frac{\ln\!\left(\frac{2|x_{21}|}{|x_{20}| (1+\cos\theta)}\!\right)}{2 |x_{12}|} \,.
\eeq
If we define
\begin{align}
    B_1 & = -\frac{x_{20,\delta}}{|x_{20}|^3} (\gamma^{\delta}\gamma^{\beta}\gamma^3)_{22}\p_{x_2^\beta} V_2 =  \frac{x_{20}\cdot \p_{x_2} V_2}{|x_{20}|^3}\,, \\
    B_2 & = -\frac{\ii\, x_{20,\delta}}{2 |x_{20}|^3} \varepsilon_{3\nu\rho} (\gamma^{\delta}\gamma^{\beta}\gamma^{\nu}\gamma^3)_{22} \p_{x_2^\beta} [x^{\rho}_2 V_1] = \frac{x_{20}^3 (V_1 + x_2^- \p_{x_2^-} V_1) - x_2^+ x_2^- \p_{x_2^3} V_1}{|x_{20}|^3}\,,\nn\\
    B_3 & = -\frac{\ii\, x_{20,\delta}\, x_{12,\alpha}}{2 |x_{20}|^3 |x_{12}|}  \varepsilon_{3\nu\rho}(\gamma^{\delta}\gamma^{\beta}\gamma^{\nu}\gamma^{\alpha})_{22} \p_{x_2^\beta}\left[x^{\rho}_2V_1\right] = \frac{x_{12}^3}{|x_{12}|} B_2 + \frac{x_2^+ x_2^- \left[2V_1 + x_{20}\cdot \p_{x_2} V_1\right]}{|x_{20}|^3 |x_{21}|}\,,\nn
\end{align}
then the Feynman integral $I_2$ is equal to
\begin{equation}
    I_2 = \frac{\lambda}{(4\pi)^2} \left[\frac{\cos\theta\ln\!\left(\frac{\epsilon}{|x_{10}|}\right)}{2  |x_{20}|^2 |x_{21}|^2} + B_1 + B_2 + B_3\right] + O(\epsilon)\,.
\end{equation}
Putting everything together, we find
\begin{equation}
    I_2 = \frac{\lambda}{(4\pi)^2}\!\! \left[\frac{\cos\theta\ln\!\left(\frac{\epsilon |x_{20}| (1+\cos\theta)}{2|x_{10}| |x_{21}|}\right) + \cos\theta - 1}{2\, |x_{20}|^2 |x_{21}|^2} + \frac{(|x_{21}| + x^3_{21}) |x_{10}|}{4\, |x_{20}|^3 |x_{21}|^3} + \frac{(1 - \cos\theta)}{4\, |x_{20}|^2 |x_{21}| |x_{10}|}\right]\! + O(\epsilon)\,.
\end{equation}

The contribution associated to the second diagram is 
\begin{equation}
    I_3 = \frac{\ii\lambda\, x_{21,\alpha}}{2 (4\pi)^3 |x_{12}|^3} \varepsilon_{\mu\nu\rho}(\gamma^{\delta}\gamma^{\nu}\gamma^{\beta}\gamma^{\alpha})_{22}\int\limits_0^1\dd \sigma\, \dot{x}_{\sigma}^{\mu} \int\dd^3 y\frac{(x_0-y)_{\delta} (y-x_2)_{\beta} (x_\sigma-y)^{\rho}}{|y - x_0 + \epsilon\hat{e}_3|^3 |y-x_2|^3 |x_{\sigma}-y|^3}\,.
\end{equation}
A computation very similar to the previous one yields
\begin{equation}
    I_3 = \frac{\lambda}{(4\pi)^2}\!\! \left[\frac{\cos\theta\ln\!\left(\frac{\epsilon |x_{21}| (1+\cos\theta)}{2|x_{10}| |x_{20}|}\right) + \cos\theta - 1}{2\, |x_{20}|^2 |x_{21}|^2} + \frac{(|x_{20}| + x^3_{02}) |x_{10}|}{4\, |x_{20}|^3 |x_{21}|^3} + \frac{(1 - \cos\theta)}{4\, |x_{20}| |x_{21}|^2 |x_{10}|}\right]\! + O(\epsilon)\,.
\end{equation}

\subsubsection{Propagator Corrections}

We have two fermion propagator corrections via gluon exchange. See one of them in figure \ref{diagrams}(c). They are simply
\begin{equation}
    I_4 = \frac{\lambda\, x^3_{02}}{2(4\pi)^2 |x_{20}|^3 |x_{21}|^2}\quad \text{and}\quad I_5 = \frac{\lambda\, x^3_{21}}{2(4\pi)^2 |x_{20}|^2 |x_{21}|^3}\,.
\end{equation}

\subsubsection{Full Conformal Result}

Let us first recall that we expect, from \eqref{correl total spin pos} and \eqref{sbs}, that
\begin{equation}
    \< \mathcal{M}^{(\frac12,-\frac12)}_{10} J_0(x_2) \> = \frac{f_{\frac12,-\frac12}(1)}{|x_{20}|^2 |x_{12}|^2} \left(\frac{\epsilon}{|x_{10}|}\right)^{\lambda}  {}_2 F_1\!\left(2,-1-\lambda; 1-\frac{\lambda}2; \frac{1-\cos\theta}2\right)\,,
\end{equation}
where we used that $\Delta = 1 + \frac{\lambda}2$, see \cite{Gabai:2022vri,Gabai:2022mya}. Expanding this around $\lambda = 0$ yields
\begin{equation}\label{prediction one loop}
    \langle \mathcal{M}^{(\frac12,-\frac12)}_{10} J_0(x_2) \rangle = \frac{f_{\frac12,-\frac12}(1)}{|x_{20}|^2 |x_{12}|^2} \left(\cos\theta + \lambda\left[\cos\theta \ln\!\left(\frac{\epsilon(1+\cos\theta)}{2 |x_{10}|}\right) + \cos\theta - 1 \right]\right) + O(\lambda^2)\,.
\end{equation}

Although we have not been able to compute $I_1$ rigorously, our conjecture for its evaluation is consistent with the prediction \eqref{prediction one loop} since we find
\begin{equation}
    \langle \mathcal{M}^{(\frac12,-\frac12)}_{10} J_0(x_2) \rangle_1 = \sum_{i=1}^5 I_i = \frac{\lambda}{(4\pi)^2 |x_{20}|^2 |x_{12}|^2}\left[\cos\theta \ln\!\left(\frac{\epsilon(1+\cos\theta)}{2 |x_{10}|}\right) + \frac32\cos\theta - 1\right]\,,
\end{equation}
which, together with the tree-level result, implies
\begin{equation}
    f_{\frac12,-\frac12}(1) = \frac1{(4\pi)^2}\left[ 1+\frac{\lambda}2+O(\lambda^2)\right]\,.
\end{equation}

\subsection{Spins of the Same Sign} \label{constfunc}

In the previous subsection, we chose the boundary operators to be $\bar\psi_2$ and $\psi_2$. However, if we choose $\bar\psi_1$ and $\psi_2$ or $\bar\psi_2$ and $\psi_1$, we find respectively  
\begin{align}
    \<\mathcal{M}_{10}^{(-\frac12,-\frac12)} J_0(x_2)\> &= \frac{-\sqrt2\,x_2^+|x_{10}|}{(4\pi)^2 |x_{21}|^3 |x_{20}|^3} \left[1+\lambda\ln\frac{|x_{21}|}{|x_{20}|}\right] + O(\lambda^2)\,,\\
    \<\mathcal{M}_{10}^{(\frac12,\frac12)} J_0(x_2)\> &= \frac{\sqrt2\, x_2^- |x_{10}|}{(4\pi)^2 |x_{20}|^3 |x_{21}|^3} \left[1+\lambda\ln\frac{|x_{20}|}{|x_{21}|}\right] + O(\lambda^2)\,,
\end{align}
for arbitrary $x_2$. Comparing this result to \eqref{corrsbars}, we extract
\begin{align}\label{fm12m12}
    f_{-\frac12,-\frac12}(\cos\theta) &= f_{-\frac12,-\frac12}(1) = -\frac{\sqrt2}{(4\pi)^2} + O(\lambda^2)\,,\\
    f_{\frac12,\frac12}(\cos\theta) &= f_{\frac12,\frac12}(1) = \frac{\sqrt2}{(4\pi)^2} + O(\lambda^2)\,.
\end{align}
It agrees with our prediction that the functions of $\cos\theta$ are actually constant when the spins of the boundary operators are of the same sign, see equations \eqref{ssbpositive J0} and \eqref{ssbnegative J0}.

\subsection{One-Loop Check of \eqref{csc0}} \la{Appcsc0}

We test in this appendix one of the equations \eqref{csc0}, which states that
\begin{equation}
    \frac{d_1^2}{\mathcal{N}_1} = 16 \cos^2(\pi\lambda/2)\frac{d_0^2}{\mathcal{N}_0}\,.
\end{equation}
We find that it indeed holds through one-loop because the various constants are given by
\begin{align}
    \mathcal{N}_0 & = \frac2{(4\pi)^2} + O\left(\lambda^2\right),\qquad\mathcal{N}_1 = \frac1{2(4\pi^2)} + O\left(\lambda^2\right),\\
    d_0 & = - \frac{\sqrt2}{(4\pi)^2} + O\left(\lambda^2\right),\qquad d_1= \frac{2\sqrt2}{(4\pi)^2} + O\left(\lambda^2\right)\,.
\end{align}
The details of the computations are presented in the following.

\subsubsection{$J_0$ Two-Point Function}

At tree level,
\begin{equation} \label{treeJ0}
    \langle J_0(x)J_0(y)\rangle_0 =  \frac2{(4\pi)^2|x-y|^4}\,.
\end{equation}
There are three possible one-loop corrections: a bulk-to-bulk gluon exchange between the two propagators and the self-energy corrections to the propagators. However, all these corrections vanish. 
For instance, the bulk-to-bulk diagram is
\begin{equation}
    \frac{\lambda\,\varepsilon_{\mu\nu\rho}}{2\ii (4\pi)^4}\!\! \int\dd^3 y_1 \dd^3 y_2 {\rm Tr} (\gamma^{\beta}\gamma^{\nu}\gamma^{\delta}\gamma^{\eta}\gamma^{\mu}\gamma^{\alpha}) \frac{\left(x-y_1\right)_{\alpha} (y_2-x)_{\beta} (y-y_2)_{\delta} (y_1-y)_{\eta} (y_1-y_2)^{\rho}}{|x-y_1|^3 |y_2-x|^3 |y-y_2|^3 |y_1-y|^3 |y_1-y_2|^3}  = 0\,,
\end{equation}
because the numerator simply vanishes. This means that
\begin{equation}
    \< J_0(x)J_0(y)\>_1 = 0\,,
\end{equation}
and that the normalisation, as defined in \eqref{normalization Js}, is indeed $\mathcal{N}_0=\frac2{(4\pi)^2}+O(\lambda^2)$.

\subsubsection{$J_1$ Two-Point Function}

We recall that $J^\mu_1=\bar\psi\gamma^\mu\psi$. At tree level, we have
\begin{equation} \label{treeJ1}
    \< J^\mu_1(x)J^\nu_1(y)\>_0 = 2\frac{2(x-y)^\mu (x-y)^\nu -\delta^{\mu\nu} |x-y|^2}{(4\pi)^2 |x-y|^6}\,.
\end{equation}
Examining one-loop corrections, we find three possible contributions: a bulk-to-bulk gluon exchange between the two tree-level propagators and the two self-energy corrections (one for each of these propagators). Focusing on $\< J^3_1(0)J^3_1(x)\>_1$ in the kinematics $x^1 = x^2 = 0$, it is easy to see that the three possible contributions vanish separately. This is trivial for the self-energy corrections. The bulk-to-bulk correction is an integral over two variables $y_1$ and $y_2$, and the sign of its integrand changes under the simple change of variables $y_1^1\to - y_1^1$, $y_2^1\to - y_2^1$, so the integral must also vanish.
We thus conclude that at separated points
\begin{equation}
    \langle J^3_1(x)J^3_1(y)\rangle_1 = 0\qquad\text{for}\qquad x\ne y\,,
\end{equation}
and that the $J_1$ normalisation, as defined in \eqref{normalization Js}, is simply $\mathcal{N}_1 = \frac1{2(4\pi)^2}+O(\lambda^2)$.

\subsubsection{$d_0$}

As stated below \eqref{ssbnegative J0 bis}, the constant $d_0$ is related to $f_{-\frac12,-\frac12}(1)$ through
\begin{equation}
    d_0 = \frac{f_{-\frac12,-\frac12}(1)}{(2\Delta-1)(3-2\Delta)} = \frac{f_{-\frac12,-\frac12}(1)}{1-\lambda^2}\,.
\end{equation}
Using equation $\eqref{fm12m12}$, we find that $d_0=\frac{-\sqrt2}{(4\pi)^2}+O(\lambda^2)$.

\subsubsection{$d_1$}

We focus on $\< M_{10}^{(-\frac12,-\frac12)} J_1^-(x_2)\> $ when $x^1_2=x^2_2=0$. According to \eqref{conformal J1} and \eqref{eq1j1}, the correlator should be of the form
\begin{equation} \label{nonpertcorrj1}
    \langle M_{10}^{(-\frac12,-\frac12)} J_1^-(x_2)\rangle =\frac{-d_1}{2\,|x_{20}|^{2+\lambda} |x_{21}|^{2-\lambda}}\,,
\end{equation}
where we used that $\Delta = 1+\lambda/2$. At tree level, we get
\begin{equation}
     \langle M_{10}^{(-\frac12,-\frac12)} J_1^-(x_2)\rangle_0=\frac{-\sqrt2}{(4\pi)^2 |x_{20}|^2 |x_{21}|^2}\,.
\end{equation}
At one-loop, the bulk-to-bulk diagram\footnote{Same integrand as in $\eqref{bulktobulkdiagram}$ up to the insertion of $\gamma^-$ between $\gamma^\delta$ and $\gamma^\beta$, and the fact that we now want the $21$ coefficient of the matrix.} yields (using numerical integration)
\begin{equation}
    L_1 = \frac{\lambda}{\sqrt2 (4\pi)^2}\begin{cases}
    \frac1{|x_{20}|^3 |x_{21}|}+\frac1{|x_{20}|^2 |x_{21}|^2}  & x_2^3>x_1^3\\
    -\frac1{|x_{20}| |x_{21}|^3} -\frac1{|x_{20}|^2 |x_{21}|^2}& x_0^3>x_2^3
    \end{cases}\,.
\end{equation}
When $x_2^1= x_2^2 = 0$, the bulk-to-line diagrams are actually the same as those computed in appendix \ref{bulktolineJ0} up to trivial pre-factors: namely, we get $\sqrt2I_2$ and $-\sqrt2I_3$. Lastly, the sum of the two self-energy corrections is
\begin{equation}
    L_2 = \frac{\lambda\sqrt2\, \text{sign}(x_{12}^3)}{(4\pi)^2 |x_{20}|^2 |x_{21}|^2}\,.
\end{equation}
The full correction at order $\lambda$ is thus 
\begin{equation}
    \<\mathcal{M}_{10}^{(-\frac12,-\frac12)} J_1^-(x_2) \>_1 =  L_1 + \sqrt2I_2 - \sqrt2I_3 + L_2 = \frac{\lambda\sqrt2 \ln\frac{\left|x_{20}\right|}{\left|x_{21}\right|}}{\left(4\pi\right)^2\left|x_{20}\right|^2\left|x_{21}\right|^2}\,.
\end{equation}
Combining it with tree level, we obtain
\begin{equation}
    \<\mathcal{M}_{10}^{(-\frac12,-\frac12)} J_1^-(x_2)\> = \frac{-\sqrt2}{(4\pi)^2 |x_{20}|^2 |x_{21}|^2}\left[1+\lambda\ln\frac{|x_{21}|}{|x_{20}|}\right]+O(\lambda^2)\,.
\end{equation}
Comparing with $\eqref{nonpertcorrj1}$, we find $d_1=\frac{2\sqrt2}{(4\pi)^2}+O(\lambda^2)$.

\section{Conformal Structures in Embedding Space}
\label{app:embedding}

We use here the embedding space formalism to determine the allowed structures $Q_i$ of \eqref{conf structures}. This formalism is, by now, relatively standard, see \cite{Costa:2011mg} and references therein for CFTs without defect and \cite{Billo:2016cpy} for CFTs with defects. Our aim is to constrain the function
\begin{equation}
    f^{(\bar s,s)}(x_0^3,x_1^3,x_2,\zeta) = \left(\frac{|x_{20}| |x_{21}|}{|x_{10}|\, x_2^{\pm}}\right)^{|s+\bar s|} \<\cM_{10}^{(\bar s,s)}\,\zeta_\mu J_1^\mu(x_2)\>\,,
\end{equation}
knowing the way it transforms under $\mathrm{SL}(2,\mathbb{R})\times \mathrm{U}(1)$. A special conformal transformation in the third direction with parameter $b\in\mathbb{R}$ is
\begin{equation}
    x^\mu\mapsto \phi^\mu_b(x) = \frac{x^\mu - b\,\delta^\mu_3 |x|^2}{\Omega_b(x)}\,,
\end{equation}
where
\begin{equation}
    \Omega_b(x) = 1 - 2\, b\,x^3 + b^2 |x|^2\,.
\end{equation}
Under such a transformation\footnote{To be precise, we should restrict ourselves to transformations that map $x_0^3<x_1^3$ to $\phi^3_b(x_0)<\phi^3_b(x_1)$.}, covariance of $f^{(\bar s,s)}$ reads
\begin{equation}
    f^{(\bar s,s)}(\phi_b(x_0), \phi_b(x_1),\phi_b(x_2), I_b(x_2)\zeta) = \Omega^{\Delta_{s}}_b\!(x_0)\, \Omega^{\Delta_{-\bar s}}_b\!(x_1)\,  \Omega^2_b(x_2)\, f^{(\bar s,s)}(x_0^3,x_1^3,x_2,\zeta)\,,
\end{equation}
where
\begin{equation}
    I^{\mu\nu}_b(x) = \left(\delta^\mu_\rho - 2\frac{(x^\mu - b\,\delta^\mu_3 |x|^2) (x_\rho - b\,\delta_{\rho 3} |x|^2)}{|x|^2 \Omega_b(x)}\right)\left(\delta^\rho_\nu - 2\frac{x^\rho x_\nu}{|x|^2 }\right)\,.
\end{equation}
Written this way, it is difficult to solve this non-linear constraint. However, there is an equivalent representation of the conformal group in which the constraint is linearly realised. The points $x_0$, $x_1$, and $x_2$ are promoted to five-dimensional variables $X_0$, $X_1$, $X_2$ that are restricted to lie on the null cone $\left\{X\cdot X = 0, X^5>0\right\}$,
where the metric is
\begin{equation}
    X\cdot Y = X^\mu Y_\mu + X^4 Y^4 - X^5 Y^5\,.
\end{equation}
Moreover, we define
\begin{equation}
    X^{\perp} = (X^1,X^2)\,,\quad X^{\parallel} = (X^3,X^4,X^5)\,,
\end{equation}
and we further require $X_0^\perp = X_1^\perp = 0$. The full three-dimensional conformal group is now linearly realised as $O(4,1)$ and the $\mathrm{SL}(2,\mathbb{R})\times \mathrm{U}(1)$ subgroup we are interested in is the following set of block-diagonal matrices:
\begin{equation}
    \left\lbrace \begin{pmatrix}
        R & 0\\
        0 & M
    \end{pmatrix}\Big|\, R\in SO(2)\,,\,M\in SO(2,1)\right\rbrace \subset O(4,1)\,.
\end{equation}

We now look for a function $F^{(\bar s,s)}(X_0,X_1,X_2,Z)$ that is
\begin{itemize}
    \item defined on $X_i\cdot X_{i} = X_2\cdot Z  = X_0^{\perp} = X_1^{\perp} = 0$,
    \item homogeneous of degrees $(-\Delta_{s},-\Delta_{-\bar s},-2,1)$ in $(X_0,X_1,X_2,Z)$ respectively,
    \item transverse: $F^{(\bar s,s)}(X_0,X_1,X_2,Z+aX_2) = F^{(\bar s,s)}(X_0,X_1,X_2,Z)$ for arbitrary $a\in\mathbb{C}$.
\end{itemize}
Moreover, the function has to be $\mathrm{SL}(2,\mathbb{R})\times \mathrm{U}(1)$-invariant:
\begin{equation}
    F^{(\bar s,s)}(MX_0^{\parallel}, MX_1^{\parallel}, RX_2^{\perp}, MX_2^{\parallel}, RZ^{\perp}, MZ^{\parallel}) = F^{(\bar s,s)}(X_0,X_1,X_2,Z)\,.
\end{equation}
for all $R\in SO(2)$ and $M\in SO(2,1)$.

Focusing first on the $Z$-dependence, there are only three linearly independent combinations that satisfy all the requirements, they are
\begin{equation}
    \mathcal{Q}_1 = \frac{Z\cdot X_1}{X_2\cdot X_1} - \frac{Z\cdot X_0}{X_2\cdot X_0}\,,\quad \mathcal{Q}_2 = \frac{2 Z^\perp\cdot X_2^\perp}{X_2^{\perp}\cdot X_2^{\perp}} - \frac{Z\cdot X_0}{X_2\cdot X_0} - \frac{Z\cdot X_1}{X_2\cdot X_1}\,,\quad  \mathcal{Q}_3 = \frac{Z^1 X^2_2 - Z^2 X^1_2}{X_2^{\perp}\cdot X_2^{\perp}}\,.
\end{equation}
And the $\mathrm{SL}(2,\mathbb{R})\times \mathrm{U}(1)$-invariant quantity is
\begin{equation}
    U = \frac{\det( X_0^{\parallel}, X_1^{\parallel}, X_2^{\parallel})}{\sqrt{-2\,(X_0\cdot X_1)(X_1\cdot X_2)(X_2\cdot X_0)}}\,,\quad 1-U^2 = \frac{(X_2^\parallel\cdot X_2^\parallel) (X_0\cdot X_1)}{2\, (X_0\cdot X_2)(X_1\cdot X_2)}\,,
\end{equation}
where the two relations are equivalent only because $X_0^\perp = X_1^\perp = 0$. We point out that the allowed structures
\begin{equation}
    \mathcal{Q}_4 = \det( Z^{\parallel}, X_1^{\parallel}, X_2^{\parallel})\quad \text{and}\quad \mathcal{Q}_5 = \det( X_0^{\parallel},Z^{\parallel}, X_2^{\parallel})
\end{equation}
are related to the the previous ones through
\begin{equation}
     \frac{\mathcal{Q}_5}{X_2\cdot X_0} - \frac{\mathcal{Q}_4}{X_2\cdot X_1}= \sqrt{\frac{2\,(X_1\cdot X_2)(X_2\cdot X_0)}{-X_0\cdot X_1}} U \mathcal{Q}_1
\end{equation}
and
\begin{equation}
    \frac{\mathcal{Q}_4}{X_2\cdot X_1} + \frac{\mathcal{Q}_5}{X_2\cdot X_0} = \sqrt{\frac{2\,(X_1\cdot X_2)(X_2\cdot X_0)}{-X_0\cdot X_1}} \frac{U^2-1}{U} \mathcal{Q}_2\,.
\end{equation}
These relations can be proved as follows. First, one notices that the four three-dimensional vectors $Z^\parallel$, $X_0^{\parallel}$, $X_1^{\parallel}$, and $X_2^{\parallel}$ are trivially linearly related through
\begin{equation}
    \det( X_0^{\parallel}, X_1^{\parallel}, X_2^{\parallel}) Z^\parallel = \mathcal{Q}_4\, X_0^{\parallel} + \mathcal{Q}_5\, X_1^{\parallel} + \det( X_0^{\parallel}, X_1^{\parallel}, Z^{\parallel}) X_2^{\parallel}\,.
\end{equation}
Second, taking the scalar product of this equation with $X^{\parallel}_0$, $X^{\parallel}_1$, and $X^{\parallel}_2$ gives
various equations that are easily manipulated to yield the two relations we are looking for. 

Once the $Z$-dependence is understood, it suffices to satisfy the homogeneity requirements. This is immediate and we obtain
\begin{equation}\label{conformal J1 embedding}
    F^{(\bar s,s)}(X_0,X_1,X_2,Z) = \frac{g_{\bar s,s}(U) \mathcal{Q}_1 + h_{\bar s,s}(U) \mathcal{Q}_2 + k_{\bar s,s}(U) \mathcal{Q}_3}{(-X_0\cdot X_1)^{\frac{\Delta_{-\bar s} + \Delta_{s} - 1}2} (-X_2\cdot X_0)^{\frac{1 + \Delta_{s} - \Delta_{-\bar s} }2} (-X_2\cdot X_1)^{\frac{1+\Delta_{-\bar s} - \Delta_{s}}2}}\,.
\end{equation}

In order to return to physical, three-dimensional space, one should simply evaluate at
\begin{equation}
    X_i = \left(x_i^\mu,\frac{1-|x_i|^2}2,\frac{1+|x_i|^2}2\right)\,,\quad Z = \left(\zeta^\mu,-\zeta\cdot x_2, \zeta\cdot x_2\right)\,.
\end{equation}
Evaluating \eqref{conformal J1 embedding}, one obtains \eqref{conformal J1}.

\paragraph{Higher Spins} The only difference when considering correlators involving higher-spin operators is that now we need to find a function that is homogeneous of degrees $\tilde{s}$ in $Z$ and $-\tilde{s}-1$ in $X_2$. It is easy to convince oneself that the only way to satisfy the requirements on the $Z$ dependence is by taking products of the previous structures. Moreover, the additional relations
\begin{equation}
    \mathcal{Q}^2_3 = \frac{(Z^\perp\cdot Z^\perp)(X_2^\perp\cdot X_2^\perp) - (Z^\perp\cdot X_2^\perp)^2}{(X_2^\perp\cdot X_2^\perp)^2} = \frac{\mathcal{Q}^2_1}{4(U^2 - 1)} - \frac{\mathcal{Q}^2_2}{4 U^2}
\end{equation}
restricts the number of independent structures to $2\tilde{s} +1$ as in \eqref{conformal Js}.

\section{Correlators with $J_1$} \label{app:corrwithj1}

In this appendix we bootstrap correlators with the spin-one current (\ref{conformal J1}) for all boundary operators. We do so by imposing the invariance of the correlator under transverse translations and the divergence-less condition $\p_\mu J^\mu_1=0$. As for $J_0$, we find that the correlator with $J_1$ is completely fixed in terms of the parameters $\Delta$ and $a_2$, (up to a normalisation-dependent constant).

First, we impose the divergence-less condition on \eqref{conformal J1}. This reads
\begin{equation}\label{divpos}
    u\, h'_{\bar s,s}(u) + \left[1-\frac{|\bar s + s| u^2}{1-u^2}\right] h_{\bar s,s}(u) = (\Delta_s - \Delta_{-\bar s}) g_{\bar s,s}(u) + \ii \frac{(\bar s + s)}{1-u^2} k_{\bar s,s}(u)\,,
\end{equation}
where $\Delta_s$ is defined in $\eqref{spectrumR}$.

\subsection{Spins of the Same Sign} \label{eqlsignJ1}

For $\bar s,s>0$ (\ref{conformal J1}) takes the form
\begin{equation} \label{corrpos}
    \langle \mathcal{M}_{10}^{(\bar{s},s)} J_1^{\mu}(x_2) \rangle = \frac{ (x_2^-)^{\bar{s}+s} \left[g_{\bar{s},s}(u) Q_1^{\mu} + h_{\bar{s},s}(u) Q_2^{\mu} + k_{\bar{s},s}(u) Q_3^{\mu}\right]}{|x_{20}|^{3-2\Delta+2s} |x_{21}|^{2\Delta+2\bar{s}-1}}\,.
\end{equation}
The invariance of this correlator under constant translation in the plus transverse direction reads
\begin{equation}\label{varpos}
    \delta_+ \langle \mathcal{M}_{10}^{\left(\bar{s},s\right)} J_1^{\mu}(x_2) \rangle = \langle \mathcal{M}_{10}^{\left(\bar{s}+1,s\right)} J_1^{\mu}(x_2) \rangle + \langle \mathcal{M}_{10}^{\left(\bar{s},s+1\right)} J_1^{\mu}(x_2)\rangle+\langle\mathcal{M}_{10}^{\left(\bar{s},s\right)}\p_+J_1^{\mu}(x_2)\rangle = 0\,.
\end{equation}
The general solution of $\eqref{varpos}$ and $\eqref{divpos}$ takes the form
\begin{equation}
    g_{\bar{s},s}(u) = \frac{\widetilde w_{\bar s, s} (s+\bar s+1)}{(s-\bar s + 2 -2\Delta) }\,,\quad
     h_{\bar{s},s}(u) = \widetilde{w}_{\bar s,s} + \frac{w_{\bar s,s}}{u}\,,\quad 
 k_{\bar s, s}(u)  = \ii\, h_{\bar s, s}(1/u)\,,
\end{equation}

with
\begin{align}\la{rec1}
    \widetilde w_{\bar{s},s} &= 2^{s+\bar s-2}\frac{(s-\bar s + 2-2\Delta)  \Gamma(\frac{3}{2}+s-\Delta) \Gamma(\Delta+\bar s -\frac{1}{2})}{(1-\Delta) \Gamma(2-\Delta) \Gamma(\Delta)} \widetilde w_{\frac{1}{2},\frac{1}{2}}\,,\\
    w_{\bar{s},s} &= 2^{s+\bar s-1}\frac{   \Gamma (2+s-\Delta) \Gamma (\bar s+\Delta)}{\Gamma(\frac{5}{2}-\Delta) \Gamma(\frac{1}{2}+\Delta)} w_{\frac{1}{2},\frac{1}{2}}\,,\nn
\end{align}
where $w_{\frac{1}{2},\frac{1}{2}}$ and $\widetilde w_{\frac{1}{2},\frac{1}{2}}$ are two numerical coefficients. 
The solution with $\bar s,s<0$ is related to this one by the symmetry \eqref{Deltasym}. In this case we have
\begin{equation}
\langle\mathcal{M}_{10}^{(\bar{s}<0,s<0)}J_1^{\mu}\left(x_2\right)\rangle=\frac{\left(x_2^+\right)^{-\bar{s}-s}\left[g_{\bar{s},s}(u) Q_1^{\mu} + h_{\bar{s},s}(u)Q_2^{\mu}+k_{\bar{s},s}(u)Q_3^{\mu}\right]}{\left|x_{20}\right|^{2\Delta-2s-1}\left|x_{21}\right|^{3-2\Delta-2\bar{s}}}\,,
\end{equation}
with
\begin{align}\la{solmm}
    g_{\bar s<0,s<0} &= g_{-\bar s ,-s}\,,\nn\\
    h_{\bar s<0,s<0} &= h_{-\bar s ,-s}\,,\quad (\Delta \rightarrow 2-\Delta\,, \ w_{\frac12,\frac12}\rightarrow w_{-\frac12,-\frac12}\,, \widetilde w_{\frac12,\frac12}\rightarrow \widetilde w_{-\frac12,-\frac12})\,,\\ 
    k_{\bar s<0,s<0}&=- k_{-\bar s ,-s} \,. \nn
\end{align}

\subsection{Spins of Opposite Signs}
\label{uneqlsignJ1sp}

For $\bar s+ s \geqslant 0 $ and $\bar s>0 $  we have
\begin{align} \label{j1corr}
\langle\mathcal{M}_{10}^{(\bar{s}>0,s<0)}J_1^{\mu}\left(x_2\right)\rangle & =\frac{\left(x_2^-\right)^{\bar{s}+s}\left[g_{\bar{s},s}(u)Q_1^{\mu}+h_{\bar{s},s}(u)Q_2^{\mu}+k_{\bar{s},s}(u)Q_3^{\mu}\right]}{\left|x_{10}\right|^{2\Delta-2s-2}\left|x_{20}\right|\left|x_{21}\right|^{1+2\bar{s}+2s}}\,.
\end{align}
The variation equation $\delta_-\langle\mathcal{M}_{10}^{\left(\bar{s}+1,s\right)}J_1^{\mu}\left(x_2\right)\rangle=0$
takes the form
\begin{equation}\la{varj1}
    \langle \delta_- \overline{\mathcal{O}}_{\bar{s}+1} (x_1) \mathcal{W} \mathcal{O}_{s}(x_0) J_1^{\mu}(x_2) \rangle + \langle \mathcal{M}_{10}^{\left(\bar{s}+1,s-1\right)} J_1^{\mu}(x_2) \rangle + \langle \mathcal{M}_{10}^{\left(\bar{s}+1,s\right)}\p_-J_1^{\mu}\left(x_2\right)\rangle=0\,.
\end{equation}

To solve this equation and (\ref{divpos}), we first consider small values of $S=\bar s+s$. Based on the first few cases, we give an   
ansatz for the functional form of the correlator for generic $S$. We then used the variation equations to fix the few remaining coefficients.

Consider the first term in (\ref{varj1}). It takes the form
\begin{align}\la{bdrcofs}
    \delta_- \overline{\mathcal{O}}_{\bar{s}+1} =& -\frac12 \delta^2_3\overline{\mathcal{O}}_{\bar{s}} + \sum_{k=-\bar s}^0\frac{\widetilde\alpha^{(1)}_{\bar{s},k}}{N} :\!(\p_+^{\bar{s}+k}J_1^3) \overline{\mathcal{O}}_{-k}\!:\\
    &+ \sum_{k=1 - \bar s}^0 \Big[\frac{\widetilde\alpha^{(2)}_{\bar s, k}}{N} :\!(\p_+^{\bar{s}+k-1}\p_3 J_1^-) \overline{\mathcal{O}}_{-k}\!: +\frac{\widetilde\alpha^{(3)}_{\bar s, k}}{N}:\!(\p_+^{\bar{s}+k-1}J_1^-) \delta_3\overline{\mathcal{O}}_{-k}\!:\Big] + \dots\,,\nn
\end{align}
where the coefficients $\widetilde\alpha^{(i)}_{\bar{s},s}$ are of order $O(N^0)$ in the planar limit. Here, the dots stand for terms that are subleading or that do not contribute to the correlator (\ref{j1corr}).

Let us consider first the $S=0$ case. The only terms in \eqref{bdrcofs} that will contribute to the correlator are
\begin{equation}\label{J1eom}
    \delta_- \overline{\mathcal{O}}_{\bar{s}+1} = -\frac12 \delta^2_3\overline{\mathcal{O}}_{-s} + \frac{\widetilde\alpha^{(1)}_{\bar s,s}}{N} :\!J_1^3 \overline{\mathcal{O}}_{-s}\!: + \dots\,.
\end{equation}
Hence, in this case \eqref{varj1} reduces to
\begin{multline}\label{varj1-2}
    -\frac12 \p_{x_1^3}^2\langle \mathcal{M}_{10}^{(-s,s)} J_1^{\mu}(x_2) \rangle + \frac{\widetilde\alpha^{(1)}_{-s,s}}{N} \langle J_1^3(x_1) J_1^{\mu}(x_2) \rangle \langle \mathcal{M}_{10}^{(-s,s)}\rangle\\
    + \langle \mathcal{M}_{10}^{\left(-s+1,s-1\right)} J_1^{\mu}(x_2)\rangle + \langle\mathcal{M}_{10}^{(-s+1,s)} \p_-J_1^{\mu}(x_2)\rangle = 0\, ,
\end{multline}
which involves the two-point function of $J_1$, given in our convention by
\begin{equation}
    \<J^\mu_1(x)J^\nu_1(0)\> = 4 N \mathcal{N}_1 \frac{2\,x^{\mu}x^{\nu} - \delta^{\mu\nu} |x|^2}{|x|^6}\,.
\end{equation}
To fix the functional dependence of the functions  $g_{\bar{s},s},h_{\bar{s},s}$ and $k_{\bar{s},s}$ we need to solve \eqref{varj1-2} and the divergence-less condition \eqref{divpos}. The solution that is regular at $u\rightarrow1^-$ is  
\begin{align} \label{J1sol}
g_{-s,s}(u) &= \tilde b_s \bar{F}_1(u) + 
{b}_s \bar{F}_2(u)\,,\qquad h_{-s,s}(u) =0\,, \\
k_{-s,s}(u) &= \frac{\ii}{2}\left[\tilde b_s\frac{(1-2\Delta_s)}{\Delta_s}\bar{F}_1(u)-u{b}_s + 2 (u^2-1) {b}_s \bar{F}_2(u) + \frac{\kappa_{s}\text{Arccos}(u)}{\sqrt{1-u^2}}\right],\nonumber \\
g_{-s+1,s}(u) &= (2\Delta_{s}+1) g_{-s,s}(u)\,,\quad\nonumber 
h_{-s+1,s}(u) = - \frac{\tilde b_{s} +{\kappa}_{s}(1+2\Delta_{s})}{2u} - \tilde b_s \bar F_1(u)\,,\nonumber \\
k_{-s+1,s}(u) &= \ii(2\Delta_s+1) \left[ (u^2-1) b_s \bar{F}_2(u)-\frac{b_s}{2} u - \frac{2\Delta_{s}\tilde b_s}{(2\Delta_{s}+1)}\bar{F}_1(u)\right]\,,\nonumber
\end{align}
where $\Delta_{s}=\Delta-s-\frac12$ and we have used the following shorthand notations
\begin{align}
    \bar{F}_1(u) &\equiv \frac{\Delta_s}{1-2\Delta_s}\,_2F_1\! \left(1,1-2\Delta_{s};\frac32-\Delta_{s};\frac{1-u}2\right), \quad\,\,\, \kappa_{s}\equiv-\mathcal{N}_1{\widetilde\alpha}^{(1)}_{-s,s} c_+ 2^{s+\frac32} \frac{\Gamma (2\Delta_s)}{\Delta_s \Gamma(2\Delta)}\, , \nn\\
    \bar{F}_2(u) &\equiv \frac{2\Delta_s-1}{4 (\Delta_s-1)}\,_2F_1\! \left(1,2-2\Delta_{s};2-\Delta_{s};\frac{1-u}2\right), \quad b_{s}\equiv \kappa_s + \tilde b_s\, .
\end{align}

Note that there are only two undetermined coefficients, say $\tilde b_{s}$ and ${\kappa}_{s}$. 
Solving \eqref{varj1-2} also provides us with recursion relations on these coefficients, which imply
\begin{equation}\label{j1rel}
    \frac{b_s}{b_{-\frac{1}{2}}} = \frac{\tilde{b}_s}{\tilde b_{-\frac{1}{2}}} = \frac{\kappa_s}{\kappa_{-\frac{1}{2}}} = \frac{ 2^{s+\frac{1}{2}} \Gamma (1-2 \Delta )}{\Gamma (2 s-2 \Delta +2)}\, .
\end{equation}

Next, we consider the case $S=1$. Keeping only the terms with $k = s= 1 - \bar s$ in $\eqref{bdrcofs}$, we get
\begin{multline} \la{S1case}
\delta_- \overline{\mathcal{O}}_{\bar{s}+1} = -\frac12 \delta^2_3\overline{\mathcal{O}}_{-s+1} + \frac{\widetilde\alpha^{(1)}_{-s+1,s}}{N} :\!J_1^3 \overline{\mathcal{O}}_{-s}\!: + \\\, +\frac{\widetilde\alpha^{(2)}_{-s+1, s}}{N} :\!\p_3 J_1^- \overline{\mathcal{O}}_{-s}\!: +\frac{\widetilde\alpha^{(3)}_{-s+1, s}}{N}:\!J_1^- \delta_3\overline{\mathcal{O}}_{-s}\!:+\dots\,.
\end{multline}

Note that the functions $g_{-s+1,s}(u)$, $h_{-s+1,s}(u)$ and $k_{-s+1,s}(u)$ have already been determined above. 
By additionally imposing the regularity of the solution at $u\rightarrow1^-$ we can fix the functional form of the solution. Instead of presenting it, we give the general $S$ ansatz directly. By iteratively increasing $S$ we find that $g_{-s+S,s}(u)$ is always a linear combination of $\bar F_1(u)$ and $\bar F_2(u)$ introduced above and that $h_{-s+S,s}(u)$ is a linear combination of $\bar F_1(u)$ and $1/u$. Using this as an ansatz for the generic $S$ case, we can relate the unknown coefficients. We find 
\begin{align}\la{j1Ssol}
    g_{\bar{s},s} =&\  2^{s+\bar s-1}\big[(2\Delta_{s}+s+\bar s)\widetilde C_{\bar s, s} \tilde b_s\bar{F}_1(u)+ 2 C_{\bar s,s } b_s\bar{F}_2(u)\big],\\h_{\bar{s},s} = &-2^{s+\bar s-1}\!\left[(\bar s+ s) \widetilde C_{\bar s, s} \tilde b_s \bar{F}_1(u)+\frac{1}{u}\! \left(\!\kappa_{s} ( C_{\bar s, s}-\delta_{\bar s,-s}) + \tilde b_{s} ( C_{\bar s, s}-\Delta_s \widetilde C_{\bar s, s})\!\right)\!\right], \nonumber
    \\ k_{\bar{s},s}=&-\ii\,2^{s+\bar s-1}\bigg[2 C_{\bar s, s}  \left(1-u^2\right) b_s\bar{F}_2(u)+ b_s C_{\bar s, s} u\\
    &\qquad\qquad\,\, + \widetilde C_{\bar s,s}(2\Delta_{s}+s+\bar s-1)\tilde b_s\bar{F}_1(u)-\frac{\delta_{\bar s,-s}\kappa_{s} \text{Arccos}(u)}{\sqrt{1-u^2}}\bigg]\,, \nonumber
\end{align}
where the coefficients $C_{\bar s, s}$ and $\widetilde C_{\bar s, s}$ are
\begin{equation}
    C_{\bar s,s} = \frac{\Gamma(\Delta_{-\bar s}+\frac12)}{ \Gamma(\Delta_{s}+\frac12)}\,,\quad \widetilde C_{\bar s, s}=\frac{\Gamma(\Delta_{-\bar s})}{\Gamma(\Delta_{s}+1)}\,.
\end{equation}
We point out that the coefficients $\widetilde\alpha^{(i)}_{\bar s, s}$ appearing in \eqref{bdrcofs} can also be expressed in terms of $\kappa_{-\frac12}$ and $\tilde b_{-\frac{1}{2}}$.

Next, for $\bar s + s\leqslant 0$ and $\bar s \geqslant 0$ we have
\begin{equation}
\langle\mathcal{M}_{10}^{\left(\bar{s},s\right)}J_1^{\mu}\left(x_2\right)\rangle =\frac{\left(x_2^+\right)^{-\bar{s}-s}\left[g_{\bar{s},s}(u)Q_1^{\mu}+h_{\bar{s},s}(u)Q_2^{\mu}+k_{\bar{s},s}(u)Q_3^{\mu}\right]}{\left|x_{10}\right|^{2\Delta+2\bar{s}-2}\left|x_{20}\right|^{1-2s-2\bar{s}}\left|x_{21}\right|}\,.
\end{equation}
Following a similar derivation as above, we find that the solution is related to the case where $\bar s+ s\geqslant 0 $ as
\begin{equation}
    g_{\bar s>0,s<0} = g_{-s,-\bar s}\, ,\quad h_{\bar s>0,s<0} = -h_{ -s, -\bar s}\, ,\quad \text{and}\quad k_{\bar s>0,s<0} = k_{ -s, -\bar s}\, .
\end{equation}

Similarly, using (\ref{Deltasym}), we find for $\bar s <0$ and $s>0$  that
\begin{align}
    g_{\bar s<0,s>0} &= g_{-\bar s ,-s}\,,\\ h_{\bar s<0,s>0} &=- h_{-\bar s ,-s}\,,\quad (\Delta \nonumber\rightarrow 2-\Delta\,, \tilde b_{-\frac12}\rightarrow \tilde b_{\frac12}\,, \kappa_{-\frac12} \rightarrow\kappa_{\frac{1}{2}} )\,,\\ k_{\bar s<0,s>0}&=- k_{-\bar s ,-s} \,. \nonumber
\end{align}

In summary, we have determined all the correlators with $J_1$ and boundary spins of opposite signs in terms of four undetermined coefficients, $\tilde b_{\pm\frac{1}{2}}$, $\kappa_{\pm\frac12}$. In the next sub-section we relate them to the four coefficients from the case where the boundary spins are of the same sign.

\subsection{Relating the Normalisations Using Variations with Displacement}
\label{cons-var}

Here, we consider transverse variations that involve the displacement operator. We use the corresponding constraints to fix the parameters in the solutions above ($w_{\pm\frac12 ,\pm\frac12}\,,\widetilde w_{\pm\frac12 ,\pm\frac12}$, $\kappa_{\pm \frac12}$ and $\tilde b_{\pm\frac12}$).

Consider the variation 
\begin{multline}\la{j1disp}
\delta_+\langle\mathcal{M}_{10}^{(\frac12,-\frac12)}J_1^{\mu}(x_2)\rangle=\langle\mathcal{M}_{10}^{(\frac32,-\frac12)}J_1^{\mu}\left(x_2\right)\rangle+\langle\overline{\mathcal{O}}_{\frac12}\mathcal{W}\delta_+\mathcal{O}_{-\frac12}J_1^{\mu}\left(x_2\right)\rangle+\langle\mathcal{M}_{10}^{(\frac12,-\frac12)} \p_+J_1^{\mu}\left(x_2\right)\rangle \\
+ \eta_+ |x_{10}|\! \int\limits_{\tilde{\epsilon}}^{1-\tilde{\epsilon}}\!\dd \sigma\,\langle\mathcal{M}_{1\sigma}^{(\frac12,\frac12)}J_1^{\mu}\left(x_2\right)\rangle\langle\mathcal{M}_{\sigma0}^{(\frac12,-\frac12)}\rangle+ \frac{\nu_+|x_{10}|}{N \mathcal{N}_1} \<\cM^{(\frac12,-\frac12)}_{10}\> \! \int\limits_0^1\!\dd\sigma\,\langle J^-_1(x_\sigma)J^\mu_1(x_2)\rangle  \, =0.
\end{multline}
Here, the last two terms represent the integration of the displacement operator, $\mathbb{D}_+$ in $\eqref{displace}$, along the line and $\tilde{\epsilon}$ is a point splitting cutoff. We focus here on the $\mu=+$ component as it turns out to be sufficient for fixing the coefficients. The expression for the first term can be constructed using section \ref{eqlsignJ1} and $\langle\cM_{10}^{(\frac12,-\frac12)}\rangle=c_+/|x_{10}|^{2\Delta}$. It is given by 
\begin{multline}\la{firsttint}
    \int\limits_{\tilde{\epsilon}}^{1-\tilde{\epsilon}}\!\!\dd\sigma\,\langle\mathcal{M}_{1\sigma}^{(\frac12,\frac12)} J_1^+(x_2) \rangle\langle\mathcal{M}_{\sigma0}^{(\frac12,-\frac12)}\rangle=\frac{c_+}{|x_{10}|^{2\Delta} |x_{21}|^{2\Delta}} \int\limits_{\tilde{\epsilon}}^{1-\tilde{\epsilon}}\!\,\frac{\dd\sigma}{\sigma^{2\Delta} |x_{2\sigma}|^{4-2\Delta}} \bigg[\frac{w_{\frac12,\frac12} x^3_{21}x^3_{2\sigma}}{|x_{21}| |x_{2\sigma}|}\\
    + \widetilde w_{\frac12,\frac12} \bigg(\! 1-\frac{(\Delta-2)x_2^+x_2^-}{(\Delta-1) |x_{2\sigma}|^2} - \frac{x_2^+x_2^-}{ (\Delta-1) |x_{21}|^2}\!\bigg)\bigg]\,.
\end{multline}
The second integral reads
\begin{equation}
    \int\limits_0^1\!\dd\sigma\,\langle J^-_1(x_\sigma) J^+_1(x_2)\rangle = -4N\mathcal{N}_1 \int\limits_0^1\!\dd\sigma\,\frac{(x^3_{\sigma2})^2}{|x_{2\sigma}|^6}\,,
\end{equation}
where $x_\sigma=(0,0,x^3_0+\sigma x^3_{10})$. For $\Delta\geqslant 1/2$ the first integral contains divergences in $\tilde{\epsilon}$ that originate from the $\sigma\to0$ region of integration. They are canceled by the second term in \eqref{j1disp}, which otherwise has no finite contribution. The finite part of the integral with $\Delta\geqslant 1/2$ can be obtained by analytic continuation from $\Delta<1/2$ and the result can be written in terms of hypergeometric functions. The second integral has no divergences and can be written using an inverse trigonometric function. Instead of analysing equation \eqref{j1disp} for general kinematics, it suffices to consider the limit $2x_2^+x^-_2=r^2\rightarrow\infty$ to order $O(r^{-6})$. In this way we arrive at the relations
\begin{equation}\label{normres}
   \widetilde w_{\frac12,\frac12} =  \frac{2 (\Delta-1)^2 \tilde b_{-\frac12}}{ c_+ \eta_+ }\,,
   \quad\nu_+ = \frac{-\kappa_{-\frac12}}{2c_+ }\,,\quad b_{-\frac{1}{2}} = \tilde b_{-\frac12} + \kappa_{-\frac{1}{2}}= -\frac{2 c_+ \eta_+ w_{\frac{1}{2},\frac{1}{2}}}{(2\Delta-1)(2\Delta-3)}\,,
\end{equation}
where higher orders serve as consistency checks. The variations $\delta_-\<\cM_{10}^{(\frac12,-\frac12)}\,J^\mu_1(x_2)\> = 0$ and $\delta_\pm \< \cM^{(-\frac12,\frac12)}J^\mu_1(x_2)\>=0$    are treated similarly. Altogether, we find
\begin{equation} 
\frac{w_{-\frac12,-\frac12}}{w_{\frac12,\frac12}}=\frac{\widetilde w_{-\frac12,-\frac12}}{\widetilde w_{\frac12,\frac12}} =-{\eta_+\over\eta_-}\,,\qquad   
\frac{\kappa_{-\frac12}}{\kappa_{\frac12}} = \frac{\tilde b_{-\frac12}}{\tilde b_{\frac12}} = -{c_+\over c_-}\,,\qquad\nu_+=- \nu_-\,.
\end{equation}

In the main text, we have presented the correlator with boundary spins $\bar s=s=-\frac{1}{2}$ in terms of the constants $d_1$ and $\tilde{d}_1$, see \eqref{guess j1}. These are related to the constants appearing in this appendix through
\begin{equation}
    d_1 = \frac{2 w_{-\frac12,-\frac12}}{(2 \Delta -3) (2 \Delta -1) }\,,\qquad\tilde d_1 = \frac{- \widetilde w_{-\frac12,-\frac12}}{2(\Delta-1)^2}\, .
\end{equation}

In summary, all the correlators discussed in this appendix can be parameterised in terms of $\Delta$, $\nu_+$ and $d_1$. Additionally, using the relation 
\begin{equation}
    a_2 = \frac{32\ii \pi^2\eta_- d_3}{\mathcal{N}_1\sin(2\pi\Delta)}\nu_+ \,,
\end{equation}
from \eqref{divergence J3}, \eqref{div J3 final}, one can trade $\nu_+$ for $a_2$.

\section{Fixing the Coefficients in the Correlators with $J_1$ and $J_2$}
\label{app:can norm}

In this appendix, we compute the coefficients $d_1$, $\tilde{d}_1$, and $d_2$ that parameterise the correlators in (\ref{guess j1}) and (\ref{m12m12J2}).

The operators $J_1$ and $J_2$ are the $U(1)$ current and the stress-energy tensor. Therefore, they are exactly conserved.  We normalise the current $J_1$ so that the boundary operators of the mesonic line have $U(1)$ charge $\pm1$. As a consequence, the corresponding Ward identity reads
\begin{equation}
    \p_\mu \<\mathcal{M}_{10}^{(\frac12,-\frac12)} J_1^{\mu} (x_2)\> = \big(\delta^{(3)}(x_{21}) - \delta^{(3)}(x_{20})\big)\<\mathcal{M}_{10}^{(\frac12,-\frac12)}\>\,.
\end{equation}
Integrating over $x_2$ inside a ball of radius $\epsilon<|x_{10}|$ centered on $x_1$ and using the divergence theorem gives
\begin{equation}\label{Ward int J1}
    \epsilon^2 \int_{S^2} \dd^2 n\,n_\mu\<\mathcal{M}_{10}^{(\frac12,-\frac12)} J_1^{\mu} (x_1 + \epsilon n)\> = \<\mathcal{M}_{10}^{(\frac12,-\frac12)}\> = \frac{c_+}{|x_{10}|^{2\Delta}}\,.
\end{equation}
Using \eqref{conformal J1} and the results of appendix \ref{app:corrwithj1}, we have the explicit form of the correlator and we compute
\begin{equation}
    \epsilon^2 \< \mathcal{M}_{10}^{(\frac12,-\frac12)} J_1^{\mu} (x_2=x_1+\epsilon n)\>n_\mu = \frac{g_{\frac12,-\frac12}(\cos\theta)\left(1 - \epsilon\frac{n\cdot x_{20}}{|x_{20}|^2}\right)}{|x_{20}| |x_{10}|^{2\Delta - 1}}\,,
\end{equation}
where we used that $h_{\frac12,-\frac12} = 0$, and we recall that 
\begin{multline}
    g_{\frac12,-\frac12}(u) = c_+\eta_- \Bigg[ \frac{\Delta\,\tilde{d}_1}{1-2\Delta}\,_2F_1\!\Big(1,1-2\Delta;\frac32-\Delta;\frac{1-u}2\Big)\\
    + \frac{(2\Delta - 1) d_1}{4(\Delta - 1)}\,_2F_1\!\Big(1,2-2\Delta;2-\Delta;\frac{1-u}2\Big)\Bigg]\,.
\end{multline}
Taking for simplicity the limit $\epsilon \to 0$, equation \eqref{Ward int J1} becomes
\begin{equation}
    2\pi \int_{-1}^1 g_{\frac12,-\frac12}(u)\, \dd u = c_+ \quad\Longrightarrow\quad d_1 \cos^2(\pi\Delta) + \tilde{d}_1 \sin^2(\pi\Delta) = \frac{\sin(2\pi\Delta)}{2\pi^2\eta_-}\,.
\end{equation}
Together with the relation $\eta_-(\tilde{d}_1 - d_1) = 2\nu_+$ found in appendix \ref{app:corrwithj1}, this allows to fix the two coefficients as
\begin{equation}
    \eta_-\, d_1 = \frac{\sin(2\pi\Delta)}{2\pi^2} - 2 \nu_+ \sin^2(\pi\Delta)\quad \text{and}\quad \eta_-\, \tilde{d}_1 = \frac{\sin(2\pi\Delta)}{2\pi^2} + 2 \nu_+ \cos^2(\pi\Delta)\, .
\end{equation}

Similarly, we can use the fact that the stress tensor is canonically normalised. On the one hand, taking the divergence of the stress tensor gives $0$ when $x_2$ is in the bulk, but produces a contact term corresponding to the insertion of the displacement operator $\mathbb{D}_+ = \eta_+\, \cO_{\frac12}\times\overline \cO_{\frac12}$ when $x_2$ is on the line. Explicitly, if $x_0^3<x_2^3<x_1^3$, then
\begin{align}
    \p_\mu \<\mathcal{M}_{10}^{(-\frac12,-\frac12)} J_2^{\mu-} (x_2)\> & = - \delta^{(2)}(x_\perp) \eta_+ \langle\mathcal{M}_{12}^{\left(-\frac12,\frac12\right)}\rangle\langle\mathcal{M}_{20}^{(\frac12,-\frac12)}\rangle\nn\\
    & = - \delta^{(2)}(x_\perp) \frac{\eta_+c_+c_-}{|x_{12}|^{4-2\Delta} |x_{20}|^{2\Delta}} \label{divergence displacement}\,,
\end{align}
where $\delta^{(2)}(x_\perp) = \delta(x_2^1) \delta(x_2^2)$, and the coefficients $c_{\pm}$ have been introduced in \eqref{normalization M}. On the other hand, we have determined the correlator $\<\mathcal{M}_{10}^{(-\frac12,-\frac12)} J_2^{\mu-}(x_2)\>$ explicitly up to one overall constant $d_2$, see equations \eqref{conformal Js} and \eqref{guess js} or \eqref{eq1j2}. From our explicit result, when $x_0^3<x_2^3<x_1^3$ and $r^2 = 2\,x_2^+ x_2^-\to 0$, we find that
\begin{align}
    \<\mathcal{M}_{10}^{(-\frac12,-\frac12)} J_2^{+-} (x_2)\> &= O(r)\,,\qquad \<\mathcal{M}_{10}^{(-\frac12,-\frac12)} J_2^{3-} (x_2)\> = O(r^0)\, ,\\
    \<\mathcal{M}_{10}^{(-\frac12,-\frac12)} J_2^{--} (x_2)\> &= \frac{-2d_2 (2\Delta-3)(2\Delta-2)(2\Delta-1)}{x_2^+(x^3_{12})^{4-2\Delta} (x^3_{20})^{2\Delta}} + O(r^0)\,.
\end{align}
Using the relation $\p_{x_2^-} (1/x_2^+) = 2\pi \delta^{(2)}(x_\perp)$ and comparing with \eqref{divergence displacement}, we get
\begin{equation}
    d_2 = \frac{\eta_+ c_+ c_-}{4\pi (2\Delta-3) (2\Delta-2) (2\Delta-1)} = - \frac{\sin(2\pi\Delta)}{8\pi^2\, \eta_-}\,,
\end{equation}
where we used relation \eqref{2pt displacement} for the last equality.

Notice that in this canonical normalisation, we can combine \eqref{csc0}, \eqref{d1 explicit}, and \eqref{d2 explicit} to derive
\begin{equation}\label{N1N2 can norm}
    16\mathcal{N}_2 = \frac{\mathcal{N}_1}{1+4\pi^4 \nu_+^2}\, .
\end{equation}

\section{Examples of Correlators and Useful Integrals}
\label{app:useful correlators}

For convenience, we collect here the correlators that are relevant for the computations of section \ref{constraining}. 

Among the correlators involving $J_1$, we need
\begin{equation}\label{eq1j1}
    \<\mathcal{M}_{10}^{(-\frac12,-\frac12)} J^+_1(x_2)\> =  \left[d_1 \mathcal{B}_1 + \tilde{d}_1 \left(\p_{x_0^-} - \p_{x_1^-}\right) \mathcal{B}_2\right]
\end{equation}
and
\begin{equation}\label{eq2j1}
   \< \mathcal{M}_{10}^{(-\frac12,-\frac12)} (\p_- J_1^3 - \p_3 J^+_1)(x_2)\> =  \frac{|x_{10}|}{2\,x_2^+} \left[d_1 (\p_{x_0^-} - \p_{x_1^-}) \mathcal{B}_1 - 4\, \tilde{d}_1 \p_{x^-_1} \p_{x^-_0} \mathcal{B}_2\right]\, ,
\end{equation}
where the functions $\mathcal{B}_1$ and $\mathcal{B}_2$ were defined in \eqref{usefulfunc}.

Among the correlators involving $J_2$, we need
\begin{equation}\label{eq1j2}
    \< \mathcal{M}_{10}^{(-\frac12,-\frac12)} J^{++}_2(x_2)\> = d_2\left[2(\p_{x_0^-} - \p_{x_1^-}) \mathcal{B}_1 + (\p^2_{x_0^-} - 6\p_{x_0^-}\p_{x_1^-} + \p^2_{x_1^-}) \mathcal{B}_2\right]
\end{equation}
and
\begin{multline}\label{eq2j2}
    \< \mathcal{M}_{10}^{(-\frac12,-\frac12)} \left(\p_- J_2^{3+} - \p_3 J^{++}_2\right)(x_2)\>\\
    = \frac{d_2 |x_{10}|}{2\,x_2^+} \left[(\p^2_{x_0^-} - 6\p_{x_0^-}\p_{x_1^-} + \p^2_{x_1^-}) \mathcal{B}_1 + 8(\p_{x_1^-} - \p_{x_0^-}) \p_{x^-_1} \p_{x^-_0} \mathcal{B}_2 \right]\, .
\end{multline}

We remark in passing that, defining $\mathcal{\widetilde O}_{\tilde s} = \p_-J_{\tilde{s}}^{3+\dots+}-\p_3J_{\tilde{s}}^{++\dots+}$, the last formula generalises to arbitrary spins $\tilde{s}\geqslant 2$ and $\bar s, s< 0$ in the form
\begin{equation}
    \langle\mathcal{M}_{10}^{(\bar s,s)} \widetilde\cO_{\tilde s}(x_2)\rangle = \frac{d_{\tilde{s}} |x_{10}|}{2\,x_2^+} \p^{-\frac12-\bar s}_{x^-_1} \p^{-\frac12-s}_{x^-_0}\left[\mathcal{D}_{2,\tilde{s}}(\p_{x_0^-},\p_{x_1^-}) \mathcal{B}_1 - 4 \mathcal{D}_{1,\tilde{s}}(\p_{x_0^-},\p_{x_1^-}) \p_{x^-_1} \p_{x^-_0} \mathcal{B}_2 \right]\, ,
\end{equation}
where the polynomials $\mathcal{D}_{i,\tilde{s}}$ were defined in \eqref{diff}.

In order to compute the integrals in the right-hand side of \eqref{MZ J0}, we need the following applications of the star-triangle identity
\begin{align}
    \int \frac{\dd^3 x_3}{|x_{30}|^{2\Delta - 1} |x_{31}|^{3-2\Delta} |x_{32}|^4} &= -\frac{2\pi^2 |x_{10}|\, \mathcal{B}_2}{x_2^+ \tan(\pi\Delta)}\,,\\
    \int \frac{\dd^3 x_3(\Delta - 1)\,x_3^+ }{|x_{30}|^{2\Delta} |x_{31}|^{4-2\Delta} |x_{32}|^4} &= -\frac{\pi^2 \tan(\pi\Delta) |x_{10}|\, \mathcal{B}_1}{2\,x_2^+}\,.
\end{align}
Similarly, the integrals that appear in the right-hand side of \eqref{MZ J1} are also applications of the star-triangle identity:
\begin{align}\label{star-triangle J1}
    A_1 &= (1-2\Delta)(2\Delta - 3) \int\frac{\dd^3 x_3\, x_3^+ |x_{10}|}{|x_{30}|^{2\Delta+1} |x_{31}|^{5-2\Delta} |x_{32}|^2} = \frac{4\pi^2}{\tan(\pi\Delta)} \mathcal{B}_2\,,\\
    A_2 &= 4\Delta(\Delta-1)(2-\Delta)\! \int\frac{\dd^3 x_3\, (x_3^+)^2|x_{10}|}{|x_{30}|^{2\Delta+2} |x_{31}|^{6-2\Delta} |x_{32}|^2}= \pi^2 \tan(\pi\Delta) \mathcal{B}_1\,.
\end{align}

\section{Details for Smooth Line Deformations}\label{app:smooth}

In this appendix, we present details that complement the ones in section \ref{sec:smooth} regarding the bootstrap of the correlator for smooth paths, following the setup discussion in \cite{Gabai:2023lax}. We consider the correlator
\beq\la{corrJ0}
    \<[\,\overline{\cO}_{1\over2}\cW\cO_{-{1\over2}}]\,J_0\>\,.
\eeq
Under a three-dimensional conformal transformation that takes $x\to\tilde x$ and $\cO(x)\to\widetilde\cO(\tilde x)$, $J(x_2)\to\tilde J(\tilde x_2)$, the correlator (\ref{corrJ0}) remains invariant. Expressing $\widetilde\cO(\tilde x)$ and $\tilde J(\tilde x_2)$ in terms of $\cO(x)$ and $J(x_2)$ results in the boundary and bulk conformal weight factors
\begin{equation} \label{cftran}
    \frac{\<[\,{\overline{\cO}}_{1\over2}(\tilde x_1) \cW[\tilde x(\cdot)] \cO_{-{1\over2}} (\tilde x_0)]\,J_0(\tilde x_2)\>}{\<[\,\overline{\cO}_{1\over2}( x_1)\cW[ x(\cdot)]\cO_{-{1\over2}}( x_0)]\,J_0(x_2)\>}={e^{\ii(1/2-\Delta)(\theta_0-\theta_1)}\over\Omega^{\Delta}(x_1)\Omega^{\Delta}(x_0)\Omega^{2}(x_2)}\,,
\end{equation}
where
\begin{equation} \label{tran}
    \frac{\p \tilde x^\mu}{\p x ^\nu}=\Omega(x) \Lambda^\mu_\nu(x)\,,\qquad\text{det}\,\Lambda^\mu_\nu(x) = 1\,,
\end{equation}
and $\theta_0$, $\theta_1$ are the transverse rotation angles at the two endpoints on the line, see \cite{Gabai:2023lax} for more details. The bootstrap strategy is to write an effective action for the smooth line in an expansion around the straight line and fix the coefficients using (\ref{tran}). 

We first import the coefficients that were already fixed in \cite{Gabai:2023lax}. The remaining coefficients are $\gamma_6$, $\gamma_7$, $\tilde \gamma_6$, $\tilde\gamma_7$, $\Omega_\pm$ and $\Xi_\pm$ in (\ref{sordervarations}). To fix them, we find it sufficient to start from a straight line and impose (\ref{cftran}) at second order in a conformal transformation deformation parameter. This parameter is the vector, $\v_c$ in (\ref{deform}) and $\u\equiv\tilde x_2-x_2$. At this order, the numerator of the left-hand side of \eqref{cftran} takes the form
\begin{equation}
    \delta^2 \(\overline{\cO} \mathcal{W} \cO J \) = \delta^2 \(\overline{\cO} \mathcal{W} \cO\) + \overline \cO\cW\cO\delta J + \overline \cO\delta\cW\cO\delta J + \overline \cO\cW\delta\cO\delta J + \overline \cO\cW\cO\delta^2 J\, .
\end{equation}
The first term in the right-hand side is given by \eqref{second-order-list} in the main text. The remaining four contributions are given by
\begin{multline}
\u^\mu \p_{x_2^\mu} \bigg[\v_{1}^+\! \<\cM^{(\frac32,-\frac12)}_{10}\! J_0\>  + \v_{0}^-\! \<\cM^{(\frac12,-\frac32)}_{10}\! J_0\>  - \frac{\ii \eta_+ |x_{10}|}{2\sin(2\pi\Delta)}\underset{(0,1}{\Sint}\dd \sigma\,\v_{\sigma}^+ \<\cM^{(\frac12,\frac12)}_{1\sigma}\! J_0\>\<\cM^{(\frac12,-\frac12)}_{\sigma0}\>\\
    + \frac{\ii\eta_- |x_{10}|}{2\sin(2\pi\Delta)}\underset{0,1)}{\Rint}\dd \sigma\,\v_{\sigma}^- \<\cM^{(\frac12,-\frac12)}_{1\sigma}\>\<\cM^{(-\frac12,-\frac12)}_{\sigma0}J_0\> \bigg] + \frac{\u^\mu \u^\nu}{2} \p_{x_2^\mu} \p_{x_2^\nu} \<\cM^{(\frac12,-\frac12)}_{10}J_0\>\, .
\end{multline}

The denominator on the left-hand side in \eqref{cftran} is given by the correlator in \eqref{ssbnegative} with $\bar s= -s= 1/2$. We set $x_\sigma=(0,0,\sigma)$, and parameterise the conformal deformation as
\beq \la{vc}
\v_c^{i}(\sigma) = \mathfrak{c}^i_0 + \mathfrak{c}^i_{1}\sigma+ \mathfrak{c}^i_{2}\sigma^2\,,
\eeq
where $\mathfrak{c}^\pm_i$ are six parameters and $\v^3_c= 0$. Correspondingly, we have
\begin{multline}
\u^{+}	=(\mathfrak c_{0}^{+}+x_2^{3}(\mathfrak c_{1}^{+}+x_2^{3} \mathfrak c_{2}^{+})-2(x_2^{+})^{2}\mathfrak c_{2}^{-})\\
	+\frac{1}{4}\big[x_2^{+}(\mathfrak c_{1}^{-}(\mathfrak c_{1}^{+}(8x_2^{3}-2)+8\mathfrak c_{2}^{+}x_2^{3})-16\mathfrak c_{2}^{-}\mathfrak c_{2}^{+}(x_2^{3}-1)x_2^{3}+(\mathfrak c_{1}^{-})^{2}-(\mathfrak c_{1}^{+})^{2})\\
	+16(\mathfrak c_{2}^{-})^{2}(x_2^{+})^{3}-2x_2^{-}(2\mathfrak c_{2}^{+}x_2^{3}+\mathfrak c_{1}^{+})^{2}\big]+O(\mathfrak{c}^3) \,,
\end{multline}
\begin{multline}
    \u^3=\big[(x_2^{3})^{2}(\mathfrak c_{1}^{-}\mathfrak c_{1}^{+}+2\mathfrak c_{2}^{-}\mathfrak c_{2}^{+})-2\mathfrak c_{2}^{-}\mathfrak c_{2}^{+}(x_2^{3})^{3}+2\mathfrak c_{2}^{-}(x_2^{+})^{2}(2\mathfrak c_{2}^{-}x_2^{3}+\mathfrak c_{1}^{-})+2\mathfrak c_{1}^{+}\mathfrak c_{2}^{+}(x_2^{-})^{2}\\+x_2^{3}(4(\mathfrak c_{2}^{+})^{2}(x_2^{-})^{2}-\mathfrak c_{1}^{-}\mathfrak c_{1}^{+})
	-2x_2^{-}x_2^{+}(\mathfrak c_{2}^{-}(\mathfrak c_{1}^{+}-2\mathfrak c_{2}^{+}(x_2^{3}-1))+\mathfrak c_{1}^{-}(\mathfrak c_{1}^{+}+\mathfrak c_{2}^{+}))\big] \\
    -\big(x_2^{+}(2x_2^{3}\mathfrak c_{2}^{-}+\mathfrak c_{1}^{-})+x_2^{-}(2x_2^{3}\mathfrak c_{2}^{+}+\mathfrak c_{1}^{+})\big)+O(\mathfrak{c}^3)\,,
\end{multline}
and $\u^-= (\u^+)^*$. Moreover, the spin source is given by 
\begin{align}
    e^{\ii(1/2-\Delta)(\theta_0-\theta_1)}=& 1+(1/2-\Delta)(\mathfrak c^-_1 \mathfrak c^+_2 - \mathfrak c^+_1 \mathfrak c^-_2)+ O(\mathfrak{c}^3)\,,\\
    \Omega(x_0)^{-\Delta}\Omega(x_1)^{-\Delta}=&1-2\Delta(\mathfrak{c}_1^-+\mathfrak{c}_2^-)(\mathfrak{c}_1^+ +\mathfrak{c}_2^+)+ O(\mathfrak{c}^3)\,.\nn
\end{align}
We plug all these ingredients and the ones from sections \ref{j0normrel} and \ref{sec:smooth} into \eqref{cftran}. We did not attempt to solve the constraint \eqref{cftran} for arbitrary kinematics, but only looked at the limit $2x_2^+x_2^-=r^2\rightarrow\infty$ to order $O(r^{-6})$. The equations we found fixed the solution uniquely as being the one given in \eqref{bootsol}. Higher orders in $1/r$ served as consistency check.

\section{Protected Operators on the Line}
\label{protectedop}

Recall that the stress tensor has the displacement operator as a contact term on the line, namely
\begin{equation}\label{j2div}
\p_\mu J^{\mu-}_2(x)= -\delta^{(2)}(x_\perp)\,\mathbb{D}_+\,,
\end{equation}
where $x_\perp$ is the projection of $x$ to the transverse plane. In an analogous way, the higher-spin currents $J_{\tilde s}$ also have contact terms of the form
\beq\la{contact}
\p_\mu J^{\mu--...-}_{\tilde s}(x)=-\delta^{(2)}(x_\perp)\,\mathbb{D}^{(\tilde s-1)}\,.
\eeq
Here, $\mathbb{D}^{(\tilde s)}$ is called a {\it tilt} operator. It is a $SL(2,{\mathbb R})$ primary of dimension $\tilde s+1$ and transverse spin $\tilde s$. For any spin $\tilde s>1$, there are $\tilde s$ such operators, namely $\cO_{\tilde s-n+\frac12}\times\overline\cO_{n-\frac12}$ with $n=1,\ldots,\tilde s$. One linear combination corresponds to the operator $\mathbb{D}^{(\tilde s)}$
\beq\la{triltop}
\mathbb{D}^{(\tilde s)}=\sum^{\tilde s}_{n=1}a_n^{(\tilde s)}\,\cO_{\tilde s-n+\frac12}\times\overline\cO_{n-\frac12}\,,
\eeq
while the other $\tilde s-1$ independent operators can be associated with contact terms of the conformal descendants of $J_{\tilde s}$.

To fix the coefficients $a_n^{(\tilde s)}$ we consider correlation functions of $J_{\tilde s+1}$ with a mesonic line where the signs of the spins of its boundary operators are equal. Taking the divergence of $J_{\tilde s+1}$ and using \eqref{contact} produces a product of two factorised mesonic lines expectation values. On the other hand, the divergence of the correlator can be computed from the explicit form of the correlator given in section \ref{sec:hs}. By equating the two, we obtain the coefficients in (\ref{triltop}). Because we have fixed the relative normalisation of the correlators with different boundary spins, we can read the ratios between the $a_n^{(\tilde s)}$ for different $n$'s. In this way we find, for example,
\begin{eqnarray}\la{D23}
    \mathbb{D}^{\left(2\right)} &\propto& \mathcal{O}_{\frac32} \times \overline{\mathcal{O}}_{\frac12}-\mathcal{O}_{\frac12} \times \overline{\mathcal{O}}_{\frac32}\,,\\
    \mathbb{D}^{\left(3\right)} &\propto& \mathcal{O}_{\frac52} \times \overline{\mathcal{O}}_{\frac12}-\mathcal{O}_{\frac32} \times \overline{\mathcal{O}}_{\frac32}+\mathcal{O}_{\frac12} \times \overline{\mathcal{O}}_{\frac52}\,.\nn
\end{eqnarray}

It is not too hard to extrapolate the relative coefficients for the higher tilt operators. In the next sub-section we present the details for ${\mathbb D}^{(2)}$.

\subsection{The Tilt Operator $\mathbb{D}^{(2)}$}

Recall that in the presence of the line
\beq\la{dJ3}
\p_\mu J_3^{\mu--}(x)=-\delta^{(2)}(x_\perp)\(a_1^{(2)}\mathcal{O}_{\frac32}\times\overline{\mathcal{O}}_{\frac12}+a_2^{(2)}\mathcal{O}_{\frac12}\times\overline{\mathcal{O}}_{\frac32}\)\,.
\eeq
To extract $a_1^{(2)}$ and $a_2^{(2)}$ we respectively start from the correlators $\langle\mathcal{M}_{10}^{(-\frac32,-\frac12)}J_3^{\mu\nu\rho}(x)\rangle$ and $\langle\mathcal{M}_{10}^{(-\frac12,-\frac32)}J_3^{\mu\nu\rho}(x)\rangle$ that were constructed in section \ref{sec:hs} and take the divergence of $J_3$. 

On the one hand, using (\ref{dJ3}) together with (\ref{normalization M}), we have (for $x_0^3<x^3<x_1^3$)
\begin{align}\la{insertofdis}
    \p_{\mu}\langle\mathcal{M}_{10}^{(-\frac32,-\frac12)}J_3^{\mu--}(x)\rangle & = -\delta^{(2)}(x_\perp)a^{(2)}_1\frac{c_-c_+(\Delta-2)(2\Delta-5)}{|x-x_1|^{6-2\Delta}|x-x_0|^{2\Delta}}\,,\\
    \p_{\mu}\langle\mathcal{M}_{10}^{(-\frac12,-\frac32)}J_3^{\mu--}(x)\rangle & = -\delta^{(2)}(x_\perp)a^{(2)}_2\frac{c_-c_+\Delta(1+2\Delta)}{|x-x_1|^{4-2\Delta}|x-x_0|^{2\Delta+2}}\,.\nn
\end{align}

On the other hand, from our explicit result, we find that (for $x_0^3<x^3<x_1^3$)
\begin{align}
    \langle\mathcal{M}_{10}^{(-\frac32,-\frac12)}J_3^{+--}(x)\rangle &= O(|x_\perp|)\,, \qquad \langle\mathcal{M}_{10}^{(-\frac32,-\frac12)}J_3^{3--}(x)\rangle = O(|x_\perp|^0)\,,\\
    \langle\mathcal{M}_{10}^{(-\frac32,-\frac12)}J_3^{---}(x)\rangle &= 4d_3\frac{\Gamma(6-2\Delta)}{\Gamma(1-2\Delta)}\times\frac{1}{x^+|x-x_1|^{6-2\Delta}|x-x_0|^{2\Delta}}+O(|x_\perp|^0)\,,\nn
\end{align}
and
\begin{align}
    \langle\mathcal{M}_{10}^{(-\frac12,-\frac32)}J_3^{+--}(x)\rangle  &= O(|x_\perp|)\,, \qquad \langle\mathcal{M}_{10}^{(-\frac12,-\frac32)}J_3^{3--}(x)\rangle = O(|x_\perp|^0)\,,\\
    \langle\mathcal{M}_{10}^{(-\frac12,-\frac32)}J_3^{---}(x)\rangle &= -4d_3\frac{\Gamma\left(4-2\Delta\right)}{\Gamma\left(-1-2\Delta\right)}\times\frac{1}{x^+|x-x_1|^{4-2\Delta}|x-x_0|^{2\Delta+2}}+O(|x_\perp|^0)\,.\nn
\end{align}

Using the relation $\p_{x^-}\left(1/x^+\right)=2\pi\delta^{(2)}(x_\perp)$ for the divergence of these correlators and comparing the result with \eqref{insertofdis}, we conclude that $a^{(2)}_1=-a^{(2)}_2$.

\end{appendix}

\bibliography{Refs.bib}

\providecommand{\href}[2]{#2}\begingroup\raggedright\begin{thebibliography}{10}

\bibitem{Vasiliev:1990en}
M.~A. Vasiliev, \emph{{Consistent equations for interacting gauge fields of all
  spins in 3+1 dimensions}},
  \href{https://doi.org/10.1016/0370-2693(90)91400-6}{\emph{Phys. Lett. B}
  {\bfseries 243} (1990) 378}.

\bibitem{Vasiliev:1992av}
M.~A. Vasiliev, \emph{{More on equations of motion for interacting massless
  fields of all spins in 3+1 dimensions}},
  \href{https://doi.org/10.1016/0370-2693(92)91457-K}{\emph{Phys. Lett. B}
  {\bfseries 285} (1992) 225}.

\bibitem{Klebanov:2002ja}
I.~R. Klebanov and A.~M. Polyakov, \emph{{AdS dual of the critical O(N) vector
  model}}, \href{https://doi.org/10.1016/S0370-2693(02)02980-5}{\emph{Phys.
  Lett. B} {\bfseries 550} (2002) 213}
  [\href{https://arxiv.org/abs/hep-th/0210114}{{\ttfamily hep-th/0210114}}].

\bibitem{Giombi:2011kc}
S.~Giombi, S.~Minwalla, S.~Prakash, S.~P. Trivedi, S.~R. Wadia and X.~Yin,
  \emph{{Chern-Simons Theory with Vector Fermion Matter}},
  \href{https://doi.org/10.1140/epjc/s10052-012-2112-0}{\emph{Eur. Phys. J. C}
  {\bfseries 72} (2012) 2112}
  [\href{https://arxiv.org/abs/1110.4386}{{\ttfamily 1110.4386}}].

\bibitem{Aharony:2011jz}
O.~Aharony, G.~Gur-Ari and R.~Yacoby, \emph{{d=3 Bosonic Vector Models Coupled
  to Chern-Simons Gauge Theories}},
  \href{https://doi.org/10.1007/JHEP03(2012)037}{\emph{JHEP} {\bfseries 03}
  (2012) 037} [\href{https://arxiv.org/abs/1110.4382}{{\ttfamily 1110.4382}}].

\bibitem{Maldacena:2012sf}
J.~Maldacena and A.~Zhiboedov, \emph{{Constraining conformal field theories
  with a slightly broken higher spin symmetry}},
  \href{https://doi.org/10.1088/0264-9381/30/10/104003}{\emph{Class. Quant.
  Grav.} {\bfseries 30} (2013) 104003}
  [\href{https://arxiv.org/abs/1204.3882}{{\ttfamily 1204.3882}}].

\bibitem{Gabai:2022vri}
B.~Gabai, A.~Sever and D.-l. Zhong, \emph{{Line Operators in
  Chern-Simons\textendash{}Matter Theories and Bosonization in Three
  Dimensions}},
  \href{https://doi.org/10.1103/PhysRevLett.129.121604}{\emph{Phys. Rev. Lett.}
  {\bfseries 129} (2022) 121604}
  [\href{https://arxiv.org/abs/2204.05262}{{\ttfamily 2204.05262}}].

\bibitem{Gabai:2022mya}
B.~Gabai, A.~Sever and D.-l. Zhong, \emph{{Line operators in
  Chern-Simons-Matter theories and Bosonization in Three Dimensions II:
  Perturbative analysis and all-loop resummation}},
  \href{https://doi.org/10.1007/JHEP04(2023)070}{\emph{JHEP} {\bfseries 04}
  (2023) 070} [\href{https://arxiv.org/abs/2212.02518}{{\ttfamily
  2212.02518}}].

\bibitem{Gabai:2023lax}
B.~Gabai, A.~Sever and D.-l. Zhong, \emph{{Bootstrapping smooth conformal
  defects in Chern-Simons-matter theories}},
  \href{https://doi.org/10.1007/JHEP03(2024)055}{\emph{JHEP} {\bfseries 03}
  (2024) 055} [\href{https://arxiv.org/abs/2312.17132}{{\ttfamily
  2312.17132}}].

\bibitem{Giombi:2016ejx}
S.~Giombi, \emph{{Higher Spin \textemdash{} CFT Duality}},  in
  \emph{{Theoretical Advanced Study Institute in Elementary Particle Physics}:
  {New Frontiers in Fields and Strings}}, pp.~137--214, 2017,
  \href{https://arxiv.org/abs/1607.02967}{{\ttfamily 1607.02967}},
  \href{https://doi.org/10.1142/9789813149441_0003}{DOI}.

\bibitem{Giombi:2016zwa}
S.~Giombi, V.~Gurucharan, V.~Kirilin, S.~Prakash and E.~Skvortsov, \emph{{On
  the Higher-Spin Spectrum in Large N Chern-Simons Vector Models}},
  \href{https://doi.org/10.1007/JHEP01(2017)058}{\emph{JHEP} {\bfseries 01}
  (2017) 058} [\href{https://arxiv.org/abs/1610.08472}{{\ttfamily
  1610.08472}}].

\bibitem{Maldacena:2011jn}
J.~Maldacena and A.~Zhiboedov, \emph{{Constraining Conformal Field Theories
  with A Higher Spin Symmetry}},
  \href{https://doi.org/10.1088/1751-8113/46/21/214011}{\emph{J. Phys. A}
  {\bfseries 46} (2013) 214011}
  [\href{https://arxiv.org/abs/1112.1016}{{\ttfamily 1112.1016}}].

\bibitem{Aharony:2012nh}
O.~Aharony, G.~Gur-Ari and R.~Yacoby, \emph{{Correlation Functions of Large N
  Chern-Simons-Matter Theories and Bosonization in Three Dimensions}},
  \href{https://doi.org/10.1007/JHEP12(2012)028}{\emph{JHEP} {\bfseries 12}
  (2012) 028} [\href{https://arxiv.org/abs/1207.4593}{{\ttfamily 1207.4593}}].

\bibitem{Gur-Ari:2012lgt}
G.~Gur-Ari and R.~Yacoby, \emph{{Correlators of Large N Fermionic Chern-Simons
  Vector Models}}, \href{https://doi.org/10.1007/JHEP02(2013)150}{\emph{JHEP}
  {\bfseries 02} (2013) 150} [\href{https://arxiv.org/abs/1211.1866}{{\ttfamily
  1211.1866}}].

\bibitem{Giombi:2011rz}
S.~Giombi, S.~Prakash and X.~Yin, \emph{{A Note on CFT Correlators in Three
  Dimensions}}, \href{https://doi.org/10.1007/JHEP07(2013)105}{\emph{JHEP}
  {\bfseries 07} (2013) 105} [\href{https://arxiv.org/abs/1104.4317}{{\ttfamily
  1104.4317}}].

\bibitem{Nagar:2024mjz}
I.~Nagar, A.~Sever and D.-l. Zhong, \emph{{Planar RG flows on line defects}},
  \href{https://doi.org/10.1007/JHEP06(2024)110}{\emph{JHEP} {\bfseries 06}
  (2024) 110} [\href{https://arxiv.org/abs/2404.07290}{{\ttfamily
  2404.07290}}].

\bibitem{Costa:2011mg}
M.~S. Costa, J.~Penedones, D.~Poland and S.~Rychkov, \emph{{Spinning Conformal
  Correlators}}, \href{https://doi.org/10.1007/JHEP11(2011)071}{\emph{JHEP}
  {\bfseries 11} (2011) 071} [\href{https://arxiv.org/abs/1107.3554}{{\ttfamily
  1107.3554}}].

\bibitem{Billo:2016cpy}
M.~Bill\`o, V.~Gon\c{c}alves, E.~Lauria and M.~Meineri, \emph{{Defects in
  conformal field theory}},
  \href{https://doi.org/10.1007/JHEP04(2016)091}{\emph{JHEP} {\bfseries 04}
  (2016) 091} [\href{https://arxiv.org/abs/1601.02883}{{\ttfamily
  1601.02883}}].

\end{thebibliography}\endgroup
\bibliographystyle{JHEP.bst}

\end{document}